\documentclass[twocolumn,times,tighten]{aastex63}

\hypersetup{linkcolor=blue,citecolor=blue,filecolor=cyan,urlcolor=blue}

\def\smallplot#1{\centering \leavevmode
\includegraphics[width=89.5mm]{#1} }

\def\smallplottwo#1#2{\centering \leavevmode
\includegraphics[width=89.5mm]{#1}
 \hfil \includegraphics[width=89.5mm]{#2} }

\def\smallplotfour#1#2#3#4{\centering \leavevmode
\includegraphics[width=89.5mm]{#1} \includegraphics[width=89.5mm]{#2}
\hfil \includegraphics[width=89.5mm]{#3} \includegraphics[width=89.5mm]{#4}}

\def\xsmallplottwo#1#2{\centering \leavevmode
\includegraphics[width=80mm]{#1}
 \hfil \includegraphics[width=80mm]{#2} }

\def\largeplottwo#1#2{\centering \leavevmode
\includegraphics[width=130mm]{#1}
 \hfil \includegraphics[width=130mm]{#2} }
\graphicspath{{./}{figures/}}

\submitjournal{ApJ}
\received{October 31, 2019}
\revised{December 17, 2019}
\accepted{December 28, 2019}

\shorttitle{Infrared Properties of AGB stars in our Galaxy and the Magellanic Clouds}
\shortauthors{Suh} 

\begin{document}

\title{Infrared Properties of Asymptotic Giant Branch Stars in Our Galaxy and the Magellanic Clouds}

\correspondingauthor{Kyung-Won Suh}
\email{kwsuh@chungbuk.ac.kr}

\author[0000-0001-9104-9763]{Kyung-Won Suh}
\affiliation{Department of Astronomy and Space Science, Chungbuk National University, Cheongju-City, 28644, Republic of Korea}

\begin{abstract}
We investigate infrared properties of asymptotic giant branch (AGB) stars in
our Galaxy and the Magellanic Clouds using various infrared observational
data and theoretical models. We use catalogs for the sample of 4996 AGB stars
in our Galaxy and about 39,000 AGB stars in the Magellanic Clouds from the
available literature. For each object in the sample, we cross-identify the
2MASS, WISE, and Spitzer counterparts. To compare the physical properties of
O-rich and C-rich AGB stars in our Galaxy and the Magellanic Clouds, we
present IR two color diagrams (2CDs) using various photometric data. We
perform radiative transfer model calculations for AGB stars using various
possible parameters of central stars and dust shells. Using dust opacity
functions of amorphous silicate and amorphous carbon, the theoretical dust
shell models can roughly reproduce the observations of AGB stars on various
IR 2CDs. Compared with our Galaxy, we find that the Magellanic Clouds are
deficient in AGB stars with thick dust shells. Compared with the Large
Magellanic Cloud (LMC), the Small Magellanic Cloud (SMC) is more deficient in
AGB stars with thick dust shells. This could be because the Magellanic Clouds
are more metal poor than our Galaxy and the LMC is more metal rich than the
SMC. We also present IR properties of known pulsating variable. Investigating
the magnitude distributions at MIR bands for AGB stars in the Magellanic
Clouds, we find that the SMC is more deficient in the bright AGB stars at MIR
bands compared with the LMC.
\end{abstract}



\section{Introduction} \label{sec:intro}

It is generally believed that asymptotic giant branch (AGB) stars are low to
intermediate mass stars (0.5 - 10 $M_{\odot}$: for solar metallicity; the exact
value of the upper limit depends on the treatment of convection: e.g.,
\citealt{siess2006}) in the last evolutionary phases evolving rapidly from the
red giant branch into planetary nebulae. Most AGB stars are long-period
variables (LPVs) with large amplitude pulsations and they have circumstellar
dust envelopes with high mass-loss rates ($\dot{M} \sim 10^{-8} - 10^{-4}
M_{\odot}/yr$; \citealt{loup1993}; \citealt{suh2014}; \citealt{hofner2018}).

During the thermally pulsing AGB (TP-AGB) phase, the AGB stars show higher
mass-loss rates, produce dust grains effectively, and enrich the insterstellar
medium in metals and dust. AGB stars are classified as O-rich AGB (O-AGB) or
C-rich AGB (C-AGB) based on the chemistry of the photosphere and/or the outer
dust envelope. Circumstellar dust envelopes around AGB stars produce various IR
features. The spectral energy distributions (SEDs) of O-AGB stars show 10
$\mu$m and 18 $\mu$m features due to amorphous silicate dust. Low mass-loss
rate O-AGB (LMOA; $\dot{M} \sim 10^{-8} - 10^{-6} M_{\odot}/yr$) stars with
thin dust envelopes show the emission features and high mass-loss rate O-AGB
(HMOA; $\dot{M} \sim 10^{-5} - 10^{-4} M_{\odot}/yr$) stars with thick dust
envelopes show the absorption features at the same wavelengths (e.g.,
\citealt{suh1999}). The detailed SEDs of LMOA stars can be reproduced by the
silicate dust with a mixture of amorphous alumina (Al$_2$O$_3$;
\citealt{suh2016}) and Fe-Mg oxides \citep{thposch2002}. Featureless amorphous
carbon (AMC) dust with a mixture of SiC and MgS grains can reproduce the SEDs
For C-AGB stars (e.g., \citealt{suh2000}; \citealt{hony2002}).

During the AGB phase, the abundances of C, N, and O in the stellar atmosphere
can be changed by the episodic third dredge-up process after each thermal pulse
(e.g., \citealt{iben1983}). When AGB stars of intermediate mass range (1.55
$M_{\odot}$ $\leq$ M $<$ 4 $M_{\odot}$: for solar metallicity) go through the
carbon dredge-up process and thus the C/O ratio is larger than 1, the O-AGB
stars may become C-AGB stars (\citealt{groenewegen1995}). However, high mass
stars (4 $M_{\sun}$ $\leq$ M $<$ 10 $M_{\sun}$: for solar metallicity) may
become hot enough so that hot-bottom burning converts the C into $^{14}$N by
means of the CN cycle (\citealt{bloecker2000}) and these stars may remain
O-AGB, which are typical HMOA stars (or OH/IR stars) with thick dust envelopes
and high mass-loss rates.

Various IR observational data at NIR, MIR, and FIR bands are available from the
Infrared Astronomical Satellite (IRAS), Infrared Space Observatory (ISO),
Midcourse Space Experiment (MSX), AKARI, Two-Micron All-Sky Survey (2MASS),
Wide-field Infrared Survey Explorer (WISE), and Spitzer. These data have been
very useful to identify new AGB stars and understand the nature of them.

A catalog of AGB stars for 3003 O-AGB and 1168 C-AGB objects in our Galaxy was
presented by \citet{sk2011}. \citet{sh2017} presented a revised list of 3828
O-AGB and 1168 C-AGB stars. IR two-color diagrams (2CDs) have been useful to
study the properties of central stars and dust envelopes for a large sample of
AGB stars (e.g., \citealt{sk2011}; \citealt{suh2015}). \citet{suh2018}
presented various IR 2CDs using the IRAS, 2MASS, AKARI, and WISE data for the
AGB stars in our Galaxy.

Thanks to the optical gravitational lensing experiment (OGLE) projects
(\citealt{sus09}) and Spitzer Space Telescope Legacy program `Surveying the
Agents of a Galaxy Evolution' (SAGE; \citealt{meixner2006}), a much larger
number of AGB stars in the Large Magellanic Cloud (LMC) and Small Magellanic
Cloud (SMC) are identified and studied. Using the Infrared Spectrograph (IRS)
data on the Spitzer Space Telescope, \citet{jones2014} found that amorphous
silicate dust grains with contributions from amorphous alumina and metallic
iron, which is similar to the grain mixture for LMOA stars in our Galaxy,
provides a good fit to the observed spectra for a number of O-AGB stars in the
LMC. \citet{sloan2016} found that AMC dust with mixture of SiC and MgS grains
fit C-AGB stars in the LMC and SMC.

\citet{ventura2016} studied infrared colors of C-AGB stars in the Magellanic
Clouds and found the redder infrared colors of C-AGB stars in the LMC compared
to their counterparts in the SMC. \citet{groenewegen2018} investigated mass
loss and luminosity in a sample of AGB stars in our Galaxy, the Magellanic
Clouds, and other nearby galaxies. \citet{nanni2019} investigated the mass-loss
and dust production rates of C-AGB stars in the Magellanic Clouds and found a
tail of extreme mass-losing C-AGB stars in the LMC with low gas-to-dust ratios
that is not present in the SMC.

In this work, we investigate IR properties of AGB stars in our Galaxy and the
Magellanic Clouds for a large sample of the objects and compare them with
theoretical models. We present various IR 2CDs for the large sample of AGB
stars using the 2MASS, WISE, and Spitzer data. We use theoretical dust shell
models for AGB stars and compare the theory with the observations. We present
infrared properties of known pulsating variables in our Galaxy and the
Magellanic Clouds. For AGB stars in the Magellanic Clouds, we investigate
magnitude distributions at MIR bands. And we compare the IR properties of AGB
stars in our Galaxy and the Magellanic Clouds.

\section{Sample Stars\label{sec:sample}}

We use catalogs of AGB stars in our Galaxy and the Magellanic Clouds from the
available literature. Table~\ref{tab:tab1} lists the reference, total number of
objects, and numbers of the cross-identified 2MASS, WISE, and Spitzer
counterparts for each class.

\begin{table*}
\centering
\caption{Sample of AGB stars in our Galaxy and the Magellanic Clouds \label{tab:tab1}}
\begin{tabular}{llllllll}
\hline
\hline
Class  &Reference &Total Number & 2MASS & WISE & IRAC$^1$ & MIPS$^1$  \\
\hline
O-AGB (our Galaxy) & \citet{sh2017} & 3828  & 3828 & 3822  & 591 (192) & 784 (303)  \\
C-AGB (our Galaxy) & \citet{sh2017} & 1168  & 1168 & 1167 & 51 (7) & 67 (18) \\
\hline
O-AGB (LMC-OGLE3) & \citet{sus09} & 37,203  & 37,194  & 34,871 & 36,113 & 2540 \\
C-AGB (LMC-OGLE3) & \citet{sus09} & 9264  & 9257  & 9092 & 8970 & 5238 \\
O-AGB (SMC-OGLE3) & \citet{sus11} & 2511  & 2511  & 2438 & 2477 & 164  \\
C-AGB (SMC-OGLE3) & \citet{sus11} & 2761  & 2757  & 2724 & 2743 & 1173 \\
\hline
O-AGB (LMC-SAGE)  & \citet{Riebel2012} & 26,231$^2$ & 26,231 & 25,745 & 26,080 & 7570 \\
C-AGB (LMC-SAGE) & \citet{Riebel2012} & 7306$^3$ & 7276  & 7268 & 7256 & 6743  \\
O-AGB (SMC-SAGE) & \citet{Srinivasan2016} & 3624  & 3624  & 3538 & 3624 & 160  \\
C-AGB (SMC-SAGE) & \citet{Srinivasan2016} & 2118  & 2117  & 2101 & 2118 & 1134 \\
O-AGB (LMC-SAGE-S)$^4$ & \citet{jones2017} & 77 & 77 & 75 & 74 & 75   \\
C-AGB (LMC-SAGE-S)$^4$ & \citet{sloan2016} & 151$^5$ & 135  & 145 & 145 & 143  \\
O-AGB (SMC-SAGE-S)$^4$ & \citet{kraemer2017} & 5 & 5 & 5 & 5 & 5   \\
C-AGB (SMC-SAGE-S)$^4$ & \citet{sloan2016} & 40 & 40  & 38 & 40 & 37  \\
\hline
\end{tabular}
\begin{flushleft}
$^1$the number in parenthesis for the objects in our Galaxy denotes the number of the data with small deviations in the S5[24] flux (see Section~\ref{sec:gagb}).
$^2$21 newly identified SAGE O-AGB objects from \citet{jones2017} are added.
$^3$13 newly identified SAGE C-AGB objects from \citet{jones2017} are added.
$^4$Identified from the SAGE IRS spectroscopy.
$^5$7 newly identified SAGE-S C-AGB objects from \citet{jones2017} are added.
\end{flushleft}
\end{table*}

\begin{table}
\scriptsize
\caption{IR bands and zero magnitude flux values \label{tab:tab2}}
\centering
\begin{tabular}{lllll}
\hline \hline
Band &$\lambda_{ref}$ ($\mu$m)	&ZMF (Jy) &Remark &Reference$^1$ 	\\
\hline
J[1.2]  &1.235	&	1594	&	2MASS	& \citet{cohen2003}\\
H[1.7]  &1.662	&	1024	&	2MASS	& \citet{cohen2003}\\
K[2.2]  &2.159	&	666.7	&	2MASS	& \citet{cohen2003}\\
W1[3.4]	&3.35	&	306.682	&	WISE    & \citet{jarrett2011}	\\
S1[3.6] &3.55	&	280.9	&	Spitzer & A	\\
S2[4.5] &4.493	&	179.7	&	Spitzer & A	\\
W2[4.6]	&4.60	&	170.663	&	WISE    & \citet{jarrett2011}	\\
S3[5.8] &5.731	&	115.0	    &	Spitzer & A	\\
S4[8.0] &7.872  &	64.9	&	Spitzer & A	\\
W3[12]$^2$	&12.0 (11.56)	& 28.3 (29.045)	&	WISE & \citet{jarrett2011}	\\
W4[22]	&22.08	&	8.284	&	WISE    & \citet{jarrett2011}	\\
S5[24]  &23.68  &	7.17	&	Spitzer & B	\\
\hline
\end{tabular}
\begin{flushleft}
$^1$A: \url{https://irsa.ipac.caltech.edu/data/SPITZER/docs/irac/iracinstrumenthandbook},
B: \url{https://irsa.ipac.caltech.edu/data/SPITZER/docs/mips/mipsinstrumenthandbook}.
$^2$For W3[12], we use a new reference wavelength and zero magnitude flux for theoretical models
(original values are given in parenthesis; see~\ref{sec:modelcolor}).
\end{flushleft}
\end{table}

\subsection{Infrared Photometric Data\label{sec:photdata}}

IRAS and AKARI data have been very useful for studying AGB stars in our Galaxy
(e.g., \citealt{sk2011}; \citealt{suh2018}). Though they were also useful for
studying AGB stars in the Magellanic Clouds (e.g., \citealt{jones2014}), the
number of the cross-identified objects was very limited because of the
relatively large beam sizes and weak sensitivities.

2MASS \citep{cutri2003} provided fluxes at J (1.25 $\mu$m), H (1.65 $\mu$m),
and K (2.16 $\mu$m) bands. The field of view (FOV) pixel size of the 2MASS
image is 2$\arcsec$. The WISE \citep{wright2010} mapped the sky at 3.4, 4.6,
12, and 22 $\mu$m. For the four WISE bands (W1, W2, W3, and W4), the FOV pixel
sizes are 2$\farcs$75, 2$\farcs$75, 2$\farcs$75, and 5$\farcs$5, and the
5$\sigma$ photometric sensitivities are 0.068, 0.098, 0.86, and 5.4 mJy
(\url{http://wise2.ipac.caltech.edu/docs/release/allsky/expsup/sec1_1.html}).
The WISE data have been useful for studying AGB stars in our Galaxy (e.g.,
\citealt{suh2018}) and they would be also useful for studying AGB stars in the
Magellanic Clouds.

The Spitzer Space Telescope (\citealt{gehrz2007}) had infrared array camera
(IRAC; 3.6, 4.5, 5.8, and 8.0 $\mu$m) and multiband imaging photometer (MIPS;
24, 70, and 160 $\mu$m) bands. For the four IRAC bands, and the 5$\sigma$
photometric sensitivities are 1.3, 2.7, 18, and 22 $\mu$Jy with the FOV pixel
size of 1$\farcs$2. For the MIPS band at 24 $\mu$m, the 5$\sigma$ photometric
sensitivity is 110 $\mu$Jy and the FOV pixel size is 2$\farcs$5.

Table~\ref{tab:tab2} lists the IR bands used in this work. For each band, the
reference wavelength ($\lambda_{ref}$) and zero magnitude flux (ZMF) value,
which are useful to to obtain theoretical models colors (see
Section~\ref{sec:modelcolor}), are also shown.

In this work, we use only good quality observational data at all wavelength
bands for the 2MASS and WISE photometric data (quality A for the 2MASS; quality
A or B for the WISE). For the Spitzer photometric data, we use all of the
available data because we use the SAGE catalogs with high reliability, which
were extracted from the full list by placing strict restrictions on the source
quality (see Sections~\ref{sec:gagb} and~\ref{sec:magb}).

\subsection{AGB stars in our Galaxy\label{sec:gagb}}

A catalog of AGB stars for 3003 O-AGB and 1168 C-AGB objects in our Galaxy was
presented by \citet{sk2011}. \citet{sh2017} presented a revised list of 3828
O-AGB and 1168 C-AGB stars based on the IRAS point source catalog (PSC). The
classification was based on IR and optical spectroscopy, IR photometry, and
maser observations (see \citealt{sk2011}; \citealt{sh2017}). The sample of 4996
Galactic AGB stars is composed of Mira variables (O-AGB: 1444; C-AGB: 292),
semiregular variables (SRVs; O-AGB: 167; C-AGB: 178), and other types according
to the American association of variable star observers (AAVSO) international
variable star index (VSX; \citealt{watson2019}). Among the 3828 O-AGB stars,
1520 objects are known to be OH/IR stars (see Section~\ref{sec:ohir}), from
which 271 objects are known to be Miras according to the AAVSO. Note that most
of the Galactic AGB stars with thick dust envelopes are not listed in the AAVSO
catalog, which is mainly based on optical observations.

Because IRAS has a large beam size, it is tricky to find appropriate 2MASS,
WISE, or Spitzer counterparts using the IRAS PSC position (see
\citealt{suh2018}). We find the AKARI PSC, 2MASS, and WISE counterparts as
described in \citet{suh2018}, which considered the beam sizes and compared the
fluxes. To find the Spitzer IRAC and MIPS counterparts, we use the same method
that was used for finding the WISE counterpart (\citealt{suh2018}). We make
cross identification to the Spitzer point sources by using the `A 24 and 70
Micron Survey of the Inner Galactic Disk with MIPS' (MIPSGAL) catalog, which
provides the Spitzer photometric data for 933,818 sources in our Galaxy.
Table~\ref{tab:tab1} lists the sample AGB stars in our Galaxy and numbers of
the cross-matched 2MASS, WISE, and Spitzer counterparts.

\begin{figure}
\centering
\smallplottwo{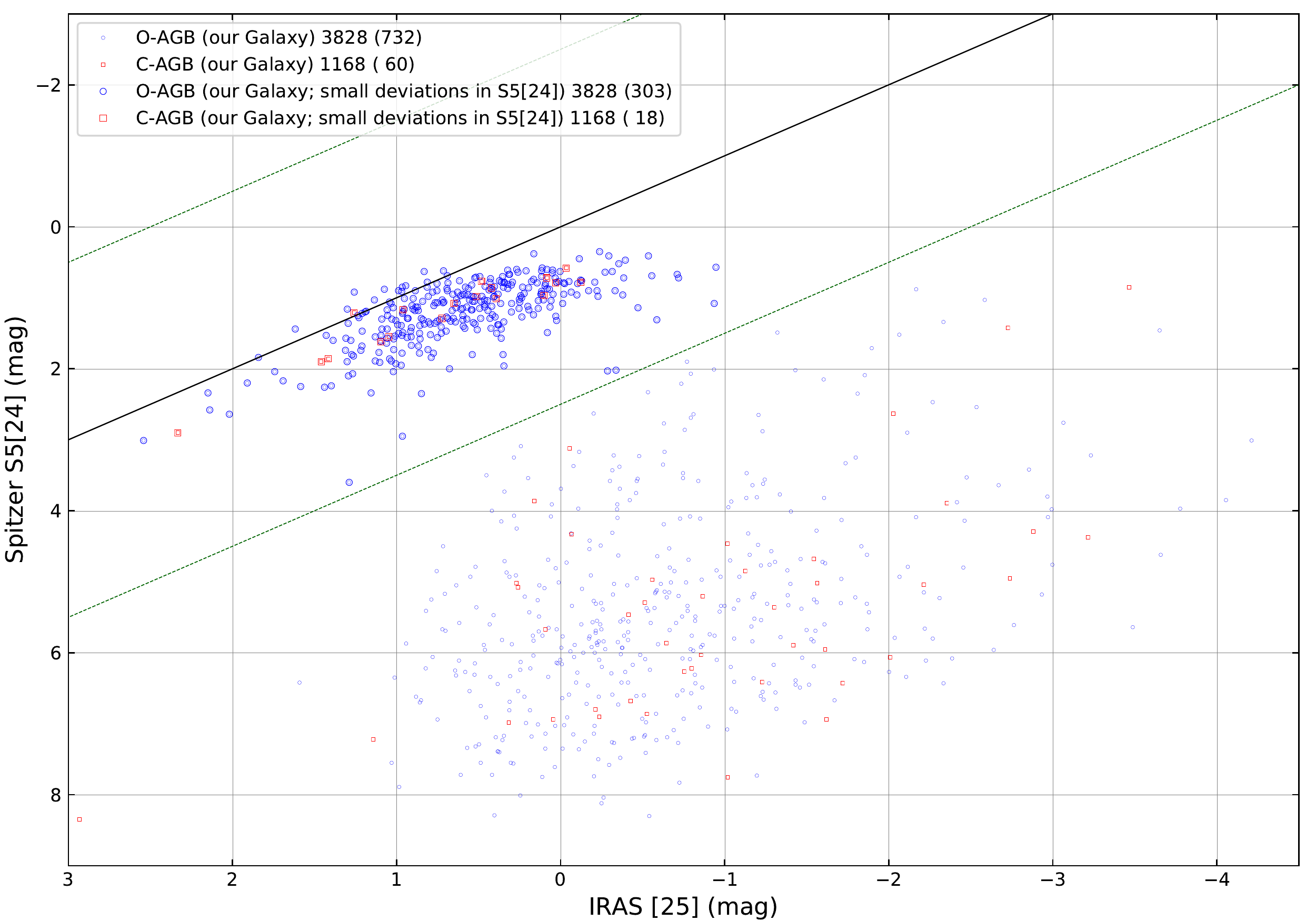}{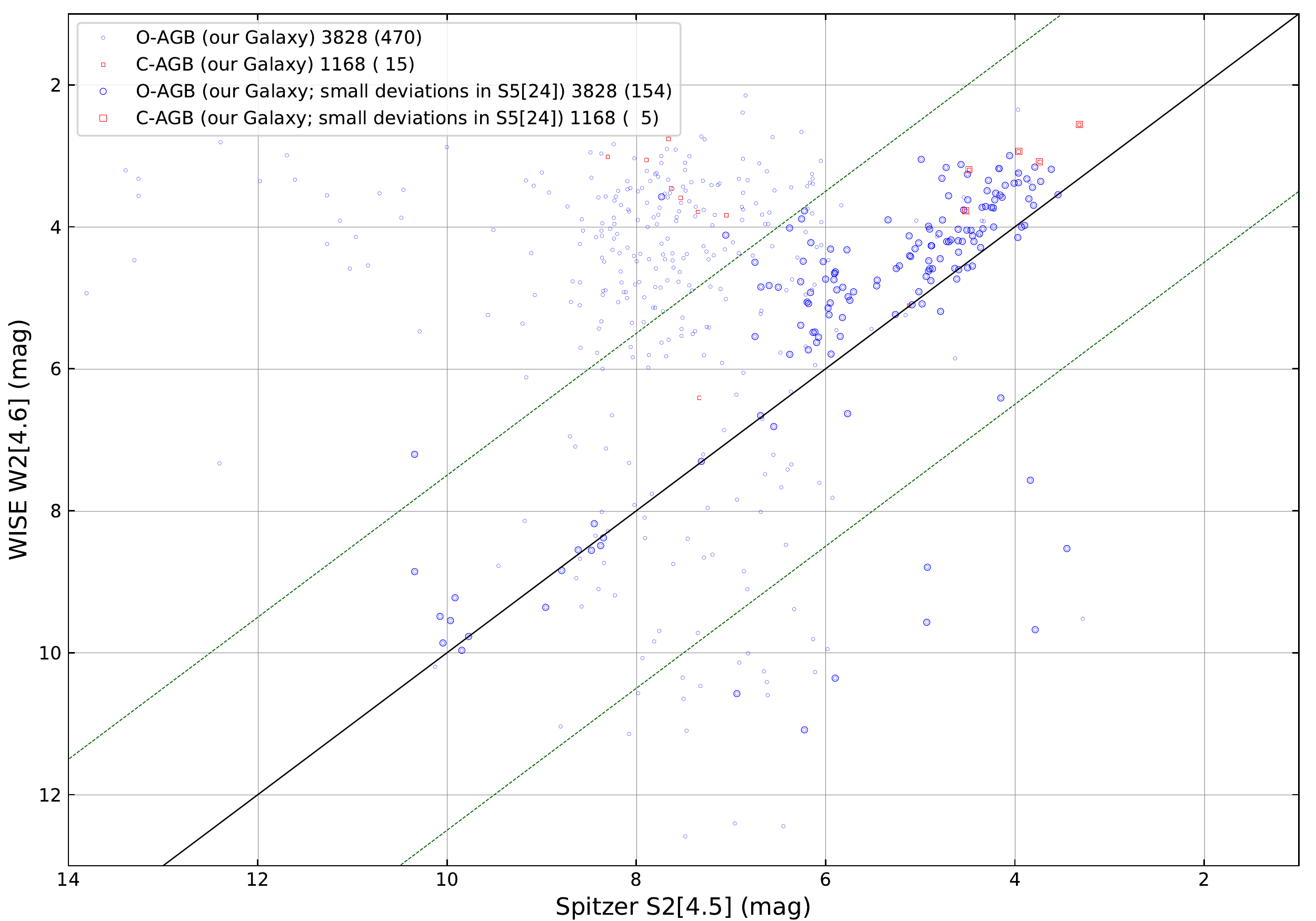}\caption{Comparison of the fluxes (in mag) at Spitzer and other bands
(Spitzer IRAC S5[24] versus IRAS PSC [25]; Spitzer IRAC S3[5.8] versus WISE W2[4.6])
for AGB stars in our Galaxy (see Table 1). For each class, the number of objects is shown.
The number in parenthesis denotes the number of the plotted objects with good quality observed data.
See Section~\ref{sec:gagb}.}
\label{f1}
\end{figure}

The upper panel of Figure~\ref{f1} shows the comparison of the IRAS [25] (25
$\mu$m) flux with the Spitzer S5[24] (24 $\mu$m) flux for AGB stars in our
Galaxy. For the Spitzer counterparts, the Spitzer fluxes drop abnormally
compared with other measurements at nearby wavelengths. This would be mainly
because of the saturation effect of the Spitzer data for the bright Galactic
AGB stars. There is a similar effect for the WISE data for AGB stars in our
Galaxy, but it is known to be minor for a considerable portion of them (see
\citealt{suh2018}). When we remove the objects with very large decreases in
Spitzer S5[24] flux from the IRAS [25] flux (more than 2.5 mag; see
Figure~\ref{f1}), more reliable data points with small deviations (see
Table~\ref{tab:tab1}) can be distinguishable. The lower panel of
Figure~\ref{f1} shows the comparison of the WISE W2[4.6] flux with the Spitzer
S3[5.8] flux for AGB stars in our Galaxy, which shows the saturation effect
only for bright objects at W2[4.6]. We also show the objects with with small
deviations in the Spitzer S5[24] flux.

\begin{figure}
\centering
\smallplot{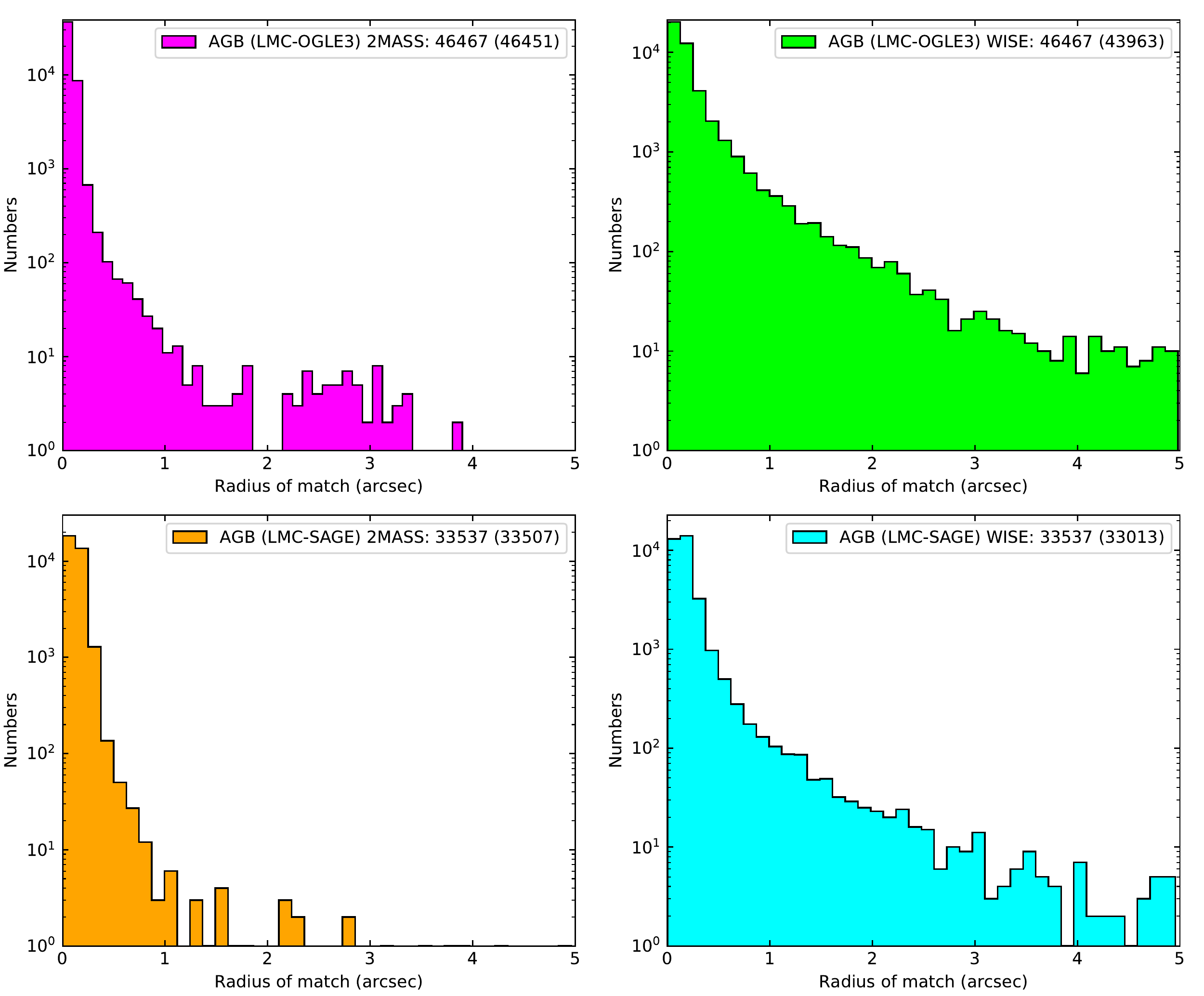}\caption{Number distributions of cross-match angular distances
for OGLE3 and SAGE AGB stars in the LMC (see Table~\ref{tab:tab1}) to the 2MASS and WISE point sources.
For each sample, the number of objects is shown.
The number in parenthesis denotes the number of the cross matches.
See Section~\ref{sec:magb-c}.}
\label{f2}
\end{figure}

\begin{figure*}
\centering
\smallplotfour{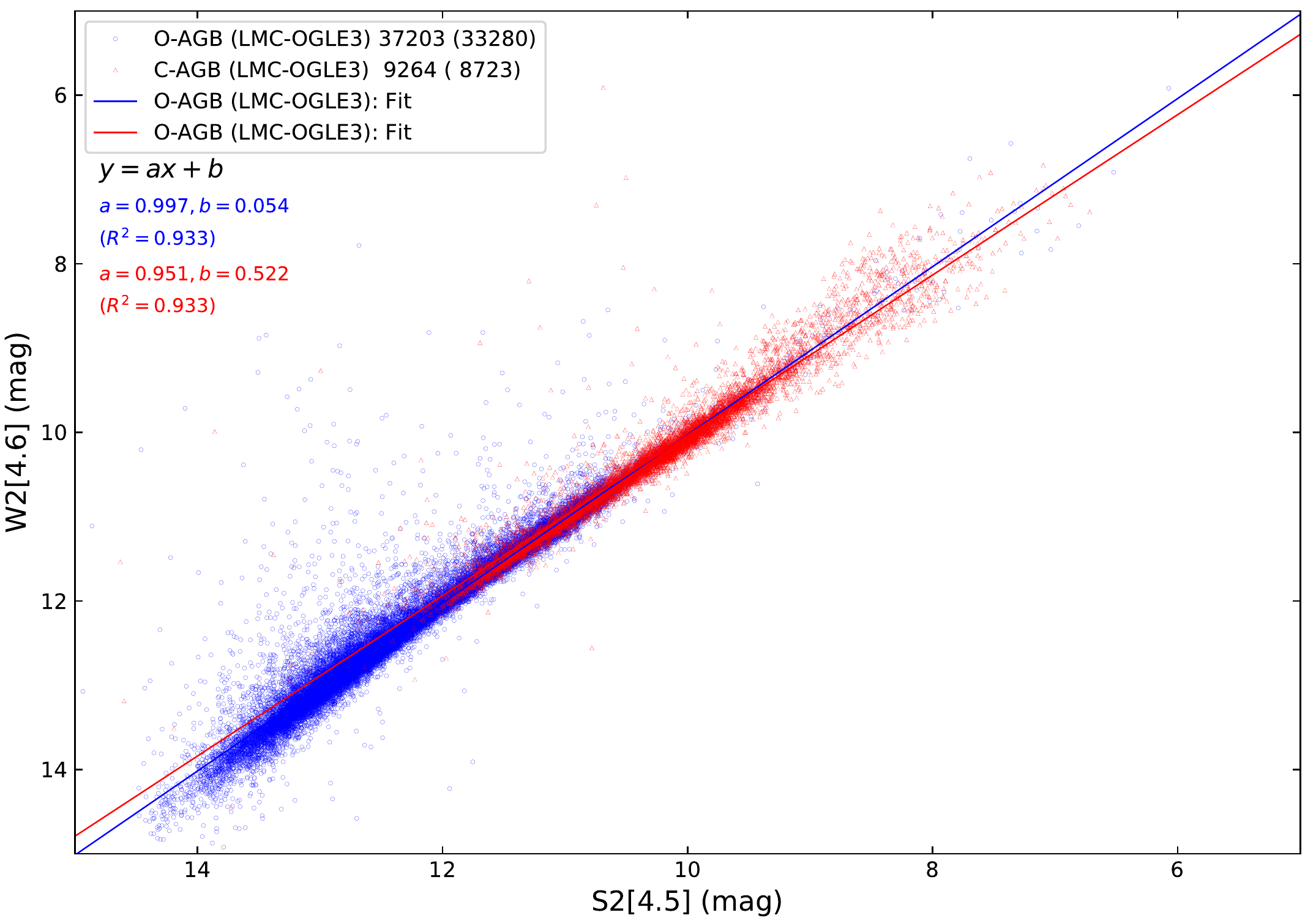}{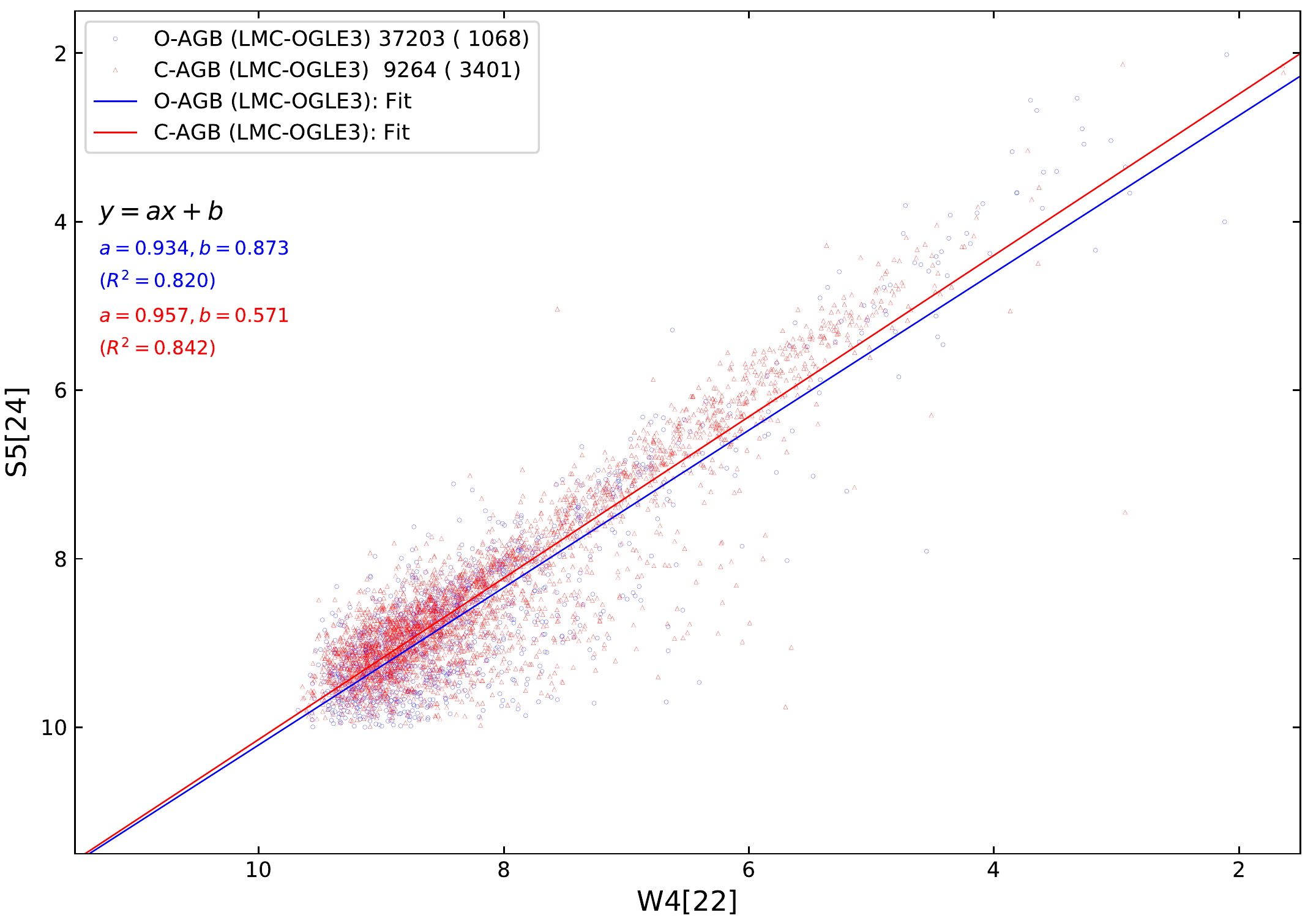}{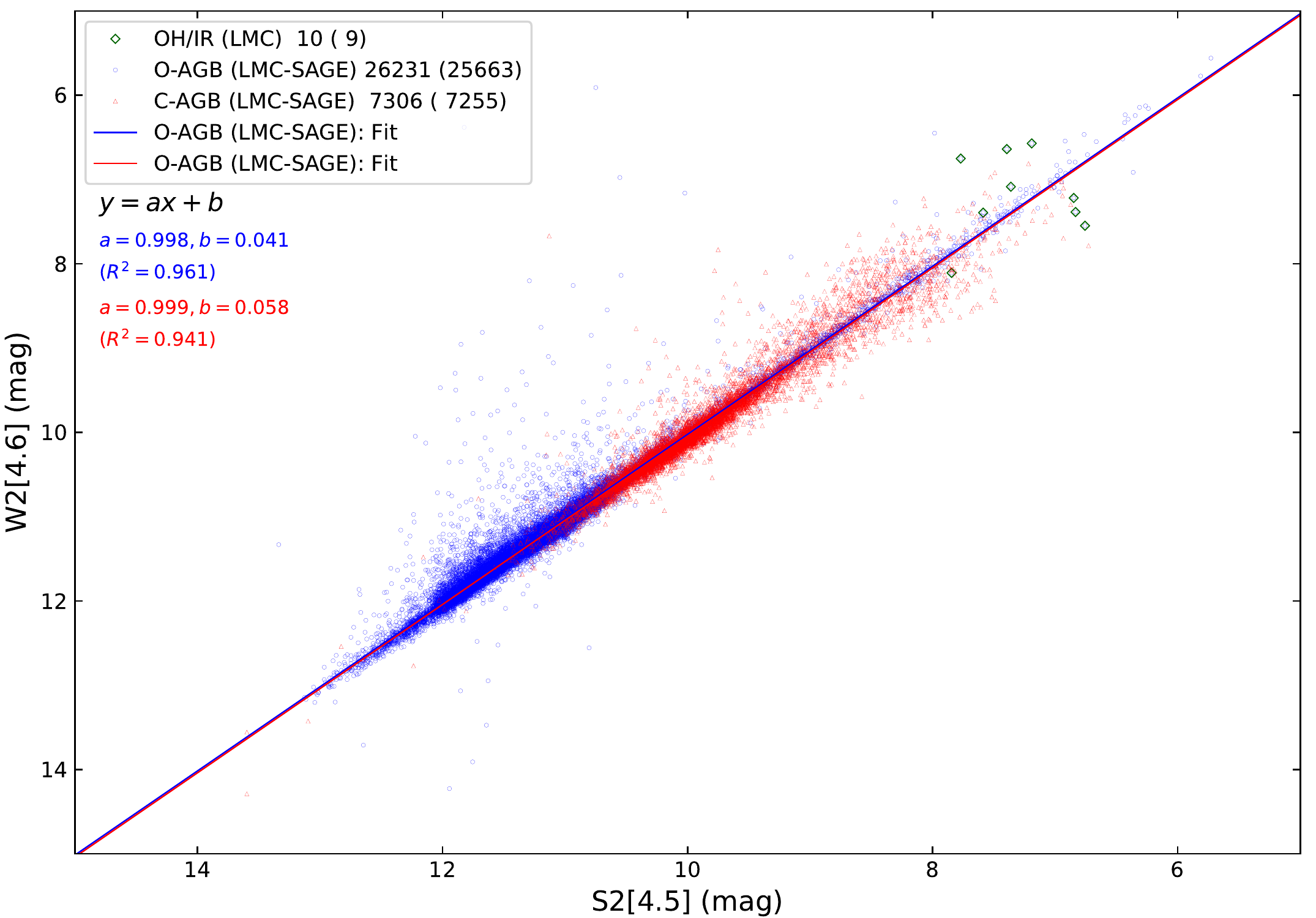}{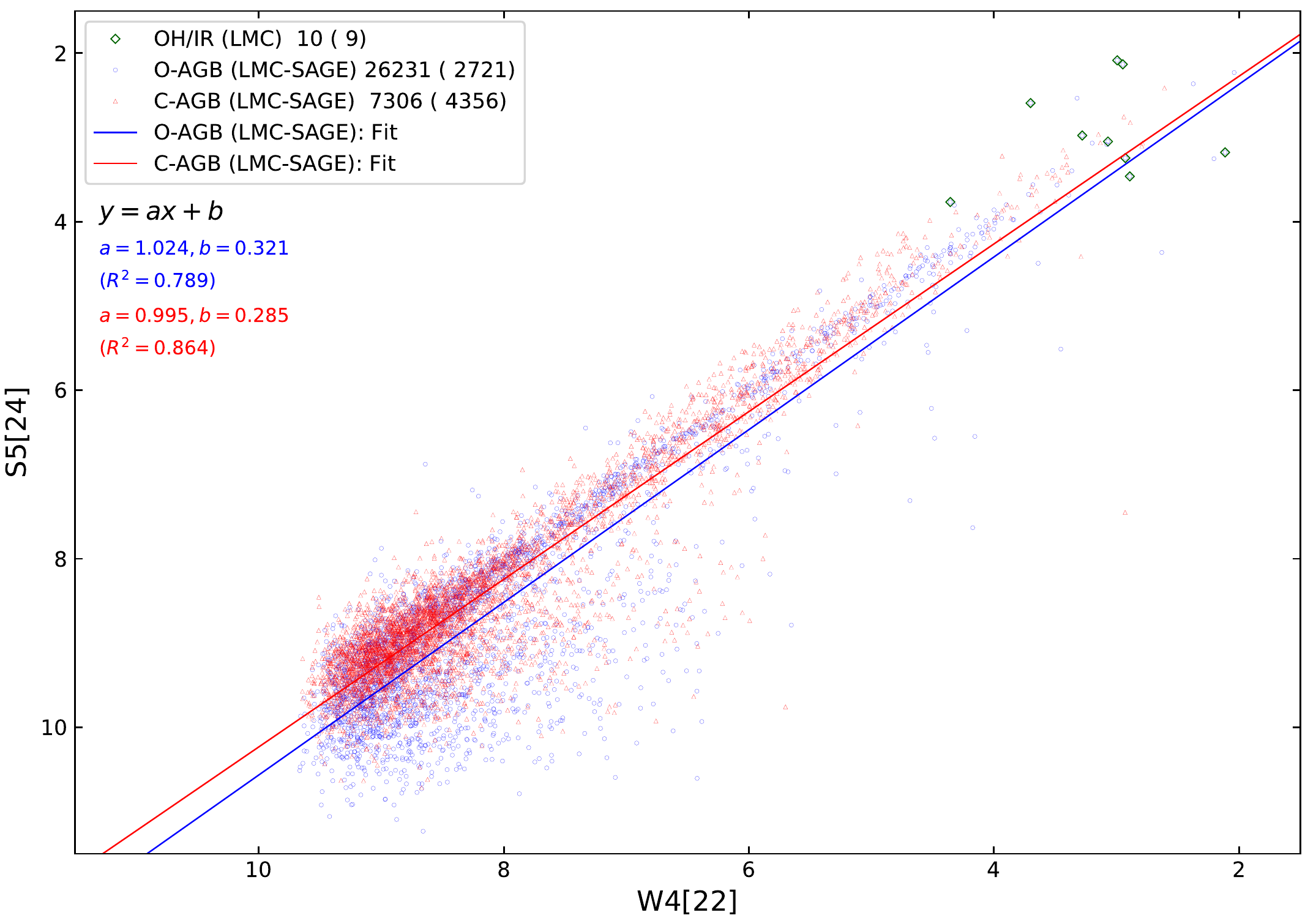}\caption{Comparison of the fluxes (in mag) at Spitzer and WISE bands
for cross-identified objects in the LMC (OGLE3 and SAGE) sample stars. For each sample, the number of objects is shown.
The number in parenthesis denotes the number of the plotted objects with good quality observed data.
The coefficients of determination ($R^2$) for the linear relations are also shown.
See Section~\ref{sec:magb-c}.}
\label{f3}
\end{figure*}

\begin{figure*}
\centering
\largeplottwo{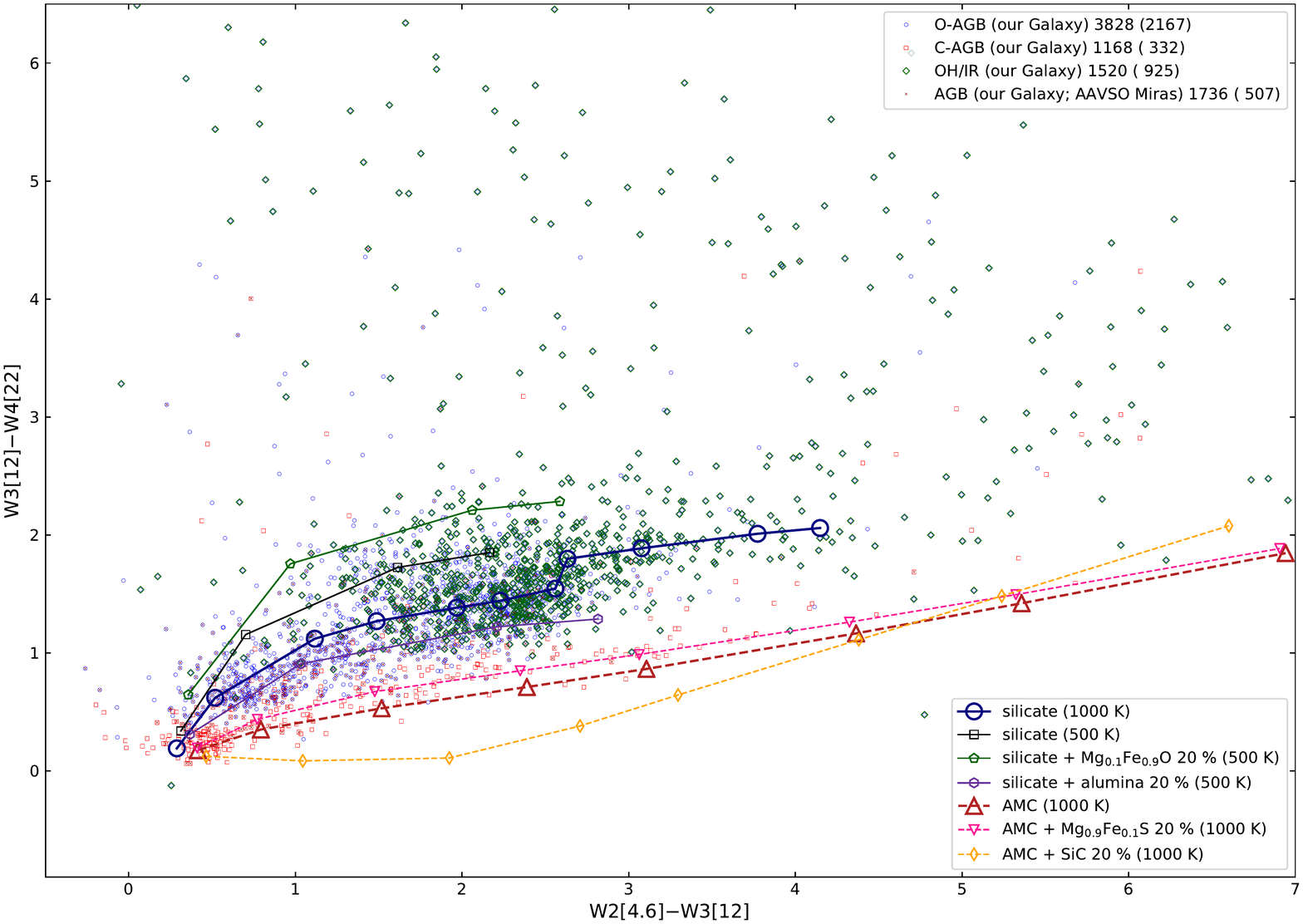}{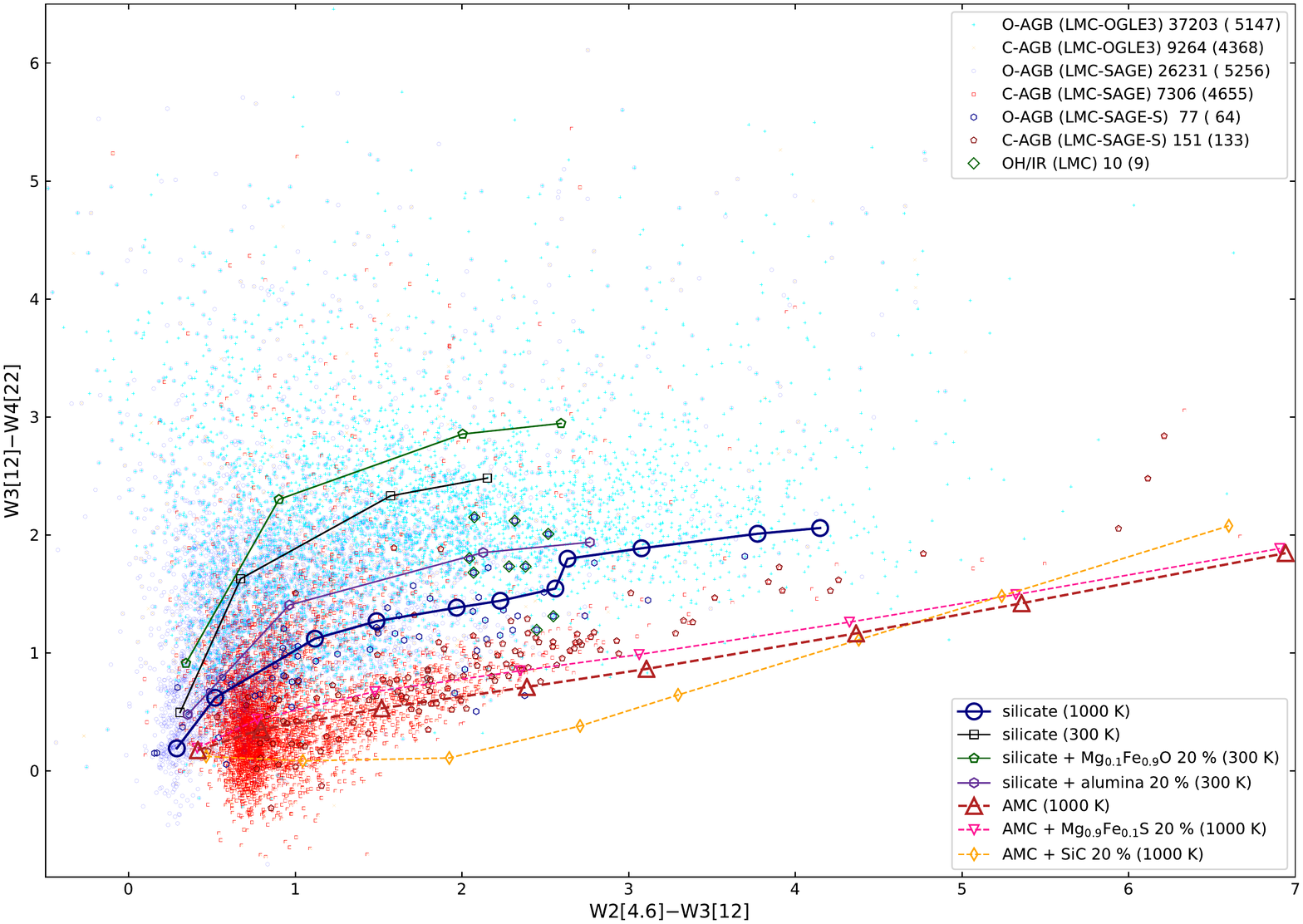}\caption{WISE 2CDs for AGB stars in our Galaxy and the LMC (OGLE3 and SAGE samples) compared with theoretical models (see Section~\ref{sec:models}).
For O-AGB models (Silicate $T_c$ = 1000 K): $\tau_{10}$ = 0.001, 0.01, 0.05, 0.1, 0.5, 1, 3, 7, 15, 30, and 40 from left to right.
For C-AGB models (AMC $T_c$ = 1000 K): $\tau_{10}$ = 0.001, 0.01, 0.1, 0.5, 1, 2, 3, and 5 from left to right.
For each class, the number of objects is shown.
The number in parenthesis denotes the number of the plotted objects on the 2CD with good quality observed colors.}
\label{f4}
\end{figure*}

\begin{figure*}
\centering
\largeplottwo{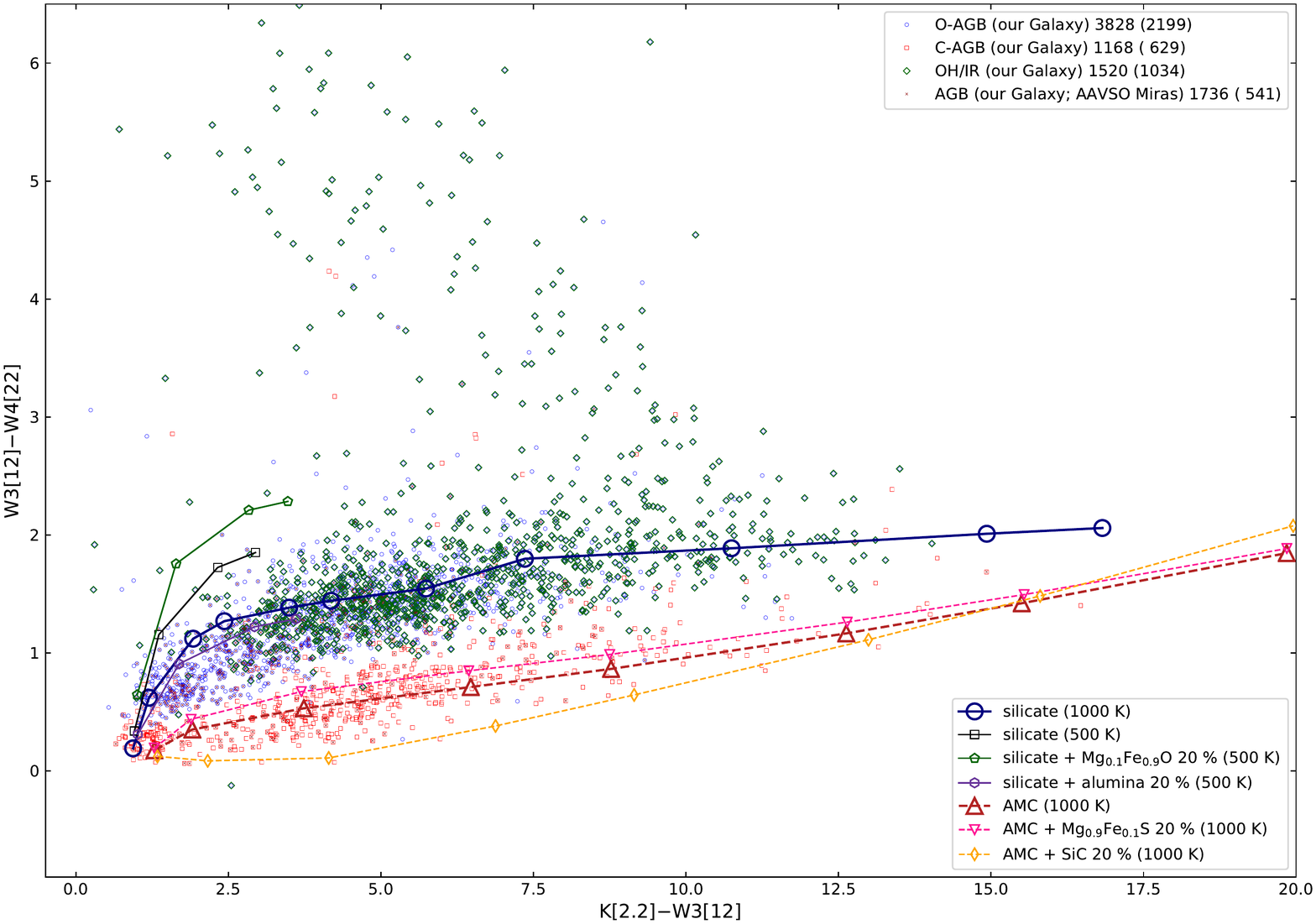}{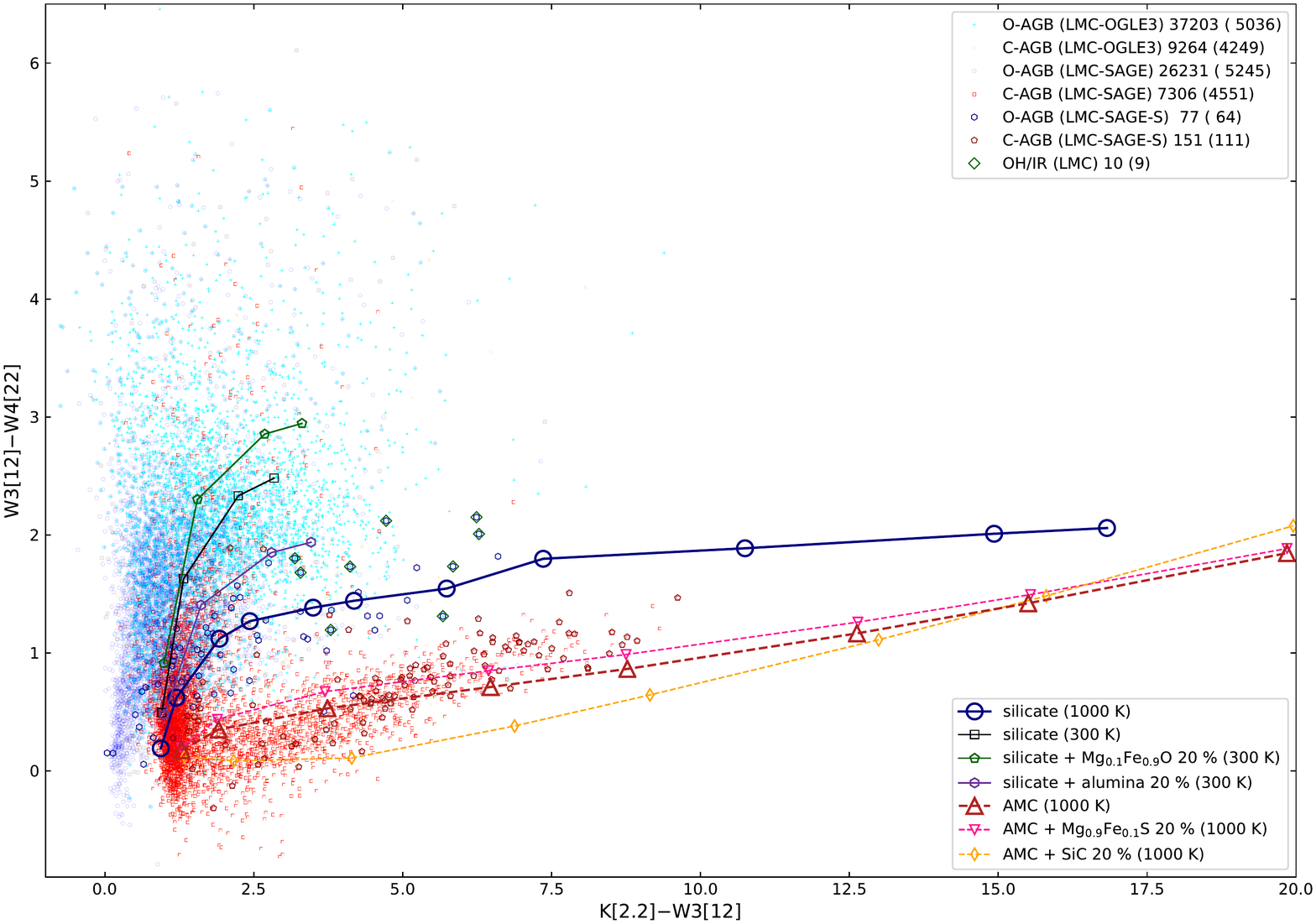}\caption{WISE-2MASS 2CDs for AGB stars in our Galaxy and the LMC (OGLE3 and SAGE samples) compared with theoretical models (see Section~\ref{sec:models}).
For O-AGB models (Silicate $T_c$ = 1000 K): $\tau_{10}$ = 0.001, 0.01, 0.05, 0.1, 0.5, 1, 3, 7, 15, 30, and 40 from left to right.
For C-AGB models (AMC $T_c$ = 1000 K): $\tau_{10}$ = 0.001, 0.01, 0.1, 0.5, 1, 2, 3, and 5 from left to right.
For each class, the number of objects is shown.
The number in parenthesis denotes the number of the plotted objects on the 2CD with good quality observed colors.}
\label{f5}
\end{figure*}

\begin{figure}
\centering
\smallplottwo{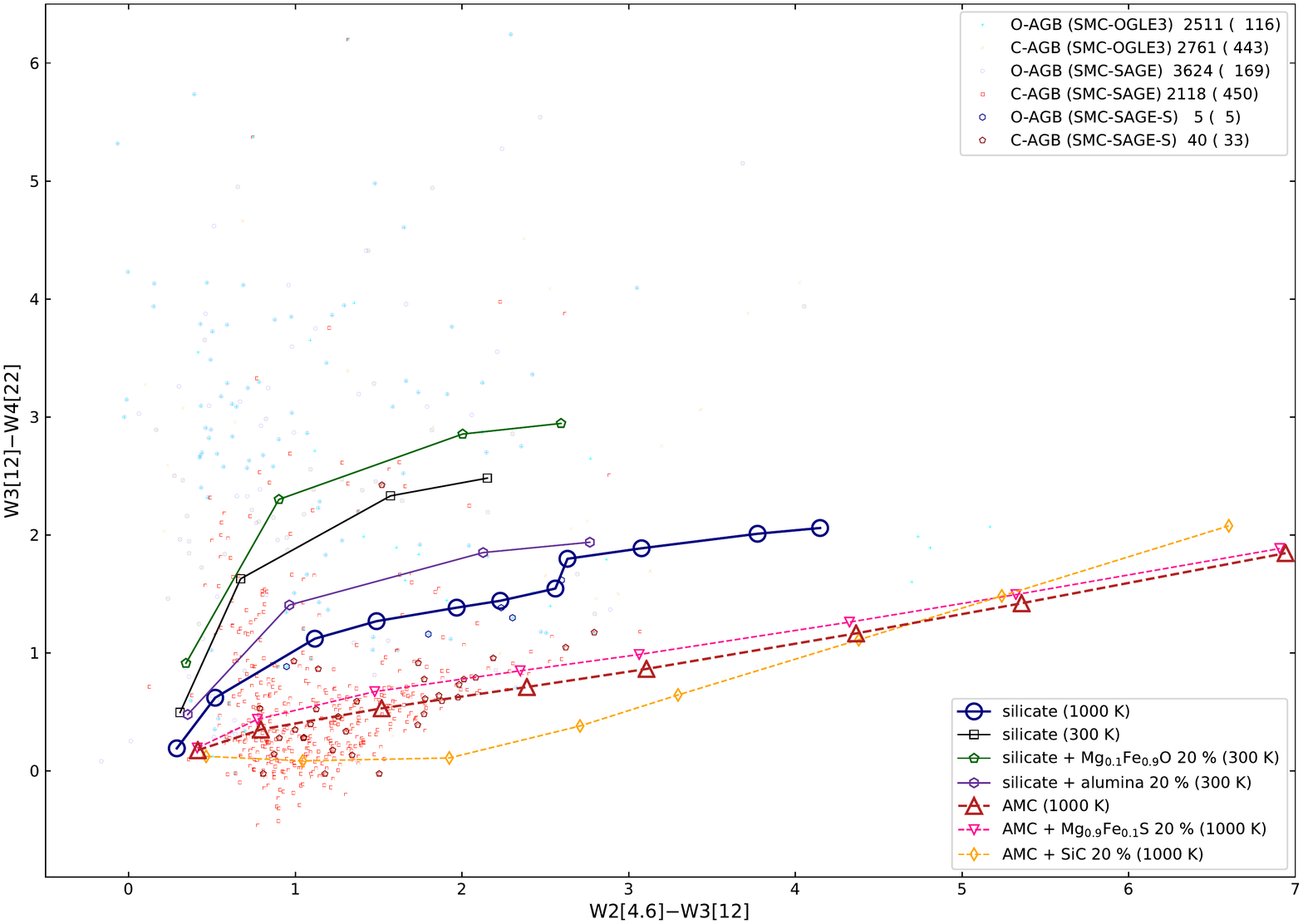}{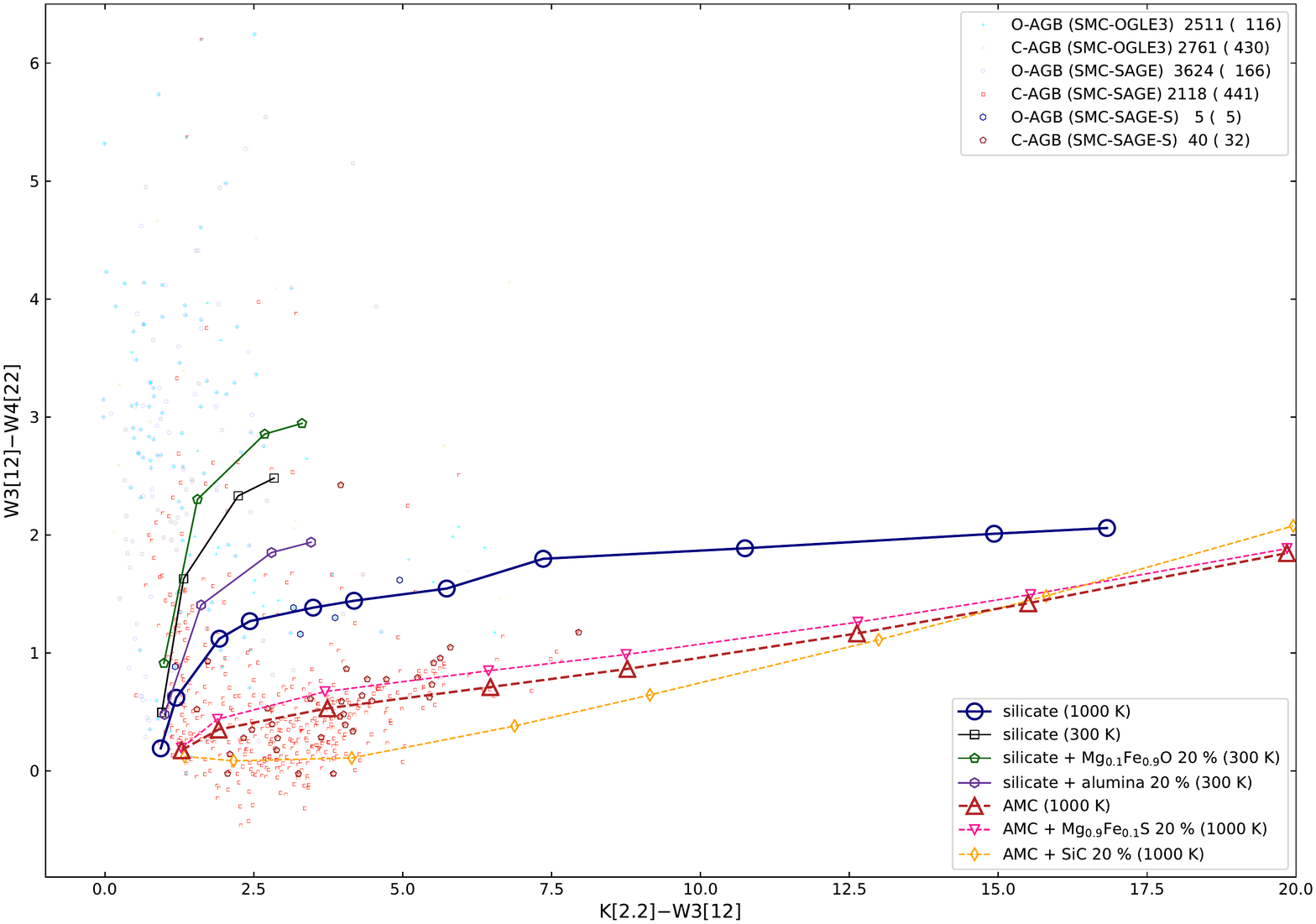}\caption{WISE-2MASS 2CDs for AGB stars in the SMC (OGLE3 and SAGE samples) compared with theoretical models (see Section~\ref{sec:models}).
For O-AGB models (Silicate $T_c$ = 1000 K): $\tau_{10}$ = 0.001, 0.01, 0.05, 0.1, 0.5, 1, 3, 7, 15, 30, and 40 from left to right.
For C-AGB models (AMC $T_c$ = 1000 K): $\tau_{10}$ = 0.001, 0.01, 0.1, 0.5, 1, 2, 3, and 5 from left to right.
For each class, the number of objects is shown.
The number in parenthesis denotes the number of the plotted objects on the 2CD with good quality observed colors.}
\label{f6}
\end{figure}

\begin{figure*}
\centering
\largeplottwo{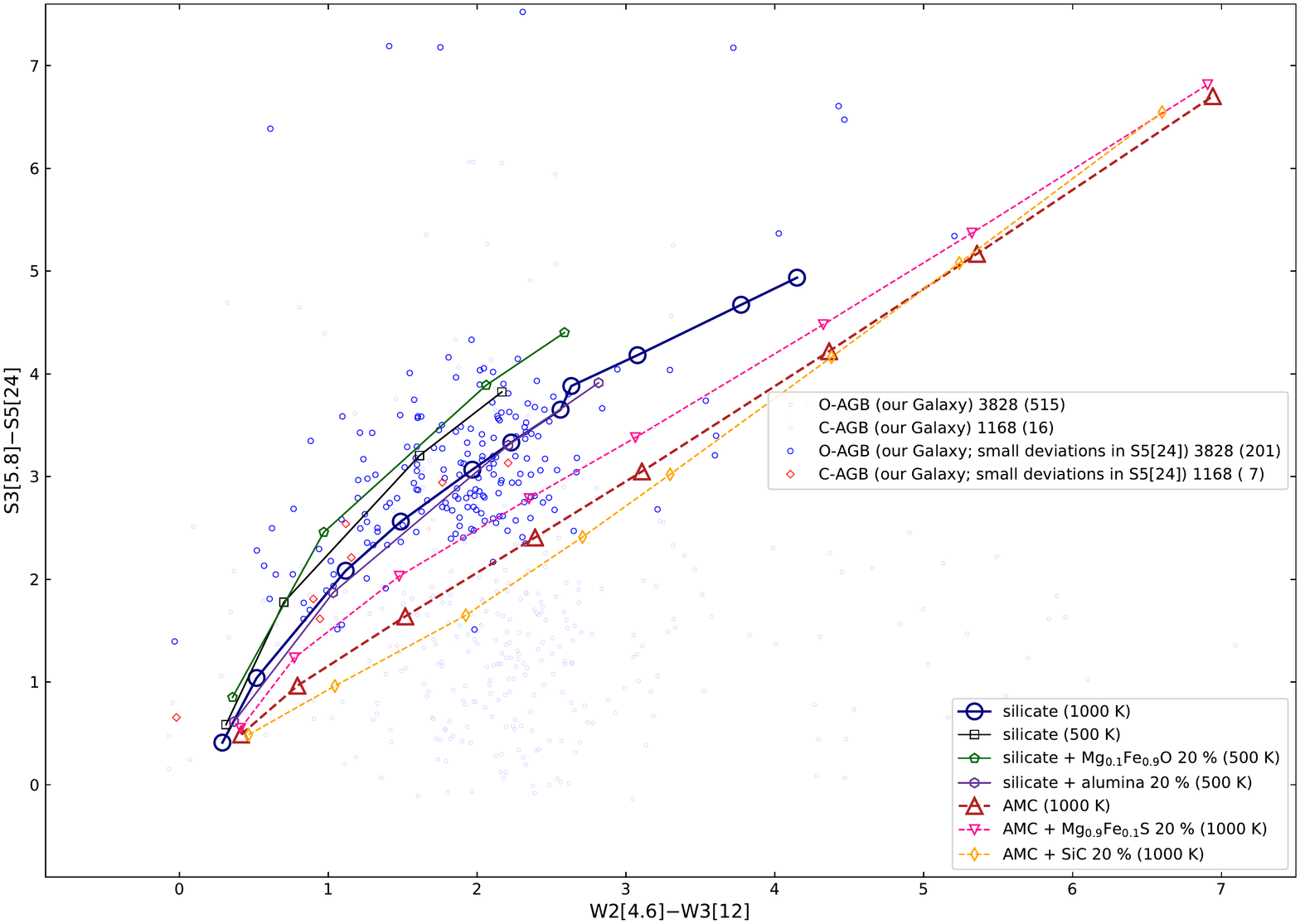}{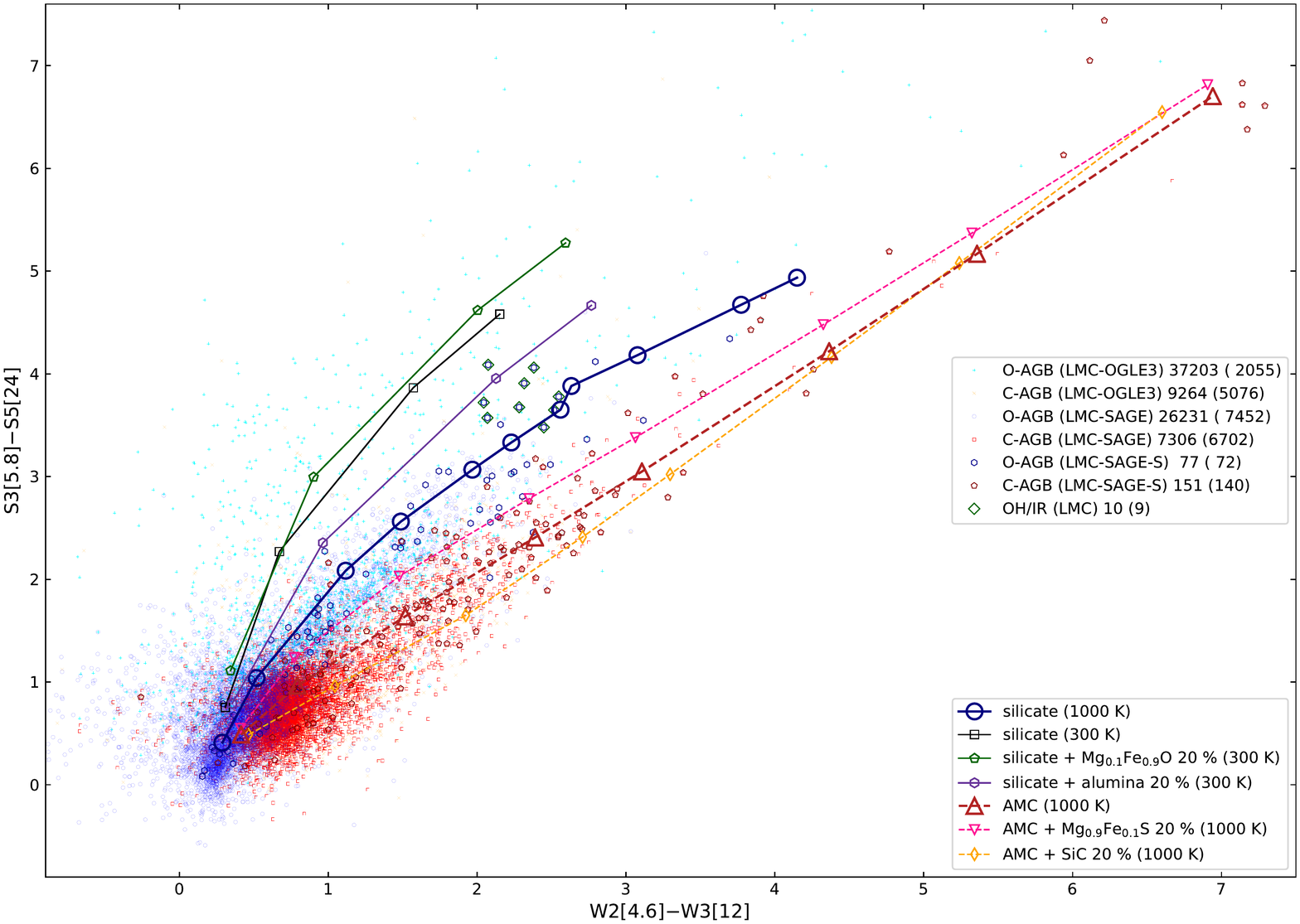}\caption{WISE-Spitzer 2CDs for AGB stars in our Galaxy and the LMC (OGLE3 and SAGE samples) compared with theoretical models (see Section~\ref{sec:models}).
For O-AGB models (Silicate $T_c$ = 1000 K): $\tau_{10}$ = 0.001, 0.01, 0.05, 0.1, 0.5, 1, 3, 7, 15, 30, and 40 from left to right.
For C-AGB models (AMC $T_c$ = 1000 K): $\tau_{10}$ = 0.001, 0.01, 0.1, 0.5, 1, 2, 3, and 5 from left to right.
For each class, the number of objects is shown.
The number in parenthesis denotes the number of the plotted objects on the 2CD with good quality observed colors.}\label{f7}
\end{figure*}

\begin{figure*}
\centering
\largeplottwo{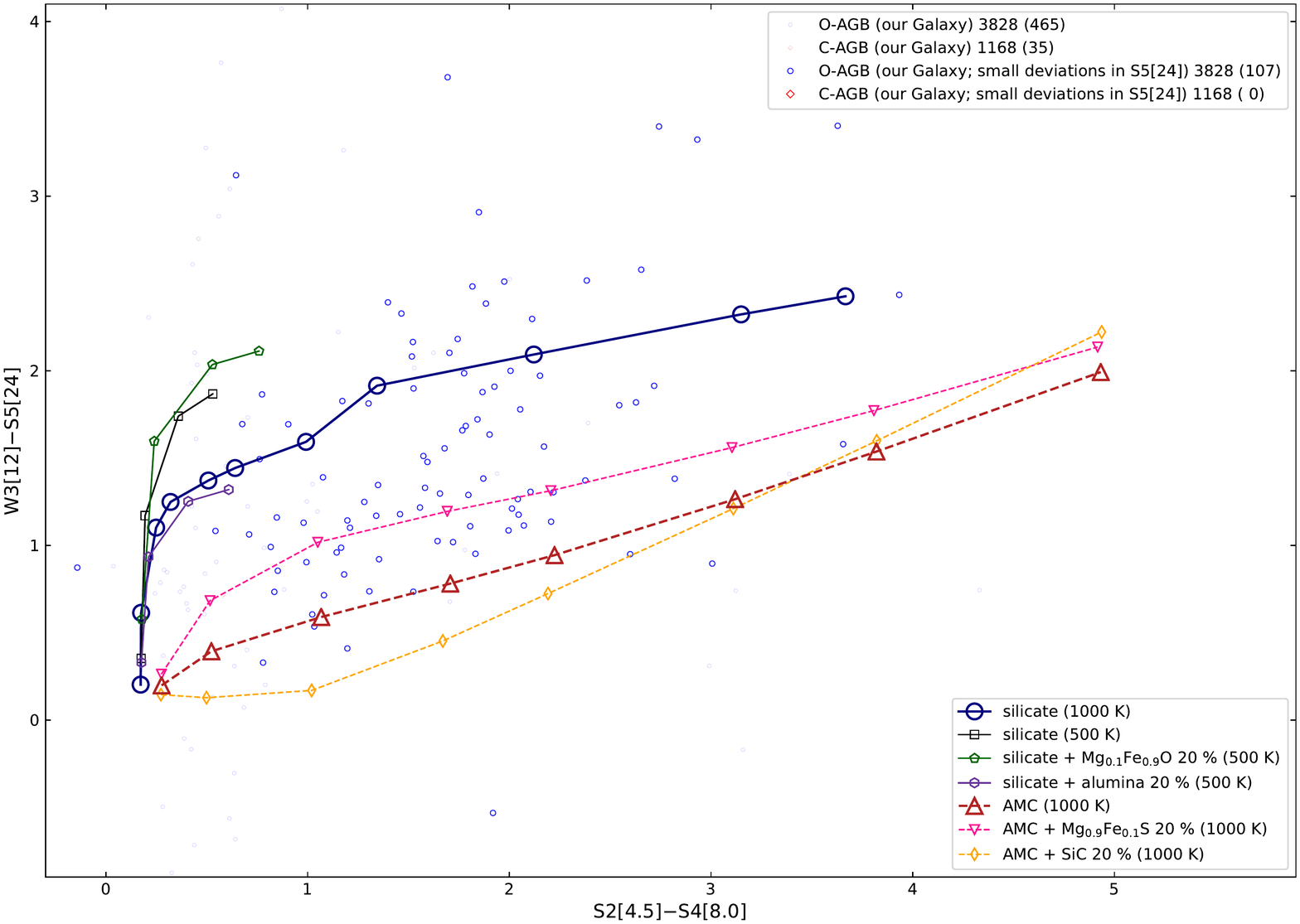}{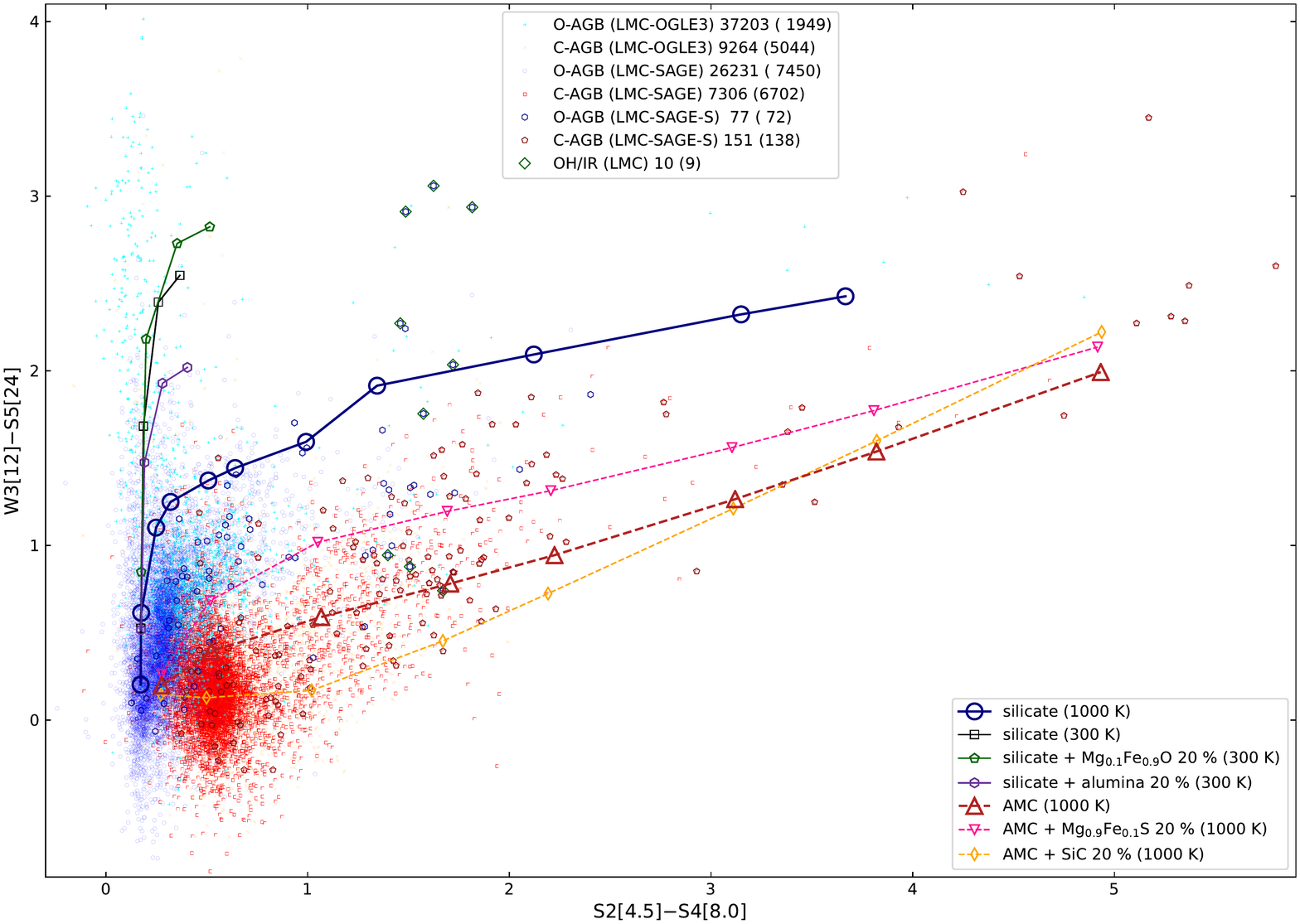}\caption{WISE-Spitzer 2CDs for AGB stars in our Galaxy and the LMC (OGLE3 and SAGE samples) compared with theoretical models (see Section~\ref{sec:models}).
For O-AGB models (Silicate $T_c$ = 1000 K): $\tau_{10}$ = 0.001, 0.01, 0.05, 0.1, 0.5, 1, 3, 7, 15, 30, and 40 from left to right.
For C-AGB models (AMC $T_c$ = 1000 K): $\tau_{10}$ = 0.001, 0.01, 0.1, 0.5, 1, 2, 3, and 5 from left to right.
For each class, the number of objects is shown.
The number in parenthesis denotes the number of the plotted objects on the 2CD with good quality observed colors.} \label{f8}
\end{figure*}

\begin{figure}
\centering
\smallplottwo{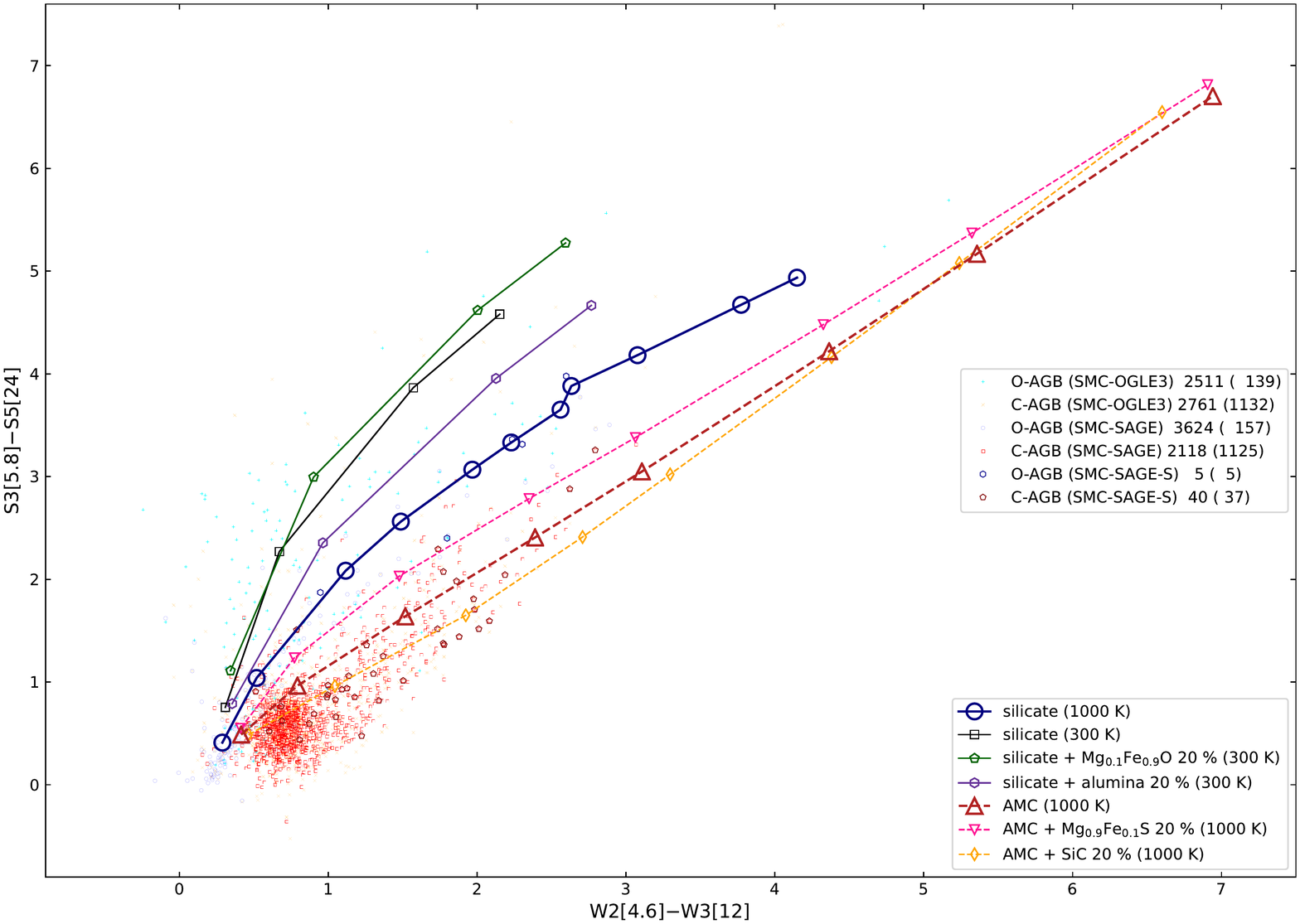}{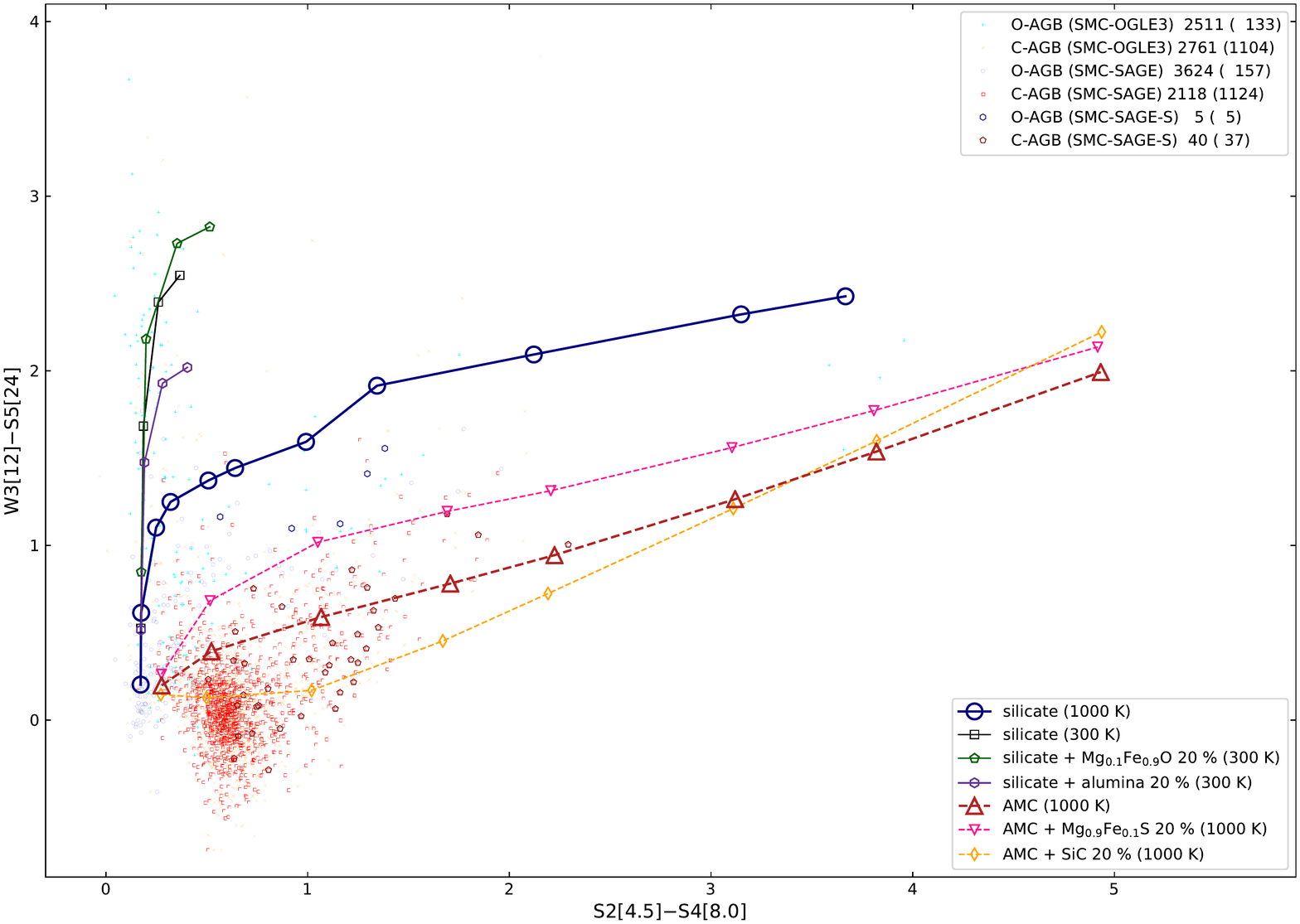}\caption{WISE-Spitzer 2CDs for AGB stars in the SMC (OGLE3 and SAGE sample) compared with theoretical models (see Section~\ref{sec:models}).
For O-AGB models (Silicate $T_c$ = 1000 K): $\tau_{10}$ = 0.001, 0.01, 0.05, 0.1, 0.5, 1, 3, 7, 15, 30, and 40 from left to right.
For C-AGB models (AMC $T_c$ = 1000 K): $\tau_{10}$ = 0.001, 0.01, 0.1, 0.5, 1, 2, 3, and 5 from left to right.
For each class, the number of objects is shown.
The number in parenthesis denotes the number of the plotted objects on the 2CD with good quality observed colors.} \label{f9}
\end{figure}

\subsection{AGB stars in the Magellanic Clouds\label{sec:magb}}

The optical gravitational lensing experiment (OGLE) projects detected many LPVs
in the Magellanic Clouds. The fourth part of the OGLE-III Catalog of Variable
Stars presented 91,995 long-period variables (LPVs) in the LMC
(\citealt{sus09}). The sample is composed of 1663 Mira variables, 11,132 SRVs,
and 79,200 small amplitude red giants (OSARGs). We use the 46,467 AGB candidate
objects (1663 Miras, 11,132 SRVs, and 33,672 bright OSARGs) as the sample AGB
stars in the LMC-OGLE3 catalog. There are 37,203 O-AGB and 9264 C-AGB objects,
which are classified based on their color selection method using the
photometric data at optical and NIR bands, in the LMC-OGLE3 sample.

The thirteenth part of the OGLE-III Catalog of Variable Stars (OIII-CVS)
contains 19,384 LPVs detected in the SMC (\citealt{sus11}). They are composed
of 352 Miras, 2222 SRVs, and 16,810 OSARGs. We use the 5272 AGB candidate
objects (352 Miras, 2222 SRVs, and 2698 bright OSARGs) as the sample AGB stars
in the SMC-OGLE3 catalog. There are 2511 O-AGB and 2761 C-AGB objects, which
are classified based on their color selection method using the photometric data
at optical and NIR bands, in the SMC-OGLE3 sample.

The LMC and SMC were imaged as a part of the SAGE program
(\citealt{meixner2006}). The SAGE program has provided a complete infrared
survey of the evolved star population in the LMC and SMC. The Spitzer Infrared
Spectrograph (IRS; $\lambda$ = 5.2-38 $\mu$m) has taken high resolution spectra
for many AGB stars in the LMC and SMC.

Analyzing the Spitzer data of the SAGE program, \citet{Riebel2012} presented a
list 33,503 candidate objects for AGB stars in the LMC. They classified them
into 26,210 O-AGB and 7293 C-AGB objects based on the comparison of the
photometric data at NIR and MIR bands with their Grid of AGB and RSG ModelS
(GRAMS). By analyzing the IRS spectral data in the SAGE program,
\citet{jones2017} identified and classified many AGB stars, from which 34
objects were new AGB stars compared with the list of \citet{Riebel2012}.
Therefore, the total number of sample AGB stars in the LMC is 33,537 (O-AGB:
26,231; C-AGB:7306).

Analyzing the data of the SAGE program for the SMC, \citet{Srinivasan2016}
presented a list of 9,621 candidate objects for evolved stars. Based on the
selection criteria presented by \citet{boyer2011}, they classified them into
2485 O-AGB, 1714 C-AGB, 1198 anomalous-AGB, 341 extreme-AGB, and other objects.
When we compare the list with the new classification by \citet{kraemer2017},
which presented a list of evolved stars in the SMC by analyzing the Spitzer IRS
spectral data, we find that four object in the list of \citet{Srinivasan2016}
are newly classified as AGB stars. When we select the AGB stars in the list, we
have 5742 AGB stars in the SMC that are classified as 3624 O-AGB and 2118 C-AGB
based on the GRAMS chemical classification.

\citet{sloan2016} presented a list 184 C-AGB stars (LMC:144; SMC: 40) by
analyzing the SAGE IRS data. When we combine the lists from \citet{sloan2016},
\citet{jones2017}, and \citet{kraemer2017}, there are 77 O-AGB and 151 C-AGB
stars in the LMC and 5 O-AGB and 40 C-AGB stars in the SMC, which are
identified from the SAGE IRS data. These SAGE IRS (SAGE-S) sample stars that
are identified by the IRS spectra would be more reliable sample of AGB stars in
the LMC and SMC.

Table~\ref{tab:tab1} lists the sample AGB stars in the LMC and SMC. Note that
all of the objects in the LMC-SAGE-S and SMC-SAGE-S samples are already
included in the LMC-SAGE and SMC-SAGE samples.

\subsection{AGB stars in the Magellanic Clouds - Cross matches\label{sec:magb-c}}

For SAGE sample objects in the LMC and SMC, we make cross identification to the
sources in the OGLE3 catalogs by finding the nearest sources within 5$\arcsec$.
From the 33,537 LMC-SAGE AGB sample objects, 22,327 objects (67 \%) are
duplicated with the LMC-OGLE3 sample (1522 Miras, 9534 SRVs, and 11,271
OSARGs). Though the chemical classification methods for the two samples (OGLE3
and SAGE) are different (see Section~\ref{sec:magb}), 19,082 objects (85 \%)
from the duplicated 22,327 objects are classified as the same class (O-AGB or
C-AGB). From the 5742 SMC-SAGE AGB sample objects, 4837 objects (84 \%) are
duplicated with the SMC-OGLE3 sample (341 Miras, 2036 SRVs, and 2460 OSARGs).

For all OGLE3 and SAGE sample objects in the LMC and SMC, we make cross
identification to the sources in the 2MASS and WISE PSC catalogs by finding the
nearest sources within 5$\arcsec$. For WISE data, multiple sample objects may
have the same cross-matched WISE point source. So we have checked all of the
duplicated cross-matches and selected only one nearest sample object for the
one WISE point source. Figure~\ref{f2} shows number distributions of the
cross-match angular distances for OGLE3 and SAGE sample objects in the LMC to
the 2MASS and WISE point sources.

For the LMC-OGLE3 objects, we make cross identification to the Spitzer point
sources by using `SAGE Winter 2008 IRAC Epoch 1 and Epoch 2 Catalog' and `SAGE
Winter 2008 MIPS 24 $\mu$m Epoch 1 and Epoch 2 Catalog' by finding the nearest
source within 5$\arcsec$. For the SMC-OGLE3 objects, we make cross
identification to Spitzer photometric data by using `SAGE-SMC IRAC Epoch 0,
Epoch 1, and Epoch 2 Catalog' and `SAGE-SMC MIPS 24 $\mu$m Epoch 0, Epoch 1,
and Epoch 2 Catalog' by finding the nearest source within 5$\arcsec$. These
SAGE catalogs were extracted from the full list by placing strict restrictions
on the source quality.

For the SAGE sample objects in the LMC and SMC, we use the Spitzer photometric
data given in the references (\citealt{Riebel2012}; \citealt{Srinivasan2016}),
which are from the SAGE Mosaic Photometry Archive. The original SAGE survey was
conducted in two epochs spaced about three months apart (\citealt{meixner2006})
and the observations from these epochs were combined into the single mosaic
photometry archive, which is deeper and has smaller photometric errors.

Therefore, the Spitzer colors for the same object in the OGLE3 and SAGE samples
can be slightly different because different catalogs were used (see IR 2CDs in
Figures~\ref{f7} -~\ref{f9}). Table~\ref{tab:tab1} lists the sample AGB stars
in the LMC and SMC and numbers of the cross-matched counterparts.

Figures~\ref{f3} shows the comparison of the fluxes (in mag) at Spitzer and
WISE bands for the cross-identified AGB stars in the LMC (OGLE3 and SAGE
samples). The overall comparison is fairly consistent for most objects.
Compared with OGLE3, there are less observed data at NIR bands but there are
more observed data at MIR bands in the SAGE sample. This could be due to the
different identification method for the SAGE sample (see
Section~\ref{sec:magb}), which would identify more optically invisible AGB
stars with thick dust shells. The larger scatters at the S5[24] band for dimmer
C-AGB stars would be due to the Mg$_{0.9}$Fe$_{0.1}$S dust feature at 28 $\mu$m
(see Section~\ref{sec:agbmodels}).

\subsection{OH/IR stars\label{sec:ohir}}

OH/IR stars are generally considered to be more massive O-AGB stars with
thicker dust envelopes and higher mass-loss rates. \citet{chen2001} presented a
list 1065 OH/IR stars in our Galaxy. The list has been corrected and updated
(\citealt{sk2011}; \citealt{ks2012}; \citealt{sh2017}), and a new list of 1520
OH/IR stars are included in the list of 3828 O-AGB stars in our Galaxy
(\citealt{sh2017}). On the IR 2CDs in Figures~\ref{f4} and~\ref{f5}, the data
for those Galactic OH/IR stars are also plotted.

Only a small number of OH/IR stars are identified in the LMC and SMC yet.
\citet{goldman2017} presented a list of positively identified 10 OH/IR stars in
the LMC. All of the 10 OH/IR objects are included in the LMC-SAGE sample (see
Table~\ref{tab:tab1}) and 6 objects from them are in the LMC-OGLE3 sample
(\citealt{sus09}), which are classified as Miras.

There is no clear identification of OH/IR stars in the SMC yet
(\citealt{goldman2018}). \citet{goldman2018} suspected that, compared with the
OH/IR stars in the Galaxy and the LMC, the lower metallicity and star formation
rate in the SMC may curtail the last dusty stellar wind phase of the most
massive O-AGB stars (see Section~\ref{sec:intro}).

\begin{figure*}
\centering
\smallplottwo{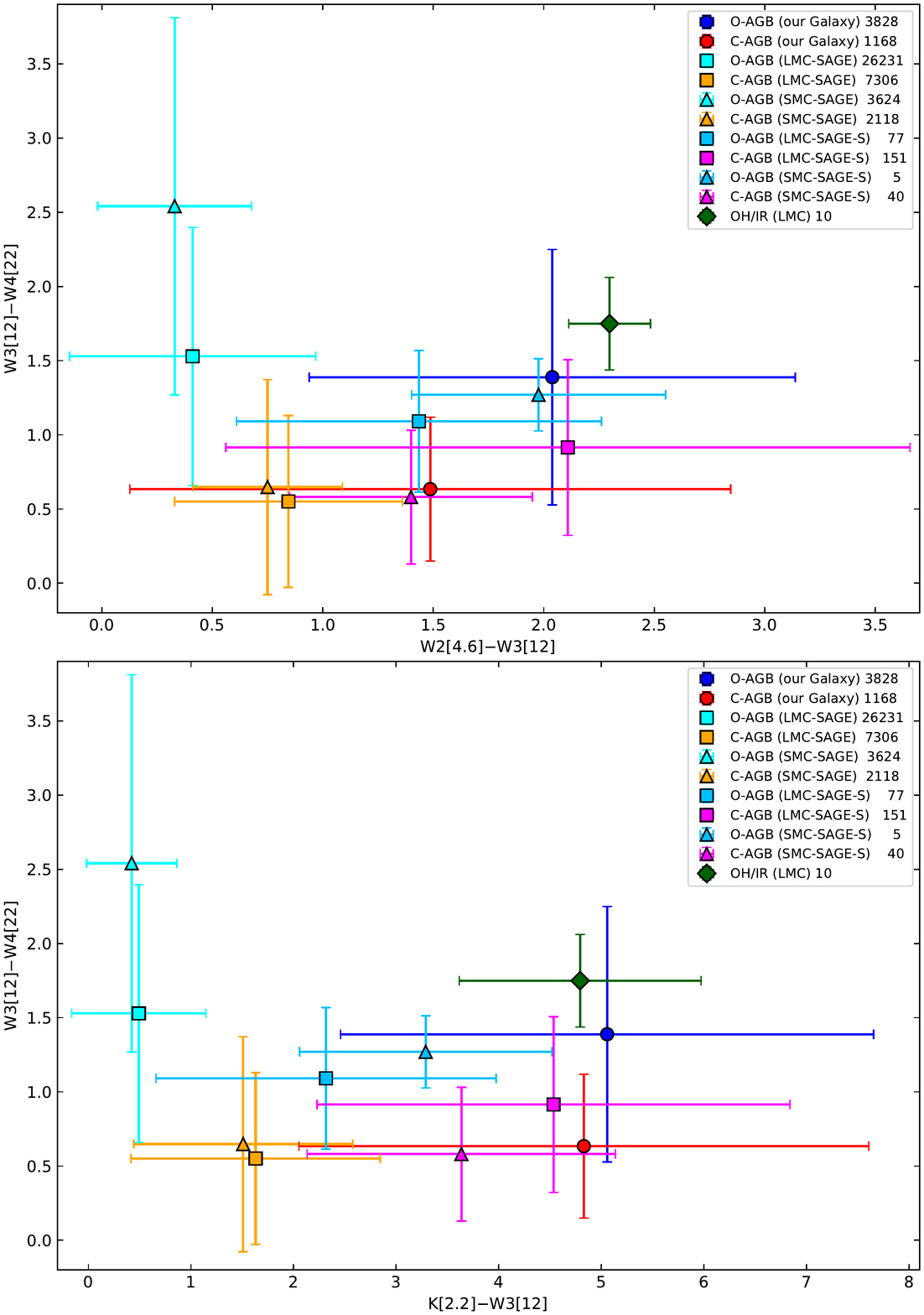}{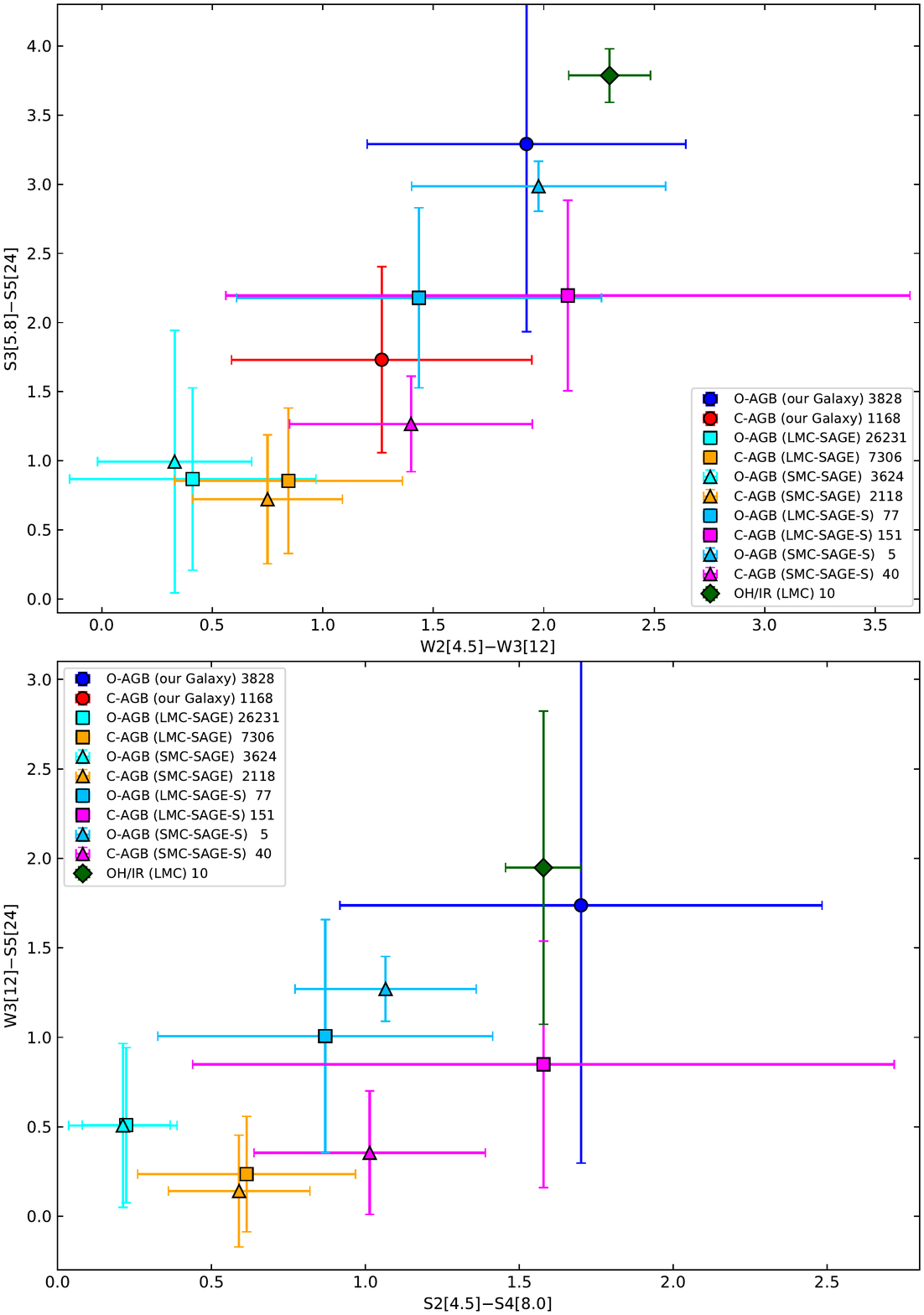}\caption{Averaged observed colors for sample stars
(see Table~\ref{tab:tab1}) used for the IR 2CDs (see Figures~\ref{f3} -~\ref{f8}).
We plot only SAGE sample stars for the Magellanic Clouds.
For the AGB stars in our Galaxy using Spitzer data, we consider only the objects
with small deviations in the S5[24] flux (see Section~\ref{sec:gagb}).}
\label{f10}
\end{figure*}

\section{Infrared Two-Color Diagrams\label{sec:2cds}}

We have complete or nearly complete SEDs from infrared spectroscopy only for a
relatively small number of stars. A large number of stars have infrared
photometric fluxes from the NIR to the FIR band thanks to the 2MASS, WISE, and
Spitzer observations. Although the photometric fluxes are less useful than a
full SED, the large number of observations at various wavelength bands can be
used to form a 2CD, which can be compared with theoretical models. IR 2CDs are
useful to statistically distinguish various properties of AGB stars and we may
use IR 2CDs to find new candidate objects for AGB stars (e.g.,
\citealt{sh2017}).

Table~\ref{tab:tab2} lists the IR bands used for the IR 2CDs presented in this
work. In this work, we ignore reddening effects at all IR bands for all objects
in our Galaxy and the Magellanic Clouds. Though the dereddening coefficient at
the K[2.2] band for objects in the LMC is known to be small (about 0.0372 mag;
\citealt{Riebel2012}), it could be larger for distant AGB stars in our Galaxy.

In this work, we use only good quality observational data at all wavelength
bands for the 2MASS and WISE photometric data (quality A for 2MASS; quality A
or B for WISE; see Section~\ref{sec:photdata}) for plotting IR 2CDs. For the
Spitzer photometric data, we use all of the available data, which are from the
good quality catalogs (see Sections~\ref{sec:gagb} and ~\ref{sec:magb}).

For IR 2CDs, we use the three IR colors at shorter wavelength bands
(W2[4.6]-W3[12], K[2.2]-W3[12], and S2[4.5]-S4[8.0]) for the horizontal axes.
These colors are mostly affected by the dust grains in the inner dust shells so
they are good measures of the overall dust optical depth. For the vertical
axes, we use the three IR colors at longer wavelength bands (W3[12]-W4[22],
S3[5.8]-S5[24], and W3[12]-S5[24]) which are affected by the dust grains in
more detached or outer dust shells.

Figures~\ref{f4} -~\ref{f9} show various IR 2CDs using the four different
combinations of observed IR colors. We compare the observations with the
theoretical dust shell models (see Section~\ref{sec:models}) for AGB stars in
our Galaxy and the Magellanic Clouds. See Section~\ref{sec:comparison} for
comparison between theory and observations.

Figure~\ref{f4} and ~\ref{f5} show the WISE 2CDs using W3[12]-W4[22] versus
W2[4.6]-W3[12] and W3[12]-W4[22] versus K[2.2]-W3[12], respectively. The upper
panels plot AGB stars in our Galaxy and the lower panels plot AGB stars in the
LMC. Figure~\ref{f6} shows the two 2CDs for AGB stars in the SMC.

Figures~\ref{f7} and~\ref{f8} show WISE-Spitzer 2CDs for AGB stars in our
Galaxy and the LMC. Figure~\ref{f7} shows 2CDs using S3[5.8]-S5[24] versus
W2[4.6]-W3[12] and Figure~\ref{f8} shows 2CDs using W3[12]-S5[24] versus
S2[4.5]-S4[8.0]. Figure~\ref{f9} shows the two 2CDs for AGB stars in the SMC.
Note that the Spitzer colors for the same object in the OGLE3 and SAGE samples
can be slightly different because different catalogs are used (see
Section~\ref{sec:magb-c}).

Generally, the stars that have thick dust shells with large dust optical depths
are located in the upper-right regions on the IR 2CDs. But there can be some
deviations. When the wavelength bands are near some dust or gas features (e.g.,
the silicate dust feature at 10 $\mu$m; see Section~\ref{sec:modelcolor}), the
IR colors can be severely affected.

Though both OGLE3 and SAGE sample stars show similar properties on the IR 2CDs,
the SAGE sample can be regarded as more reliable sample of AGB stars because
the selection method was more sophisticated and considered more NIR and MIR
photometric data (see Section~\ref{sec:magb}).

Figure~\ref{f10} shows the error bar plots of the averaged observed colors of
the sample stars for various IR colors used for the four IR 2CDs presented in
Figures~\ref{f4} - ~\ref{f9}. For the objects in the Magellanic Clouds, we
present only the SAGE sample stars.

On all of the IR 2CDs, we also plot the sequences of theoretical dust shell
models at increasing dust optical depth for AGB stars (see
Section~\ref{sec:models}). We will discuss the meanings of these 2CDs in
Section~\ref{sec:comparison} by comparing the observations with the theoretical
models.

\begin{figure}
\centering
\smallplottwo{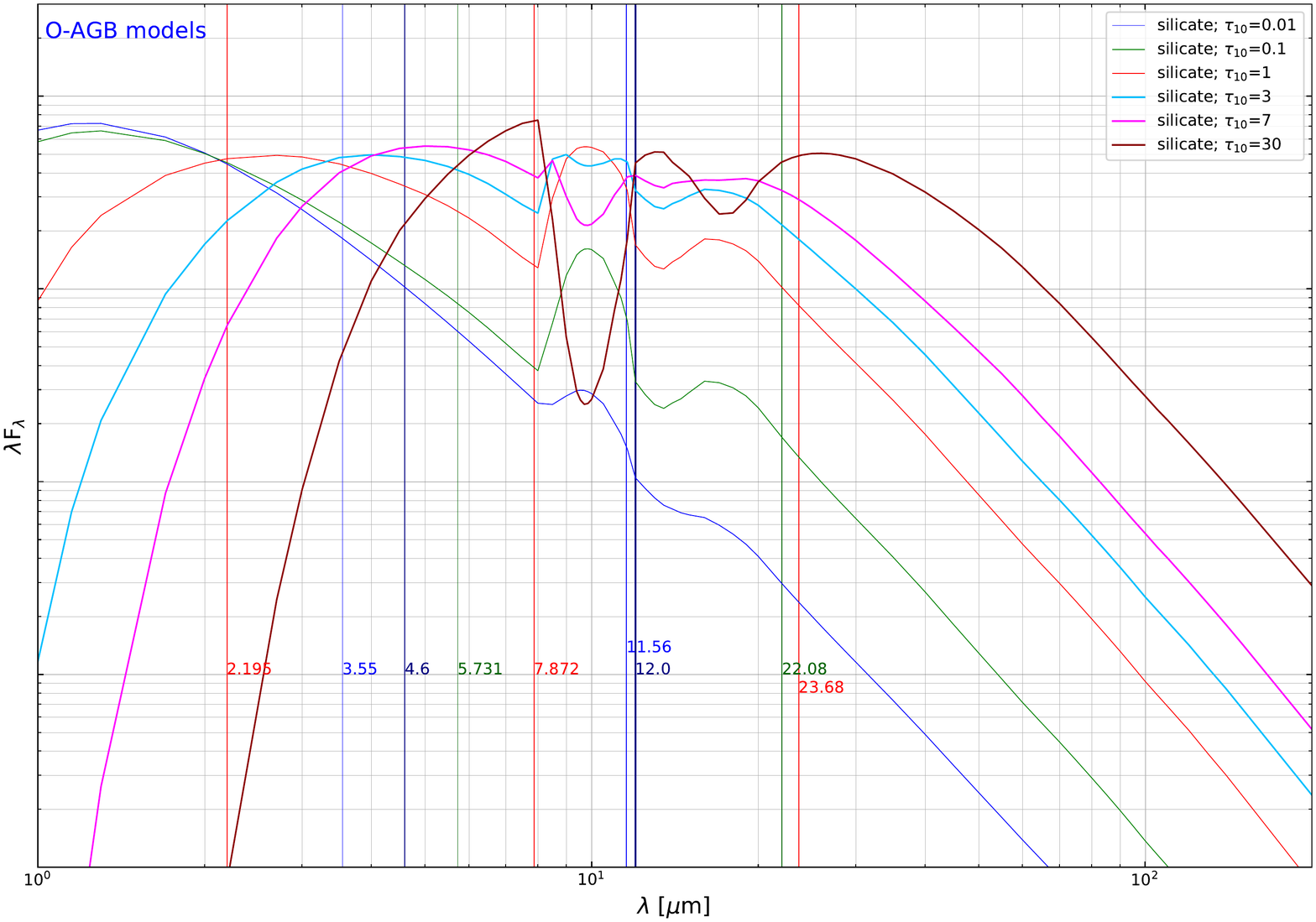}{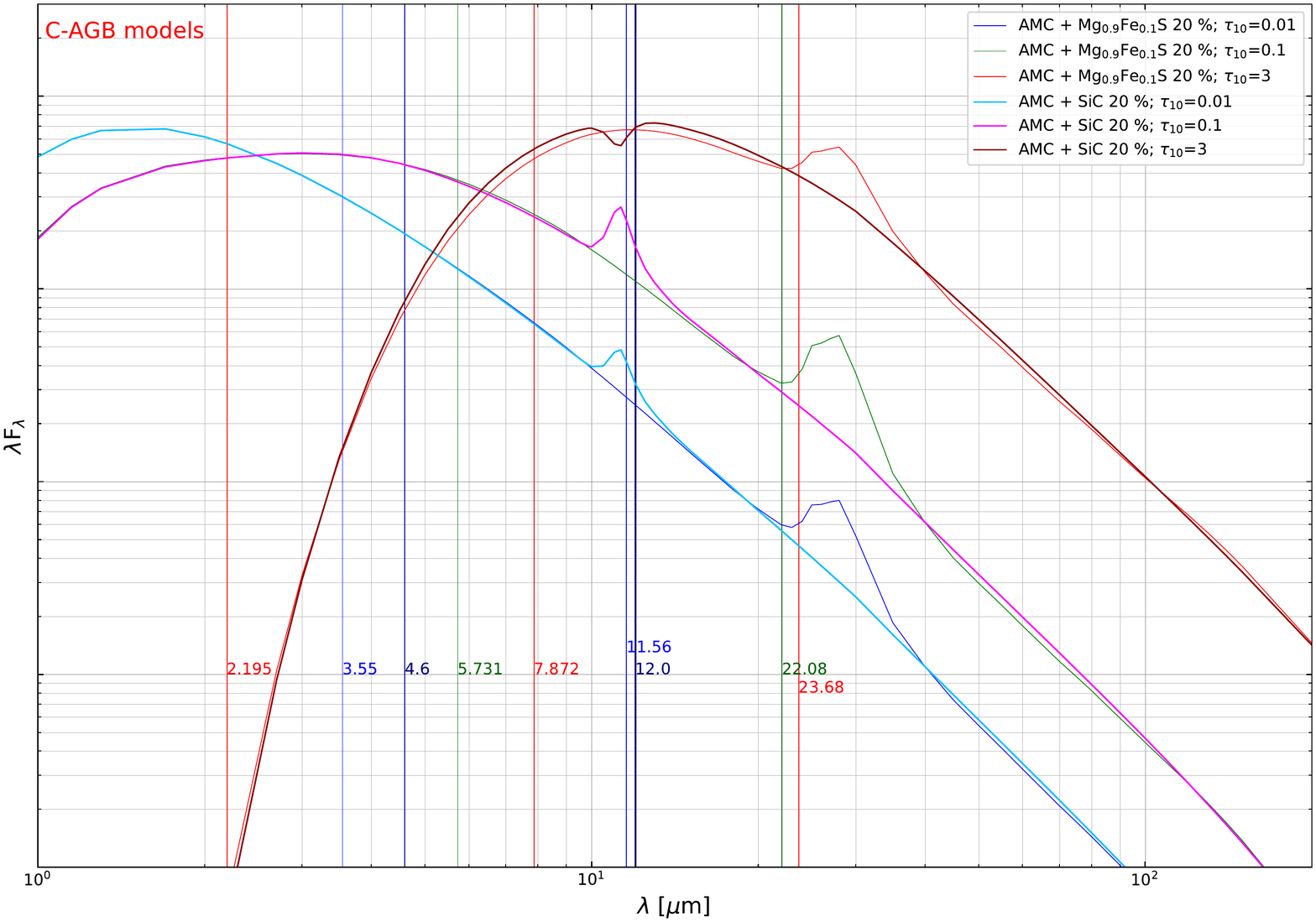}\caption{Theoretical model SEDs for O-AGB stars (silicate; $T_c$=1000 K)
and C-AGB stars (AMC; $T_c$=1000 K) for a number of dust optical depths (see~\ref{sec:agbmodels}).
The reference wavelengths for major IR bands are also indicated (see Table~\ref{tab:tab2}).}
\label{f11}
\end{figure}

\begin{figure}
\centering
\smallplottwo{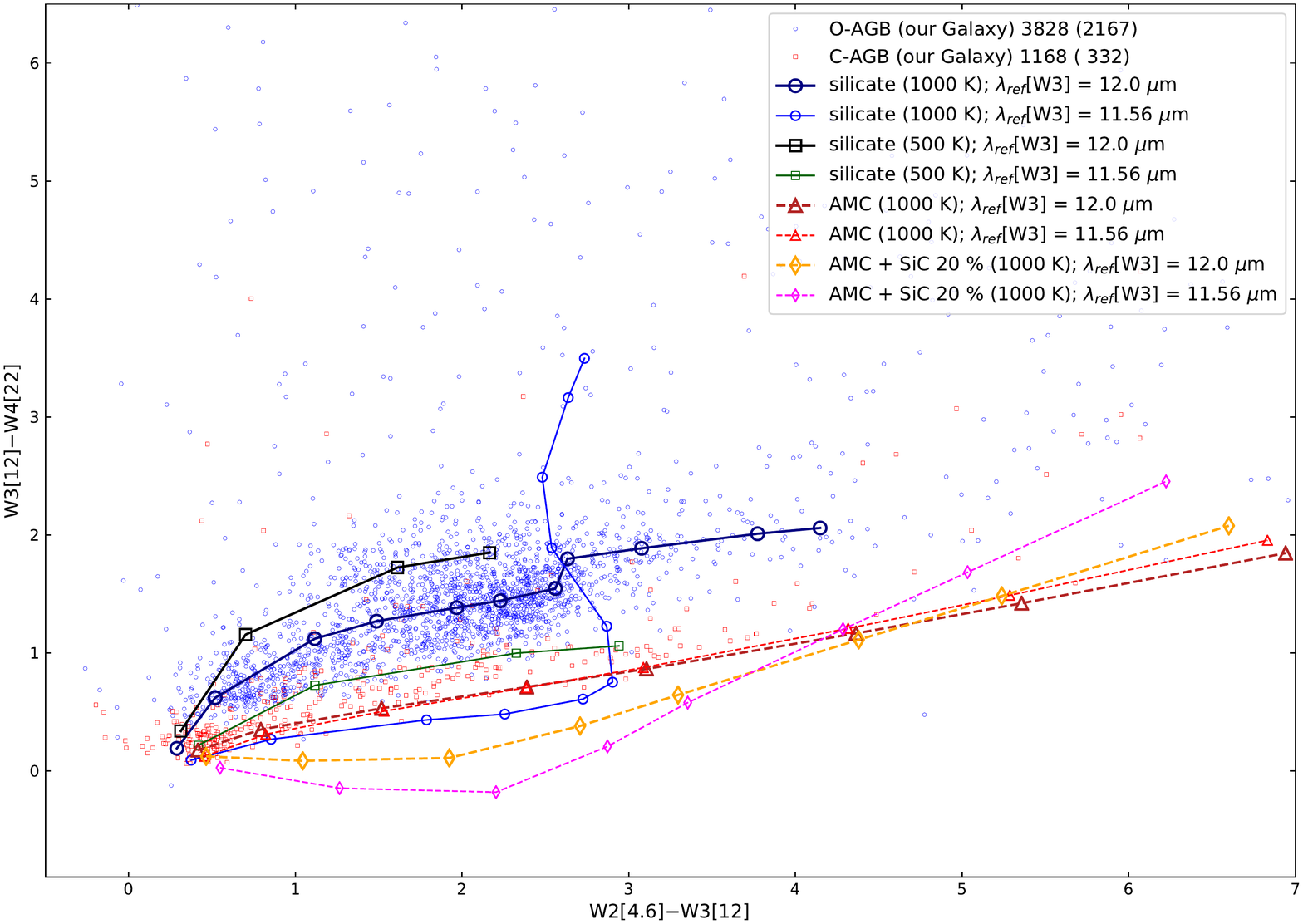}{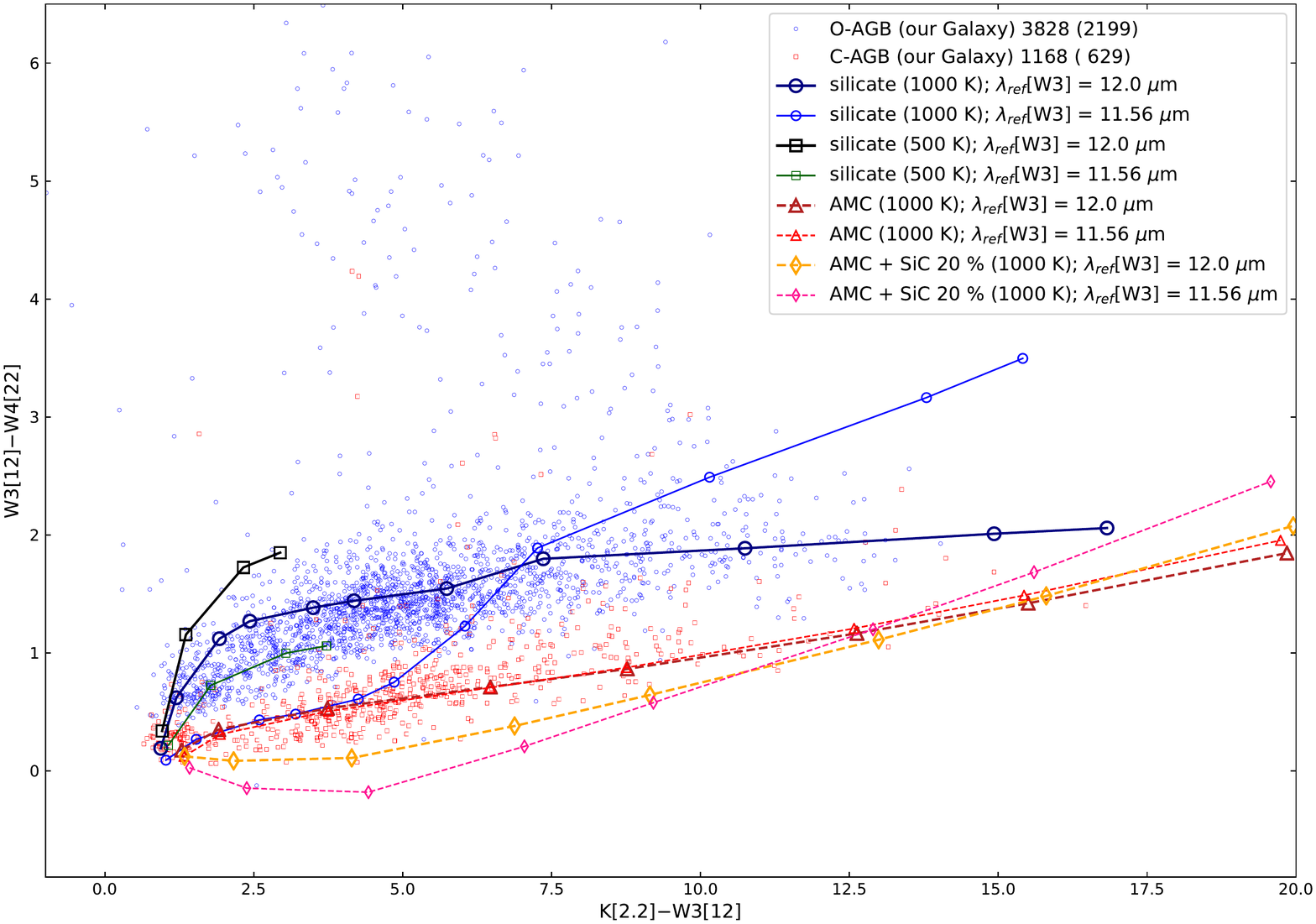}\caption{Theoretical dust shell model tracks on the WISE-2MASS 2CDs for
different reference wavelengths of the W3[12] band (see the model SEDs: Figure~\ref{f11}).
These are compared with the observations of AGB stars in our Galaxy.
In this work, we use the reference wavelength of 12 $\mu$m for the WISE W3[12] band (see Table~\ref{tab:tab2}).
This choice produces model colors that make a better fit with the observations of AGB stars in our Galaxy
and the Magellanic Clouds (compare with Figures~\ref{f4} -~\ref{f9}). See Section~\ref{sec:modelcolor}.}
\label{f12}
\end{figure}

\section{Theoretical Dust Shell Models\label{sec:models}}

On all of the 2CDs in Figures~\ref{f4} -~\ref{f9}, theoretical model tracks for
AGB stars are plotted to be compared with the observations. We use spherically
symmetric dust shell models for AGB stars.

\subsection{Dust Shell Models for AGB Stars\label{sec:agbmodels}}

We use the radiative transfer code DUSTY (\citealt{ivezic1997}) for a
spherically symmetric dust shell. For all models, we use a continuous power law
($\rho \propto r^{-2}$) dust density distribution. We assume that the dust
formation temperature ($T_c$) is 1000 K. But for LMOA stars, we also use 500 K
and 300 K because it is known that LMOA stars have lower $T_c$ (e.g.,
\citealt{suh2004}). The inner radius of the dust shell is set by the $T_c$ and
the outer radius of the dust shell is taken to be $10^4$ times the inner
radius. The radii of spherical dust grains are assumed to be 0.1 $\mu$m
uniformly. We use 10 $\mu$m as the fiducial wavelength of the dust optical
depth ($\tau_{10}$). Because the shape of the model SED is independent on the
luminosity of the central star when all other parameters are fixed, the DUSTY
code calculates the model SED only in relative scale.

For O-AGB stars, we use optical constants of warm and cold silicate dust from
\citet{suh1999}. We compute eleven models ($\tau_{10}$ $=$ 0.001, 0.01, 0.05,
0.1, 0.5, 1, 3, 7, 15, 30, and 40) with $T_c$ = 1000 K. We use warm silicate
for LMOA stars (7 models with $\tau_{10}$ $\leq 3$) and cold silicate for HMOA
stars (4 models $\tau_{10}$ $> 3$). We assume that the stellar blackbody
temperature is 3000 K for $\tau_{10}$ $\leq 0.1$ (four models), 2500 K for $0.1
<$ $\tau_{10}$ $\leq 3$ (three models), and 2000 K for $\tau_{10}$ $> 3$ (four
models). Figure~\ref{f10} shows model SEDs for O-AGB stars (silicate;
$T_c$=1000 K) for six major dust optical depths.

Only for LMOA stars with thin dust shells (the four models with $\tau_{10}$
$\leq 0.1$), we also use lower $T_c$ (500 K and 300 K). Also alumina and Fe-Mg
oxide grains as well as silicates are necessary to reproduce the observed SEDs
(see Section~\ref{sec:intro}). We use three different dust opacity models: a
simple mixture of warm silicate and Fe$_{0.9}$Mg$_{0.1}$O (20\% by number) and
a simple mixture of warm silicate and alumina (20\% by number) as well as pure
warm silicate. For Fe$_{0.9}$Mg$_{0.1}$O dust, we use the optical constants
from \citet{henning1995}. For alumina dust, we use the optical constants from
\citet{suh2016}, which were derived from the optical constants in narrower
wavelength range obtained by \citet{begemann1997}.

For C-AGB stars, we use the optical constants of AMC and SiC dust grains from
\citet{suh2000} and \citet{pegouri1988}, respectively. For MgS dust, we use the
optical constants of Mg$_{0.9}$Fe$_{0.1}$S dust, which is close to pure MgS,
from \citet{begemann1994}. We compute eight models ($\tau_{10}$ $=$ 0.001,
0.01, 0.1, 0.5, 1, 2, 3, and 5) with $T_c$ = 1000 K. We assume that the stellar
blackbody temperature is 2500 K for $\tau_{10}$ $< 1$ (four models) and 2000 K
for $\tau_{10}$ $\geq 1$ (four models). We use three different dust opacity
models: a simple mixture of AMC and Mg$_{0.9}$Fe$_{0.1}$S (20 \% by number) and
a simple mixture of AMC and SiC (10\% by number) as well as pure AMC.

Figure~\ref{f11} shows model SEDs for AGB stars ($T_c$=1000 K) for major dust
optical depths. For O-AGB models, silicate dust features at 10 and 18 $\mu$m
are shown for various dust optical depths ($\tau_{10}$). For LMOA stars,
amorphous alumina and Fe$_{0.9}$Mg$_{0.1}$O dust grains produce broad emission
features at 11.8 and 19.6 $\mu$m, respectively (see \citealt{suh2018}). For
C-AGB models, SiC dust features at 11.3 $\mu$m and Mg$_{0.9}$Fe$_{0.1}$S dust
features at 28 $\mu$m are shown for different dust optical depths.

The gas-to-dust ratio ($\Psi$) is generally estimated to be 50 - 200 in our
Galaxy (the average $\Psi$ is about 100) and $\Psi$ tends to decrease for a
higher metallicity (\citealt{draine2007}). \citet{nanni2019} found that $\Psi$
is larger in the Magellanic Clouds ($\Psi \sim$ 700) probably due to the lower
metallicity. The optical depths of the dust shells around AGB stars would be
dependent on the initial masses and metallicity (e.g., \citealt{ventura2016}).
In a galaxy with a higher metallicity (and a lower $\Psi$), the star formation
in the higher mass range would be more active and the galaxy would have a
higher ratio of AGB stars with thick dust shell.

If we assume that the stellar blackbody luminosity is $10^4$ $L_{\odot}$,
$\Psi$ = 100, and dust shell expansion velocity is 15 km/sec, the mass-loss
rates are $3.8 \times 10^{-7}$, $7.1 \times 10^{-6}$, and $6.5 \times 10^{-5}
M_{\odot}/yr$ for LMOA stars ($T_c$=1000 K; $\tau_{10}$ = 0.1), C-AGB stars
with $\tau_{10}$ = 1, and HMOA star with $\tau_{10}$ = 15, respectively.

\subsection{Model color indices\label{sec:modelcolor}}

To compare theoretical models with observations on 2CDs, we need to obtain
model colors from the model SEDs. We obtain the model colors using the
reference (or effective or isophotal) wavelength and zero magnitude flux (ZMF)
given in the reference for the 2MASS, WISE, and Spitzer photometric data (see
Table~\ref{tab:tab2}).

The color index is defined by
\begin{equation}
M_{\lambda 1} - M_{\lambda 2} = - 2.5 \log_{10} {{F_{\lambda 1} / ZMF_{\lambda 1}} \over {F_{\lambda 2} / ZMF_{\lambda 2}}}
\end{equation}
where $ZMF_{\lambda i}$ is the ZMF at given wavelength ($\lambda i$) (see
Table~\ref{tab:tab2}).

The reference wavelength for the W3[12] band largely affects the model colors
for O-AGB stars because the wavelength is very near the conspicuous 10 $\mu$m
silicate features of the model SEDs (see Figure~\ref{f10}). For the WISE W3[12]
band, the isophotal wavelength is 11.56 $\mu$m (\citealt{jarrett2011}) and the
response function weighted average wavelength is 12.33 $\mu$m. The isophotal
wavelength (11.56 $\mu$m) for the W3[12] band, which was obtained from the
observations of Vega (\citealt{jarrett2011}), could be too short for dusty AGB
stars.

The theoretical model for O-AGB stars used in this work reproduced various
spectral and photometric observations of O-AGB stars (e.g., \citealt{suh2002};
\citealt{suh2004}; \citealt{gonzalez2018}) in wide wavelength ranges reasonably
well. However, the same model produced model colors that show very large
deviations from the observed colors of O-AGB stars in our Galaxy
(\citealt{suh2018}) and the Magellanic Clouds (this work) when we use
$\lambda_{ref}$ = 11.56 $\mu$m for the W3[12] band.

For the WISE W3[12] band, we use the reference wavelength of 12 $\mu$m and ZMF
of 28.3 Jy (the same values as those for the IRAS [12] band;
\citealt{beichman1988}) to obtain the theoretical model colors (see
Table~\ref{tab:tab2}).

Figure~\ref{f12} shows IR 2CDs using two different reference wavelengths for
the W3[12] band compared with the observations of AGB stars in our Galaxy. This
small change makes large differences for the O-AGB model colors. Though the SiC
dust feature at 11.3 $\mu$m is also affected, the effect is minor for C-AGB
models without SiC dust (see Figure~\ref{f11}). With this choice
($\lambda_{ref}$ = 12 $\mu$m for the W3[12] band), the model colors fit the
observed colors much better on all IR 2CDs for AGB stars in our Galaxy and the
Magellanic Clouds (see Figures~\ref{f3} -~\ref{f8}).

\subsection{Limitations of the Theoretical Models\label{sec:limit}}

The theoretical dust shell model used in this work does not consider gas-phase
radiation processes (see Section~\ref{sec:models}). AGB stars show various
gas-phase emission or absorption features in NIR and MIR band due to
circumstellar molecules such as H$_2$O, CO, and C$_2$H$_2$ (e.g,
\citealt{lancon2000}; \citealt{LeBertre2005}; \citealt{gonneau2016}). The
deviations of the theoretical models from the observations would be larger at
the wavelength bands where gas-phase radiation processes are more active.

Also, the spherically symmetric dust shell model does not consider nonspherical
dust envelopes. The observed colors of AGB stars with nonspherical dust
envelopes can show various deviations from the theoretical models at NIR and
MIR bands.

\begin{figure*}
\centering
\smallplotfour{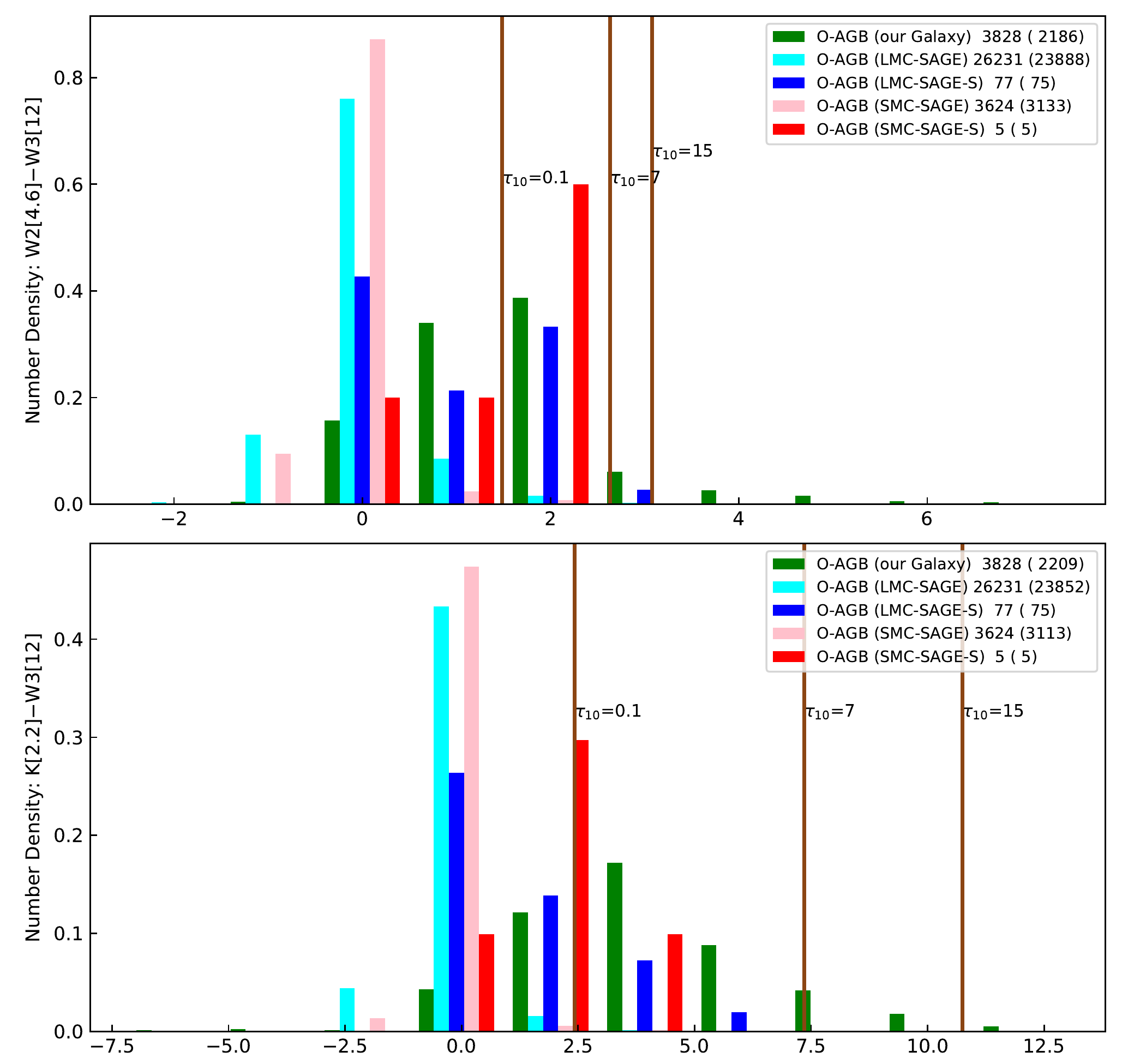}{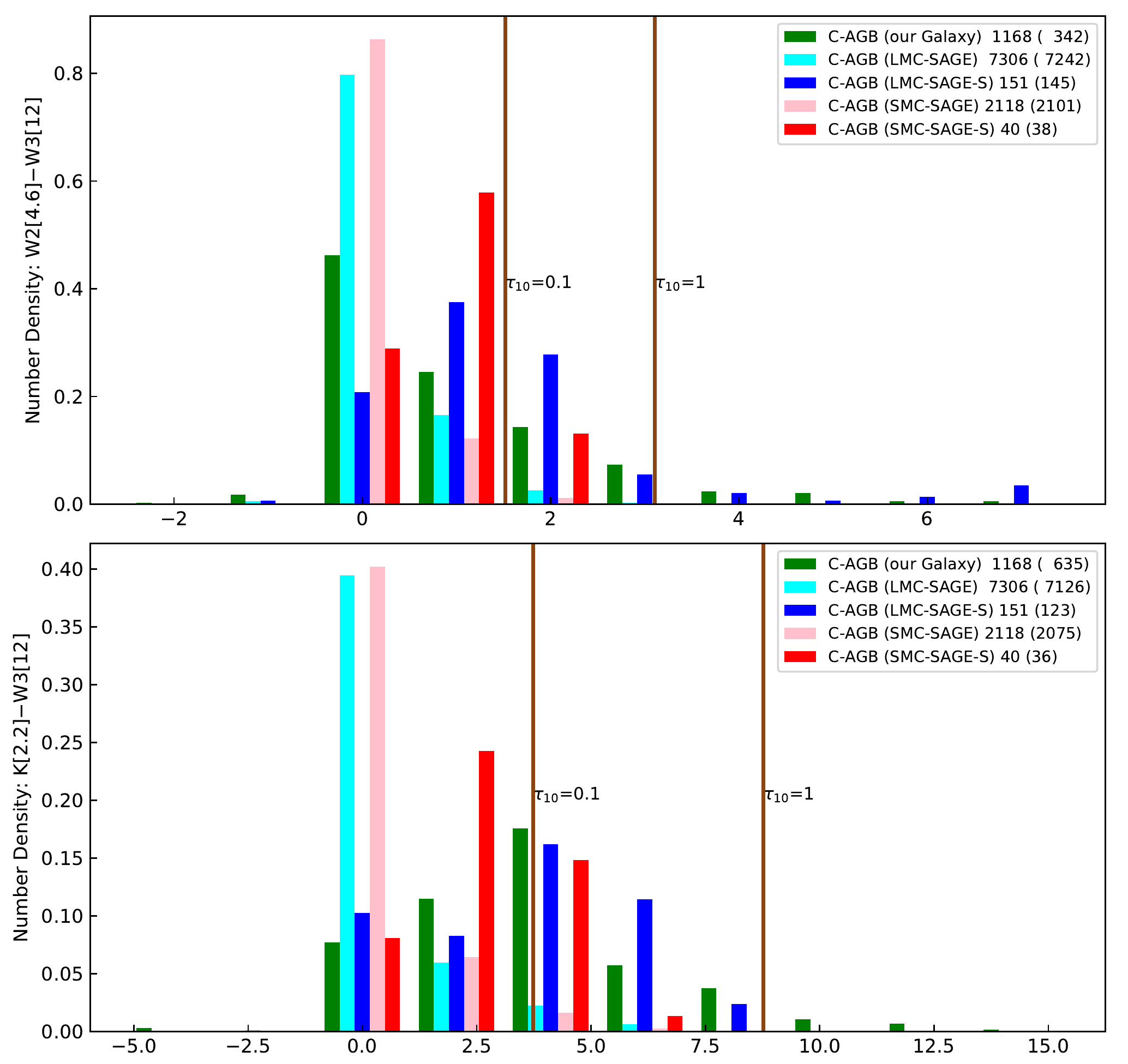}{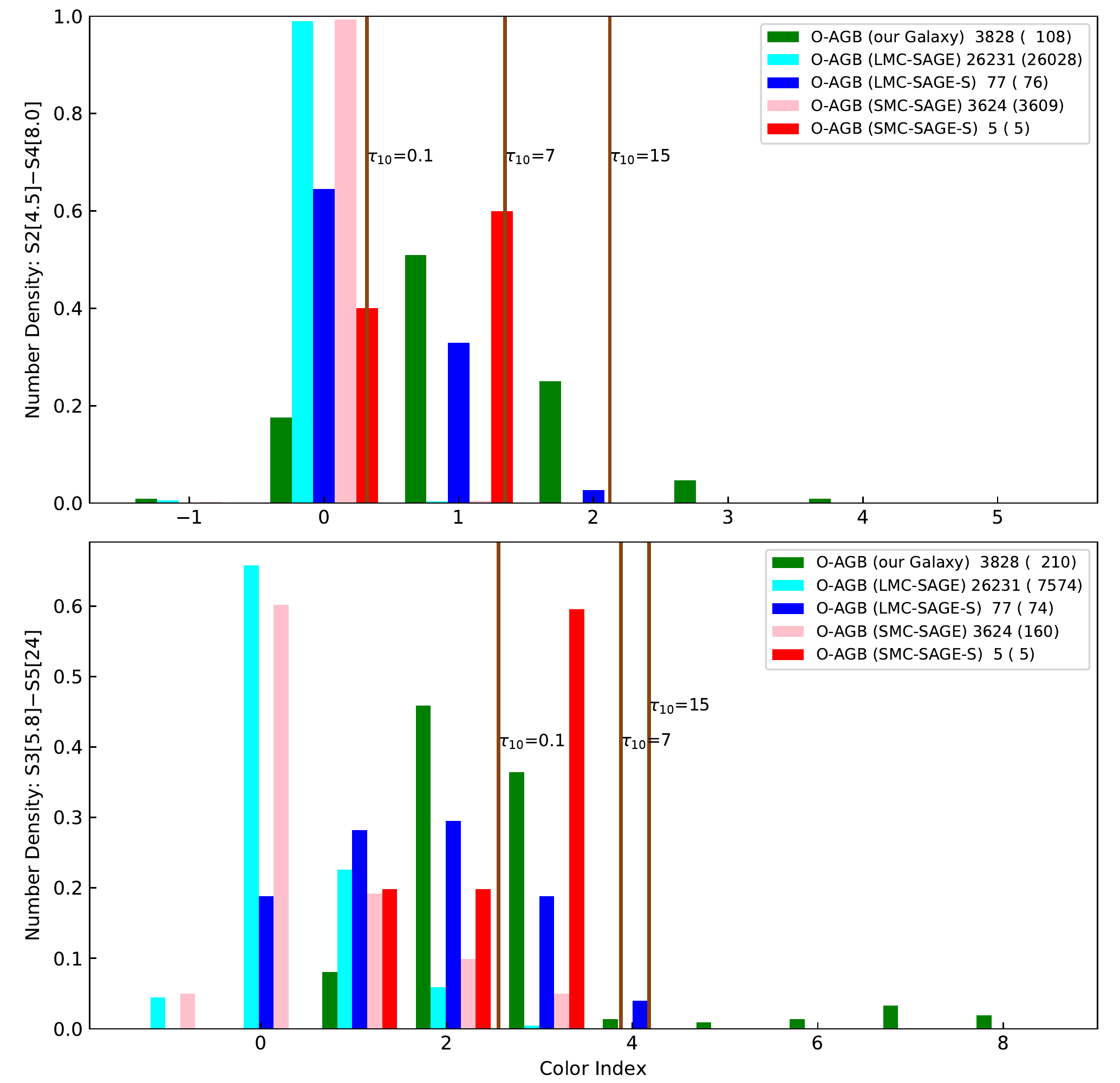}{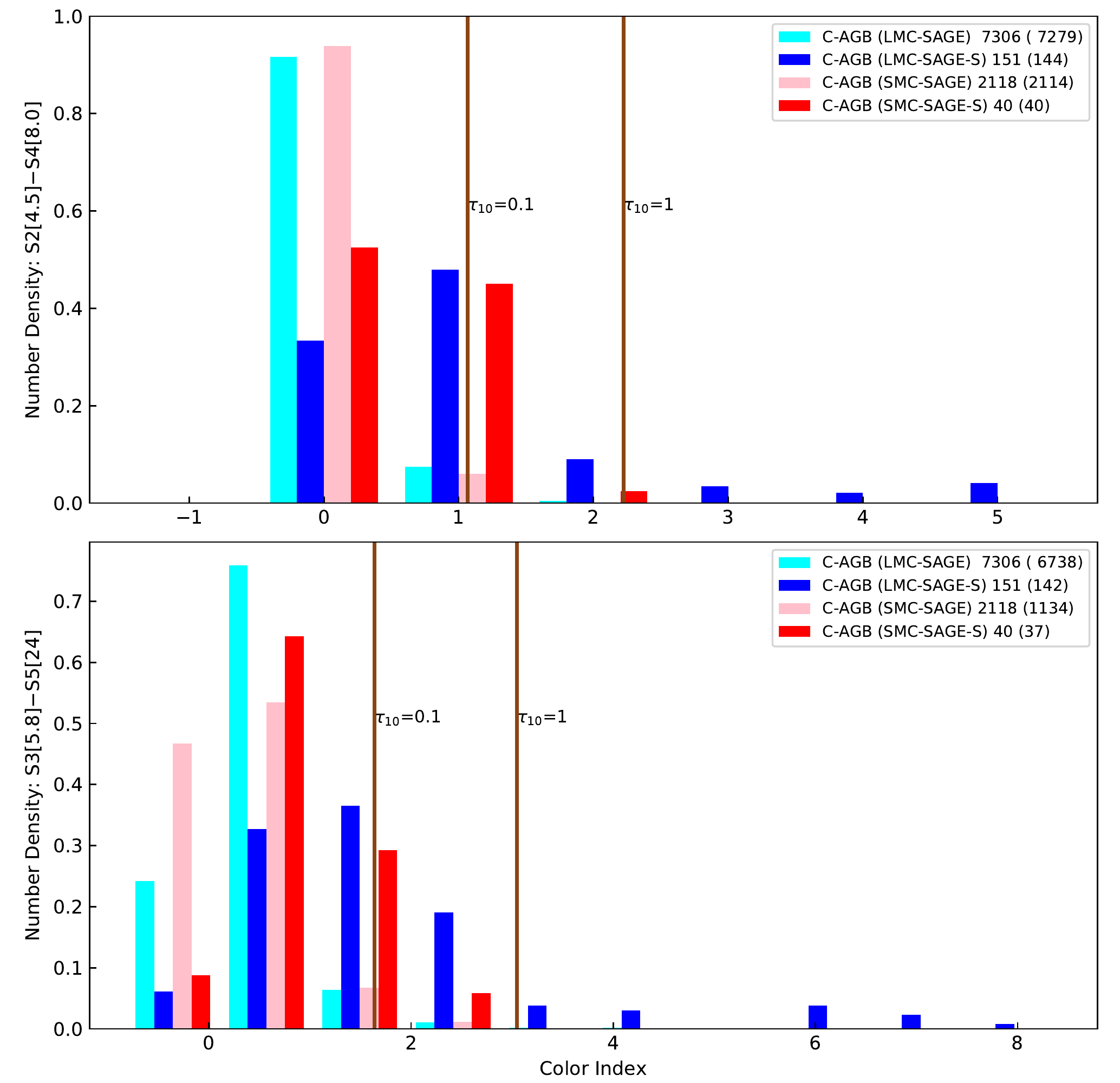}\caption{Number density distributions of observed IR colors for O-AGB and
C-AGB stars in our Galaxy and the Magellanic Clouds (LMC-SAGE and SMC-SAGE).
The vertical brown lines indicate theoretical model colors for the two or three dust shell optical depths ($\tau_{10}$).
For each class, the number of objects is also shown.
The number in parenthesis denotes the number of the plotted objects with good quality observed colors.
See Table~\ref{tab:tab3}.}
\label{f13}
\end{figure*}

\begin{table*}
\scriptsize
\centering
\caption{Percentages of the objects with thin and thick dust shells from the comparison of the observed
colors with the theoretical models. See Figure~\ref{f13}.\label{tab:tab3}}
\begin{tabular}{lllllllll}
\hline
\hline
Class$^1$        & W2[4.6]-W3[12] & K[2.2]-W3[12] &S1[3.6]-S4[8.0] &S2[4.5]-S4[8.0] &S3[5.8]-S5[24]  &[12]-[25]$^2$ &K-[12]$^2$\\
\hline
O-AGB (our Galaxy)	&29.0, 17.5 (2186)	&14.4, 17.6 (2209)	&3.7, 39.7 	(136)	&2.8, 67.6 (108)	&21.9, 13.8 (210)	&45.2, 13.8 (3568)	&8.4, 21.9 (2189)\\
O-AGB (LMC-SAGE)	&95.1, 0.7 (23888)	&98.4, 0.01 (23852)	&97.9, 0.04	(25907)	&85.2, 0.2 (26028)	&97.6, 0.1 (7574)	&-			&-		\\
O-AGB (LMC-SAGE-S)	&52.0, 5.3 (75)		&64.0, 0.0 (75)		&40.5, 5.4	(74)	&14.5, 29.0 (76)	&59.5, 8.1 (74)		&-			&-		\\
O-AGB (SMC-SAGE)	&98.3, 0.2 (3133)	&99.2, 0.0 (3113)	&98.7, 0.1 	(3610)	&90.0, 0.3 (3609)	&91.9, 0.6 (160)	&-			&-		\\
O-AGB (SMC-SAGE-S)	&20.0, 0.0 (5)		&20.0, 0.0 (5)		&0.0, 0.0 	(5)	&0.0, 20.0 (5)		&40.0, 20.0 (5)		&-			&-		\\
\hline
C-AGB (our Galaxy)	&61.7, 12.0 (342)	&33.7, 8.5 (635)	&-			&-			&-			&52.7, 23.0 (1098)	&22.3, 18.2 (687)\\
C-AGB (LMC-SAGE)	&92.2, 0.6 (7242)	&92.3, 0.06 (7126)	&93.3, 0.6 	(7247)	&92.8, 0.6 (7279)	&94.0, 0.6 (6738)	&-			&-		\\
C-AGB (LMC-SAGE-S)	&37.2, 13.1 (145)	&34.2, 1.6 (123)	&37.1, 13.3 	(143)	&36.8, 13.9 (144)	&40.9, 14.8 (142)	&-			&-		\\
C-AGB (SMC-SAGE)	&95.3, 0.0 (2101)	&93.8, 0.0 (2075)	&95.7, 0.1 	(2115)	&95.3, 0.1 (2114)	&94.4, 0.2 (1134)	&-			&-		\\
C-AGB (SMC-SAGE-S)	&60.5, 0.0 (38)		&47.2, 0.0 (36)		&55.0, 2.5 	(40)	&55.0, 2.5 (40)		&78.4, 2.7 (37)		&-			&-		\\
\hline
O-AGB (our Galaxy)$^*$	&10.1 (2186)	&2.8 (2209)	&11.0 	(136)	&22.2 (108)	&8.1 (210)	&9.2 (3568)	&4.8 (2189)\\
O-AGB (LMC-SAGE)$^*$	&0.37 (23888)	&0.004 (23852)	&0.0 (25907)	&0.008 (26028)	&0.04 (7574)	&-		&-	\\
O-AGB (LMC-SAGE-S)$^*$	&2.67 (75)	&0.0 (75)	&0.0	(74)	&1.32 (76)	&1.35 (74)	&-		&-	\\
O-AGB (SMC-SAGE)$^*$	&0.064 (3133)	&0.0 (3113)	&0.055 (3610)	&0.055 (3609)	&0.0 (160)	&-		&-	\\
O-AGB (SMC-SAGE-S)$^*$	&0.0 (5)	&0.0 (5)	&0.0 	(5)	&0.0 (5)	&0.0 (5)	&-		&-	\\
\hline
\end{tabular}
\begin{flushleft}
For each column of the IR color, the percentages of the objects with thin and thick dust shells
for O-AGB stars ($\tau_{10} <$ 0.1, $\tau_{10} >$ 7) and C-AGB stars ($\tau_{10} <$ 0.1,
$\tau_{10} >$ 1) are listed except for the last five rows.
The last five rows (marked by $^*$) list the percentages of the O-AGB stars with very thick dust shells ($\tau_{10} >$ 15).
The number in parenthesis denotes the number of the observed objects used for the IR 2CDs and histograms.
For the colors of AGB stars in our Galaxy using Spitzer data, we consider only the objects with small deviations in the S5[24] flux (see Section~\ref{sec:gagb}).
$^1$See Table~\ref{tab:tab1} for the sample information.
$^2$The data for the IRAS-2MASS colors are from \citet{sh2017}.
\end{flushleft}
\end{table*}

\begin{table}
\caption{Weighted averaged percentages$^1$ of AGB stars with thin and thick dust shells\label{tab:tab4}}
\centering
\begin{tabular}{lllll}
\hline \hline
Class    &Dust Shell & Our Galaxy	& LMC & SMC  \\
\hline
O-AGB  & silicate ($\tau_{10}) <$ 0.1  & 27.1 \% & 79.5 \% & 82.7 \% \\
O-AGB  & silicate ($\tau_{10}) >$ 7  & 17.2 \% & 3.4 \% & 1.8 \% \\
C-AGB  & AMC ($\tau_{10} >$ 1)   & 17.1 \% & 9.4 \%&  1.3 \%\\
O-AGB  & silicate ($\tau_{10} >$ 15)  & 7.0 \%  & 0.44 \% & 0.048 \% \\
\hline
\end{tabular}
\begin{flushleft}
\scriptsize
$^1$The averages are weighted by the numbers of observed objects listed in Table~\ref{tab:tab3} (see the text for details).
For the Magellanic clouds, the numbers of the observed objects in SAGE-S samples are multiplied by 100.
\end{flushleft}
\end{table}

\section{Comparison between Theory and Observations\label{sec:comparison}}

On various IR 2CDs using four different combinations of IR colors in
Figures~\ref{f4} -~\ref{f9}, we compare the observations with the theoretical
dust shell models (see Section~\ref{sec:models}) for AGB stars. We find that
the theoretical dust shell model can roughly reproduce the observations of AGB
stars on the IR 2CDs using the dust opacity functions of amorphous silicate and
amorphous carbon with a mixture of other dust species.

Compared with our Galaxy, we find that more AGB stars in the LMC and SMC are
located in the lower-left regions of any IR 2CDs. For all of the observed
colors (W2[4.6]-W3[12], W3[12]-W4[22], K[2.2]-W3[12], S2[4.5]-S4[8.0],
S3[5.8]-S5[24], and W3[12]-S5[24]), the averaged color of AGB stars in the LMC
or SMC is bluer than that of AGB stars in our Galaxy (see Figure~\ref{f10}).
Note that this difference is systematic only for the averaged K[2.2]-W3[12]
color of the O-AGB stars, and the error bars overlap for all other colors.

Though the methods of chemical classification into O or C for the two AGB
samples of the Magellanic Clouds (OGLE3 and SAGE) are different, only a minor
portion of objects are classified into different classes (see
Section~\ref{sec:magb}). The identification and chemical classification for the
SAGE samples would be more reliable because they are based on the comparison of
the more photometric data at NIR and MIR bands with the Grid of AGB and RSG
ModelS (GRAMS) from \citet{Riebel2012} and \citet{Srinivasan2016}. The SAGE-S
samples of AGB stars would be even more reliable because they were obtained
using the Spitzer IRS spectral data. For the comparison of number distributions
of AGB stars on the IR 2CDs with the theoretical models, we use only the SAGE
samples (including the SAGE-S samples) for the objects in the Magellanic
Clouds.

\subsection{Number distributions of IR colors\label{sec:cnumber}}

We may compare the number distribution of observed IR colors with the
theoretical model. Figure~\ref{f13} shows the number density distributions of
the four observed IR colors for AGB stars in our Galaxy and the Magellanic
Clouds. The four IR colors (W2[4.6]-W3[12], K[2.2]-W3[12], S2[4.5]-S4[8.0], and
S3[5.8]-S5[24]) are good measures of the dust optical depth ($\tau_{10}$) as we
can see on the IR 2CDs (see Figures~\ref{f4} -~\ref{f9}).

We mark the theoretical model colors on Figure~\ref{f13}. For O-AGB stars, the
dust shell (silicate; $T_c$=1000 K) model colors for typical LMOA stars
($\tau_{10}$=0.1) and HMOA stars ($\tau_{10}$=7 and 15) are indicated. For
C-AGB stars, the dust shell (AMC; $T_c$=1000 K) model colors for the thin dust
shell ($\tau_{10}$=0.1) and thick dust shell ($\tau_{10}$=1) model colors are
indicated. See Section~\ref{sec:agbmodels} for the detailed model parameters.

We may also obtain the percentages of the objects with thin or thick dust
shells using the information presented in Figure~\ref{f13}.
Table~\ref{tab:tab3} lists the percentages of the objects with thin (O-AGB:
$\tau_{10} <$ 0.1; C-AGB: $\tau_{10} <$ 0.1), thick (O-AGB: $\tau_{10} >$ 7;
C-AGB: $\tau_{10} >$ 1), and very thick (O-AGB: $\tau_{10} >$ 15) dust shells
from the comparison of the observed colors with the theoretical models for the
seven IR colors: W2[4.6]-W3[12], K[2.2]-W3[12], S1[3.6]-S4[8.0],
S2[4.5]-S4[8.0], S3[5.8]-S5[24], IRAS [12]-[25], and 2MASS-IRAS K-[12]. Though
the three colors (S1[3.6]-S4[8.0], IRAS [12]-[25], and 2MASS-IRAS K-[12]) were
not used for the IR 2CDs presented in this paper, they are also good measures
of the dust optical depth. The data for IRAS [12]-[25] and 2MASS-IRAS K-[12]
colors are from \citet{sh2017}, which presented IR properties of AGB stars in
our Galaxy.

Table~\ref{tab:tab4} lists weighted averaged percentages of the objects with
thick dust shells (O-AGB: $\tau_{10} >$ 7; C-AGB: $\tau_{10} >$ 1; O-AGB:
$\tau_{10} >$ 15) obtained from Table~\ref{tab:tab3}. For AGB stars in our
Galaxy, we obtain the weighted averaged percentages from the data for four IR
colors: W2[4.6]-W3[12], K[2.2]-W3[12], IRAS [12]-[25], and 2MASS-IRAS K-[12].
We do not use Spitzer colors for our Galaxy because the sample number is small
and they show large deviations.

For AGB stars in the Magellanic Clouds, we obtain the weighted averaged
percentages from the data for four IR colors: W2[4.6]-W3[12], S1[3.6]-S4[8.0],
S2[4.5]-S4[8.0], S3[5.8]-S5[24]. We do not consider K[2.2]-W3[12] color because
it shows 'bluing effect' for the Magellanic Clouds (see
Section~\ref{sec:wise-2mass}). In obtaining the weighted averaged percentages
for the objects in the Magellanic Clouds, the numbers of observed objects for
SAGE-S samples are multiplied by 100 because the SAGE-S samples are more
reliable than the SAGE samples.

Compared with our Galaxy, we find that the LMC and SMC are deficient in O-AGB
stars with thick dust shells on any IR 2CDs. The weighted averaged percentages
of HMOA stars with thick dust shells ($\tau_{10} >$ 7) for our Galaxy (17.2 \%)
is larger than the ones for the LMC (3.4 \%) and SMC (1.8 \%) SAGE sample stars
(see Table~\ref{tab:tab4}). The percentages of HMOA stars with very thick dust
shells ($\tau_{10} >$ 15) for our Galaxy (7.0 \%) is even larger than the ones
for the LMC (0.44 \%) and SMC (0.348 \%).

For C-AGB stars in our Galaxy and the LMC, the observations can be reproduced
by the C-AGB models in wide ranges of the dust optical depth ($\tau_{10}$ =
0.001 - 5) on all IR 2CDs except for the 2CD using the K[2.2]-W3[12] color, for
which AGB stars in the LMC show the `bluing' effect (see
Section~\ref{sec:wise-2mass}). The weighted averaged percentages of C-AGB stars
with thick dust shells ($\tau_{10} >$ 1) for our Galaxy (17.1 \%) is larger
than the ones for the LMC (9.4 \%) and SMC (1.3 \%) (see Table~\ref{tab:tab4}).

Compared with our Galaxy, we find that much larger portions of O-AGB stars in
the LMC and SMC have thin dust shells with smaller dust optical depths
($\tau_{10}$) on the IR 2CDs (see Figures~\ref{f4} -~\ref{f9} and histograms in
Figure~\ref{f13}). The weighted averaged percentages of LMOA stars with thin
dust shells ($\tau_{10} <$ 0.1) for the LMC (79.5 \%) or SMC (82.7 \%) are much
larger than the ones for our Galaxy (27.1 \%) (see Table~\ref{tab:tab4}). This
could be due to a selection effect. In our Galaxy, it is difficult to identify
the optically visible AGB stars (with thin dust shells) using optical or NIR
surveys because of the severe extinction by the Galactic disk. Note that the
sample of Galactic AGB stars is based on the IRAS PSC (see
Section~\ref{sec:gagb}).

\subsection{WISE 2CDs\label{sec:wise}}

Figure~\ref{f4} shows WISE 2CDs using W3[12]-W4[22] versus W2[4.6]-W3[12]. The
upper panel plots AGB stars in our Galaxy and the lower panel plots AGB stars
in the LMC.

For LMOA stars with thin dust shells in our Galaxy and the LMC, the silicate
dust with a mixture of amorphous alumina (Al$_2$O$_3$) and
Fe$_{0.9}$Mg$_{0.1}$O can explain wider regions on the IR 2CDs. Compared with
our Galaxy, we find that LMOA stars in the LMC have more detached dust shells
($T_c ~$ 300 K).

For O-AGB stars, the ratio of HMOA stars with thick dust shells ($\tau_{10} >$
7) for the LMC (SAGE-S: 5.3 \%; SAGE: 0.68 \%) is smaller than the one for our
Galaxy (17.5 \%) (see Table~\ref{tab:tab3}). Even for the most dusty O-AGB
stars (or OH/IR stars) in the LMC, $\tau_{10}$ would be about 1-7, which
produce conspicuous emission or shallow self absorption silicate features at 10
$\mu$m (see Figure~\ref{f10}; see \citealt{jones2017} for the Spitzer IRS
spectra). On the other hand, $\tau_{10}$ for many O-AGB stars in our Galaxy are
about 10-40, which produce deep silicate absorption features at 10 $\mu$m
(e.g., \citealt{suh1999}; \citealt{suh2004}).

C-AGB stars in our Galaxy and the LMC are located in the wide range of the
C-AGB model colors ($\tau_{10}$ = 0.001 - 5) on this 2CD. For C-AGB stars, the
ratio of C-AGB stars with thick dust shells ($\tau_{10} >$ 1) for the LMC
(SAGE-S: 13.1 \%; SAGE: 0.6 \%) is comparable to the one for our Galaxy (12.0
\%) (see Table~\ref{tab:tab3}). For C-AGB stars, the effect of SiC dust feature
(at 11.3 $\mu$m) on the W3[12]-W4[22] color is conspicuous while the effect of
Mg$_{0.9}$Fe$_{0.1}$S dust feature (at 26 $\mu$m) is small.

The upper panel of Figure~\ref{f6} shows the WISE 2CDs for AGB stars in the
SMC. For C-AGB stars, the ratio of objects with thick dust shells ($\tau_{10}
>$ 1) for the SMC is 0 \%, which is much smaller than those for the LMC and our
Galaxy (see Table~\ref{tab:tab3}).

\subsection{WISE-2MASS 2CDs\label{sec:wise-2mass}}

Figure~\ref{f5} shows WISE-2MASS 2CDs using W3[12]-W4[22] versus K[2.2]-W3[12].
The upper panel plots AGB stars in our Galaxy and the lower panel plots AGB
stars in the LMC.

Unlike other three IR 2CDs (compare with Figures~\ref{f4} -~\ref{f9}), we find
that the observed K[2.2]-W3[12] colors on this WISE-2MASS 2CD are bluer than
the theoretical model colors. Compared with other IR colors, the observed
K[2.2]-W3[12] colors show smaller percentages of thick dust shell for both the
O-AGB and C-AGB stars in our Galaxy and the LMC (see Table~\ref{tab:tab3}). The
effect is much stronger for the AGB stars in the LMC than for those in our
Galaxy. For the LMC-SAGE-S sample stars, the ratio of C-AGB stars with thick
dust shells ($\tau_{10}>$ 1) for the K[2.2]-W3[12] color (1.6 \%) is much
smaller than those for other colors (13.1-14.8 \%) and the ratio of O-AGB stars
with thick dust shells ($\tau_{10}>$ 7) for the K[2.2]-W3[12] color (0.0 \%) is
also much smaller than those for other colors (5.3-29.0 \%).

The cause of the `bluing' effect could be some gas-phase emission at the K[2.2]
band due to circumstellar (or interstellar) molecules (e.g., CO) and/or
inadequate dust opacity for the dust shell model. There have been some studies
on NIR spectra of AGB stars (e.g., \citealt{lancon2000};
\citealt{LeBertre2005}), but it is not yet clear whether AGB star in our Galaxy
and the LMC show similar NIR spectra at the K[2.2] band or not. Though it is
not presented in a 2CD in this paper, we have found that the observed
S1[3.6]-S4[8.0] colors for the LMC does not show the `bluing' effect (see
Table~\ref{tab:tab3}).

The lower panel of Figure~\ref{f6} shows the WISE-2MASS 2CDs for AGB stars in
the SMC. Compared with other IR colors, the observed K[2.2]-W3[12] colors show
small percentages of the AGB stars with thick dust shell in the SMC (see
Table~\ref{tab:tab3}). But it is not clear whether the AGB stars in the SMC
show the `bluing' effect or not because other IR colors also show small
percentages of the AGB stars with thick dust shells.

\begin{figure}
\centering
\smallplottwo{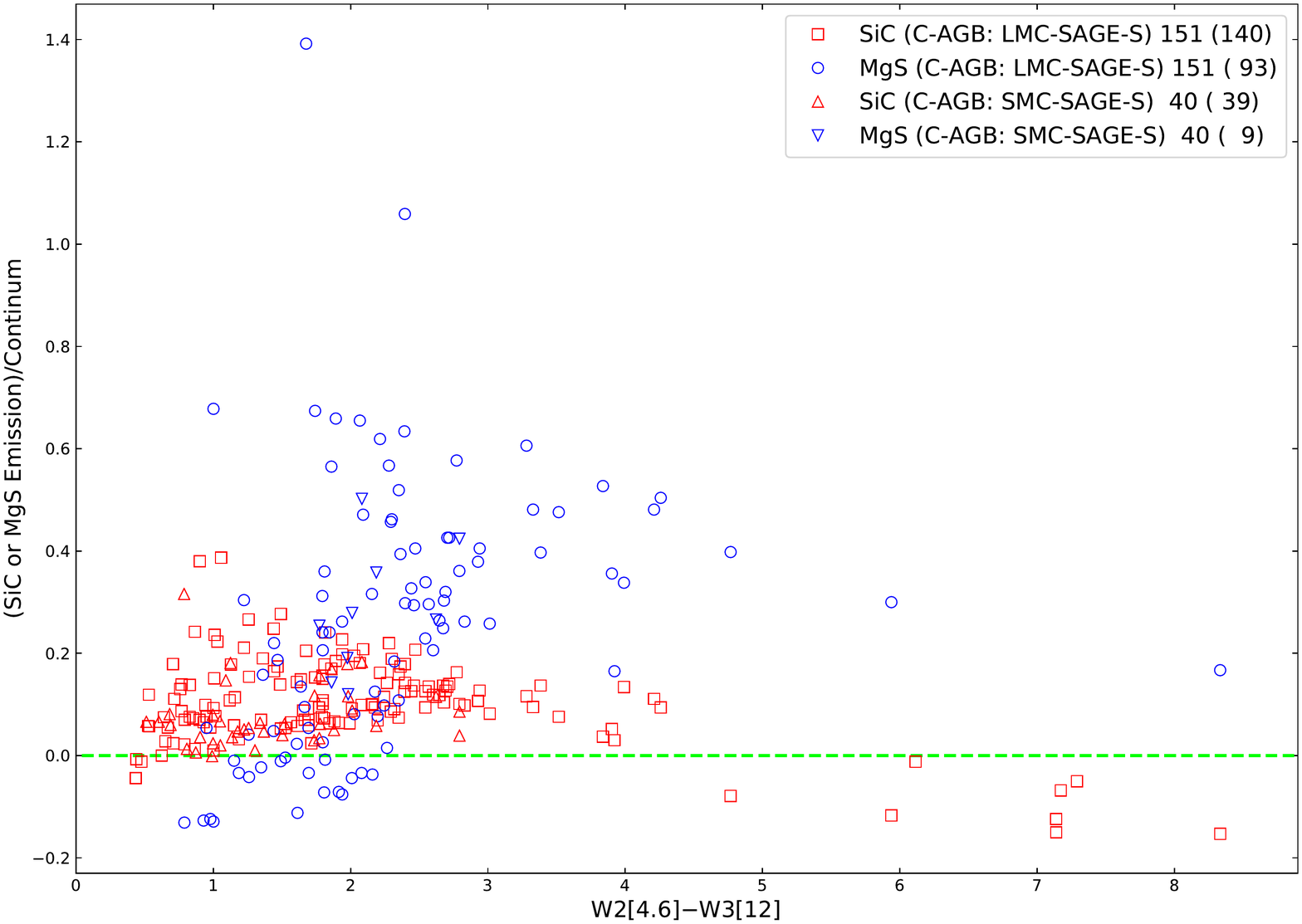}{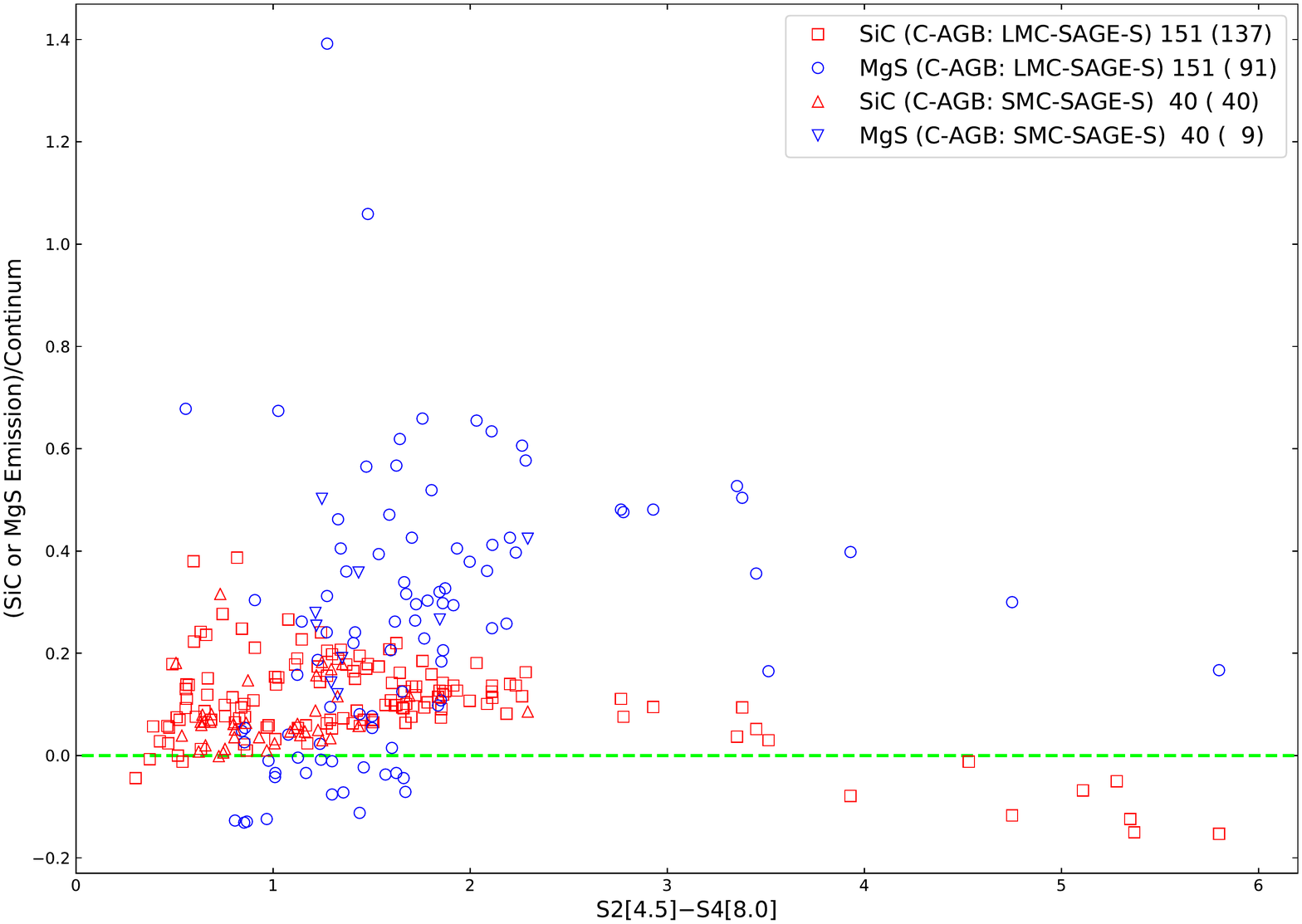}\caption{Strength of SiC and MgS line emission features
in the Spitzer IRS spectra versus IR colors for the C-AGB stars in the (LMC and SMC) SAGE-S samples.
The line emission data are from \citet{sloan2016}. For each class, the number of objects is shown.
The number in parenthesis denotes the number of the plotted objects with observed data.}
\label{f14}
\end{figure}

\begin{figure*}
\centering
\smallplottwo{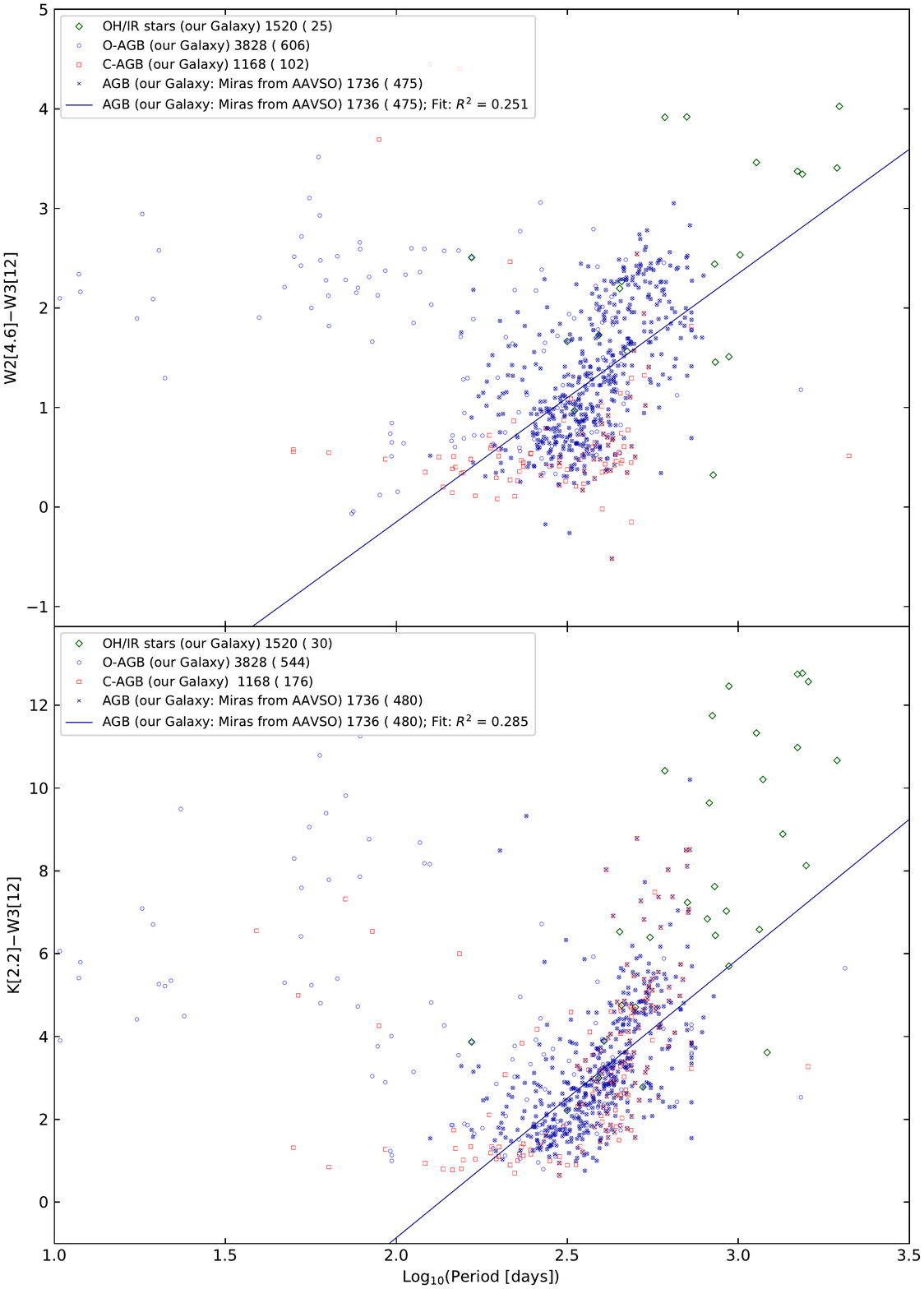}{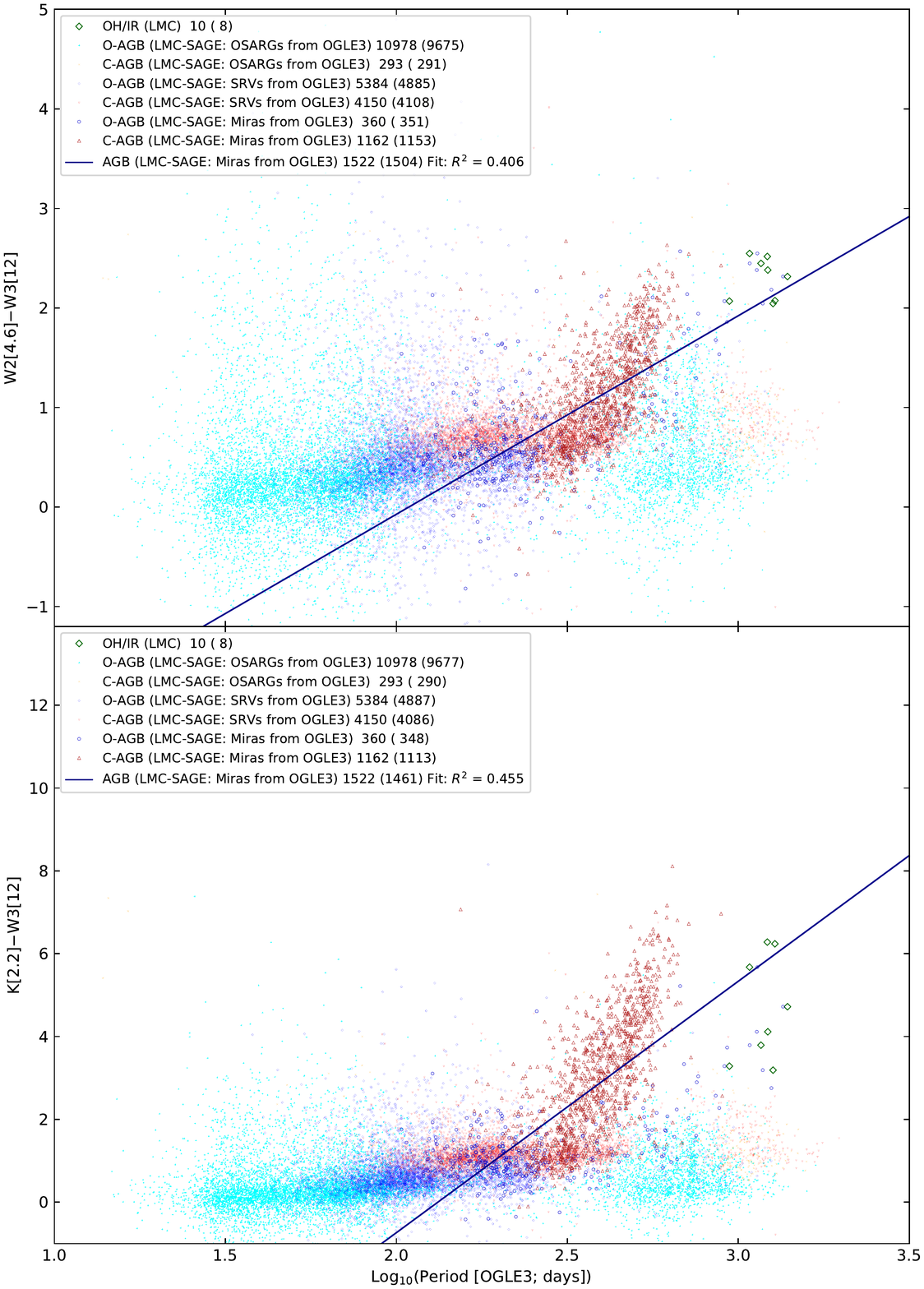}\caption{Period-color relations for
AGB stars in our Galaxy and the LMC. The periods for AGB stars in our Galaxy are from AAVSO.
The periods for OH/IR stars in our Galaxy are from \citet{chen2001}. The
periods for OH/IR stars in the LMC are from \citet{goldman2017}.
For each class, the number of objects is shown.
The number in parenthesis denotes the number of the plotted objects with good quality observed data.}
\label{f15}
\end{figure*}

\begin{figure*}
\centering \leavevmode
\xsmallplottwo{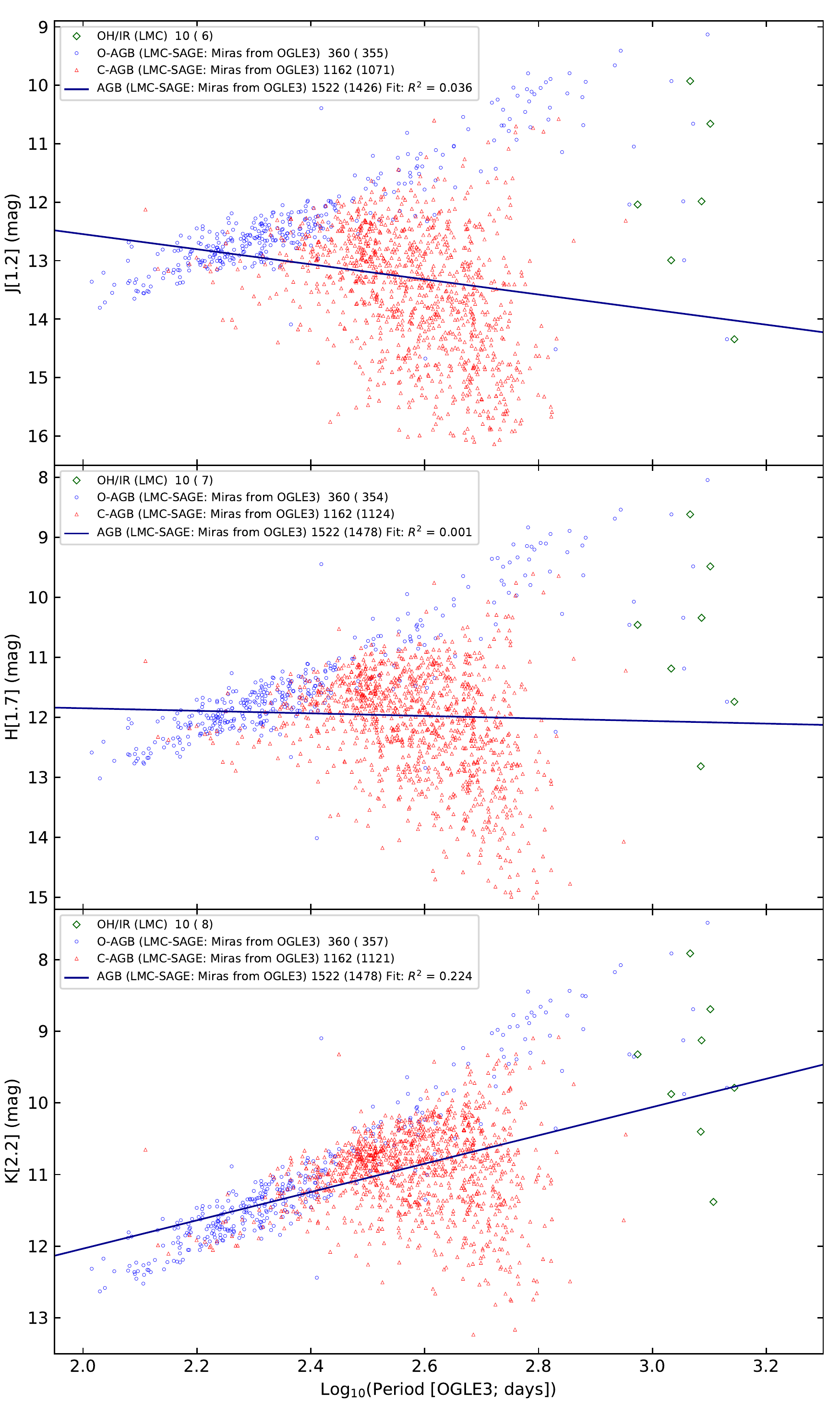}{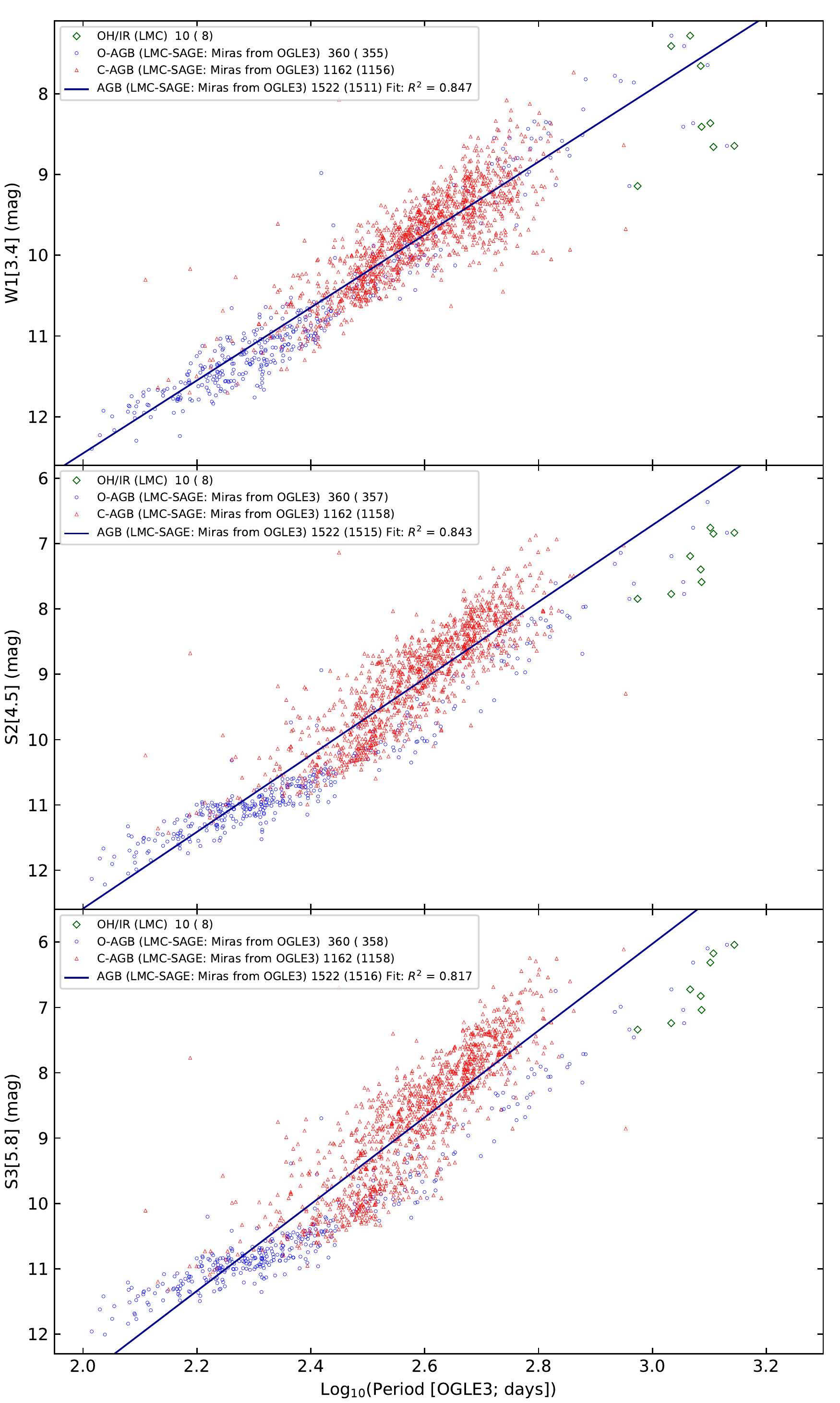} \caption{Period-magnitude relations (at NIR and MIR
bands) for Mira variables in the LMC-SAGE AGB sample. The periods for OH/IR
stars in the LMC are from \citet{goldman2017}. For each class, the number of
objects is shown. The number in parenthesis denotes the number of the
plotted objects with good quality observed flux data. The coefficient of determination
($R^2$) of the linear relationship is also shown.} \label{f16}
\end{figure*}

\begin{figure*}
\centering
\smallplottwo{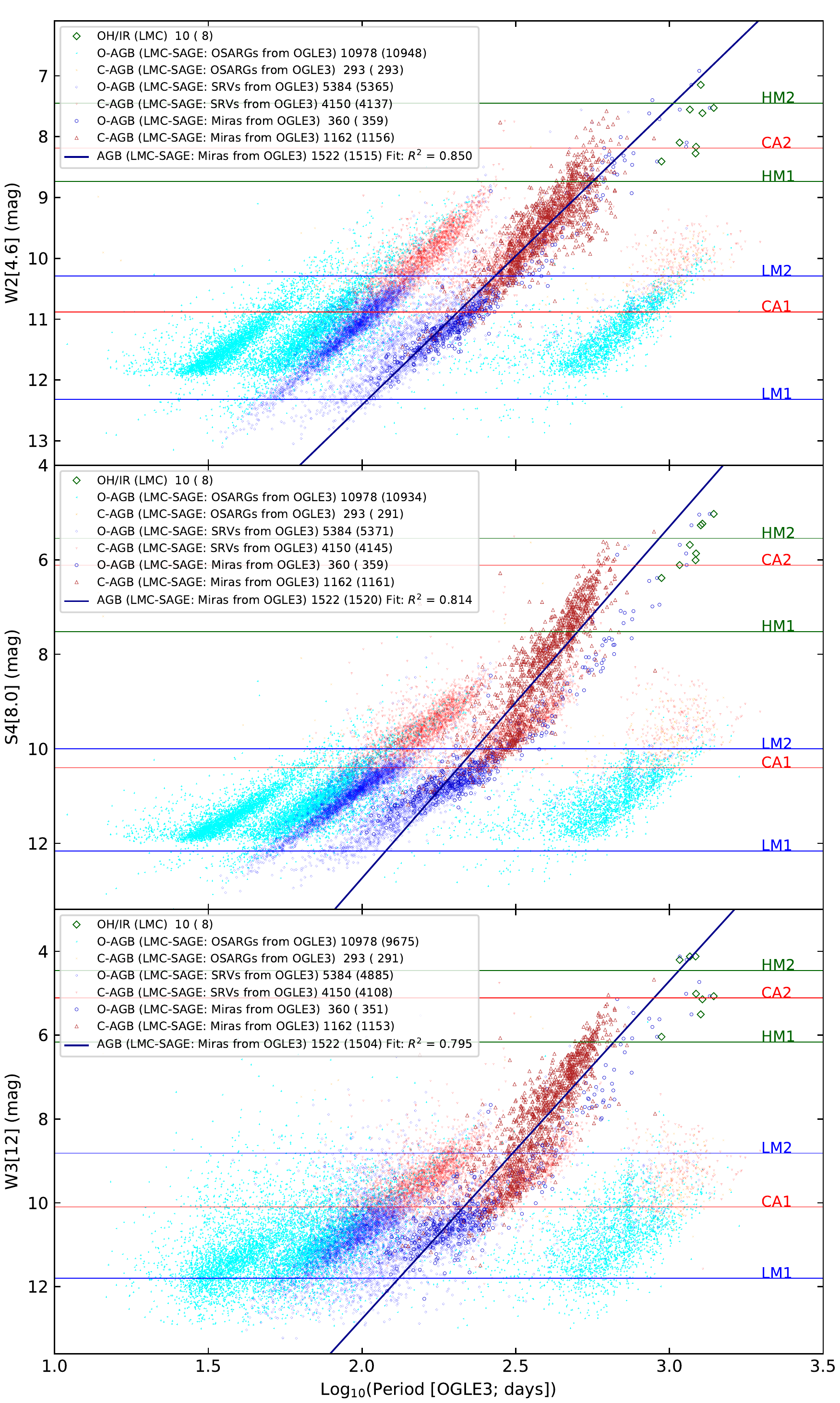}{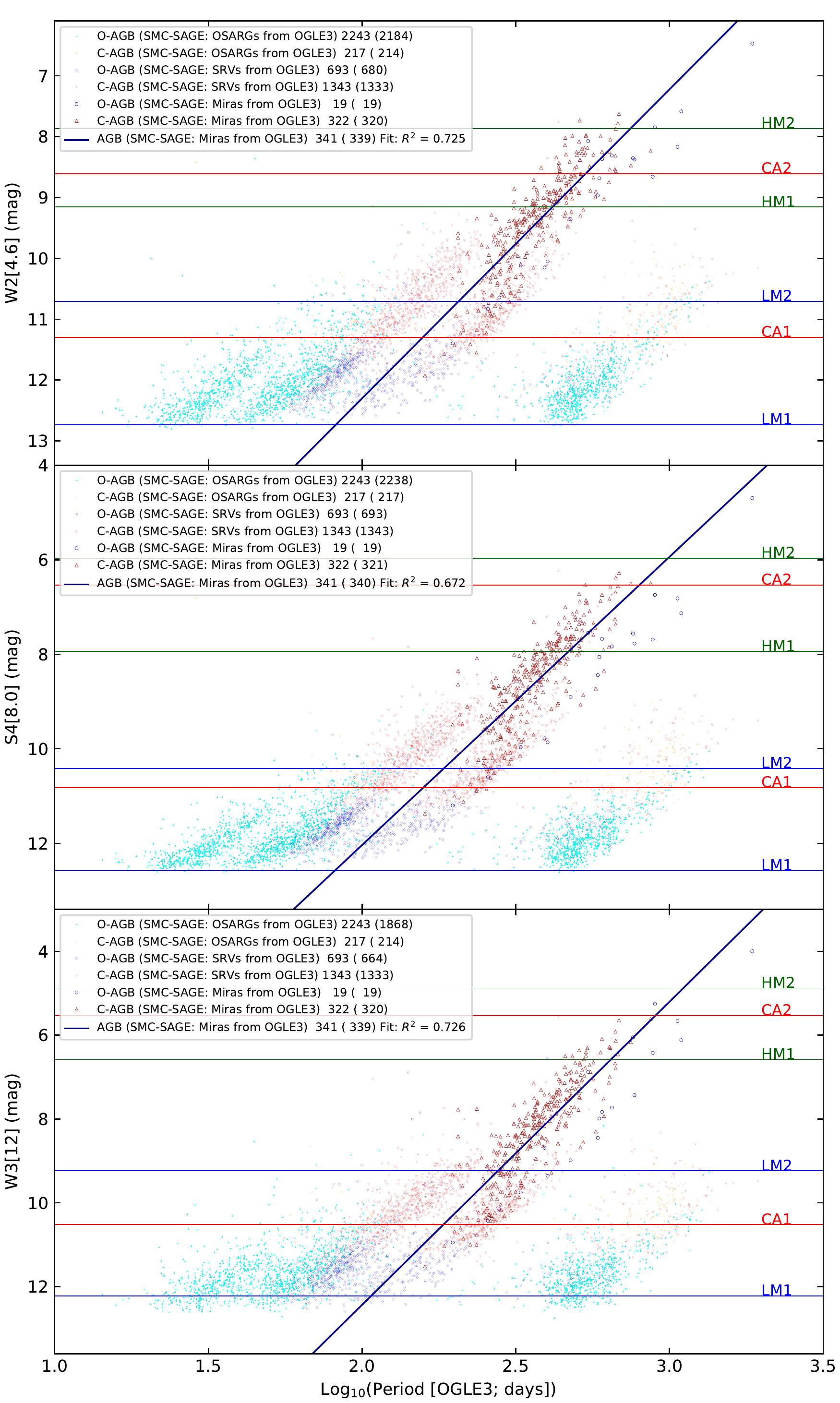}\caption{Period-magnitude relations (at NIR and MIR bands) for
all AGB stars in the SAGE samples (LMC in the left and SMC in the right panel).
The periods for OH/IR stars in the LMC are from \citet{goldman2017}.
For each class, the number of objects is shown.
The number in parenthesis denotes the number of the plotted objects with good quality observed flux data.
The coefficient of determination ($R^2$) of the linear relationship for Mira variables is shown.
The horizontal lines indicate model magnitudes for LMOA (blue), C-AGB (red), and HMOA (green) stars (see Table~\ref{tab:tab5}).}
\label{f17}
\end{figure*}

\begin{figure*}
\centering
\smallplottwo{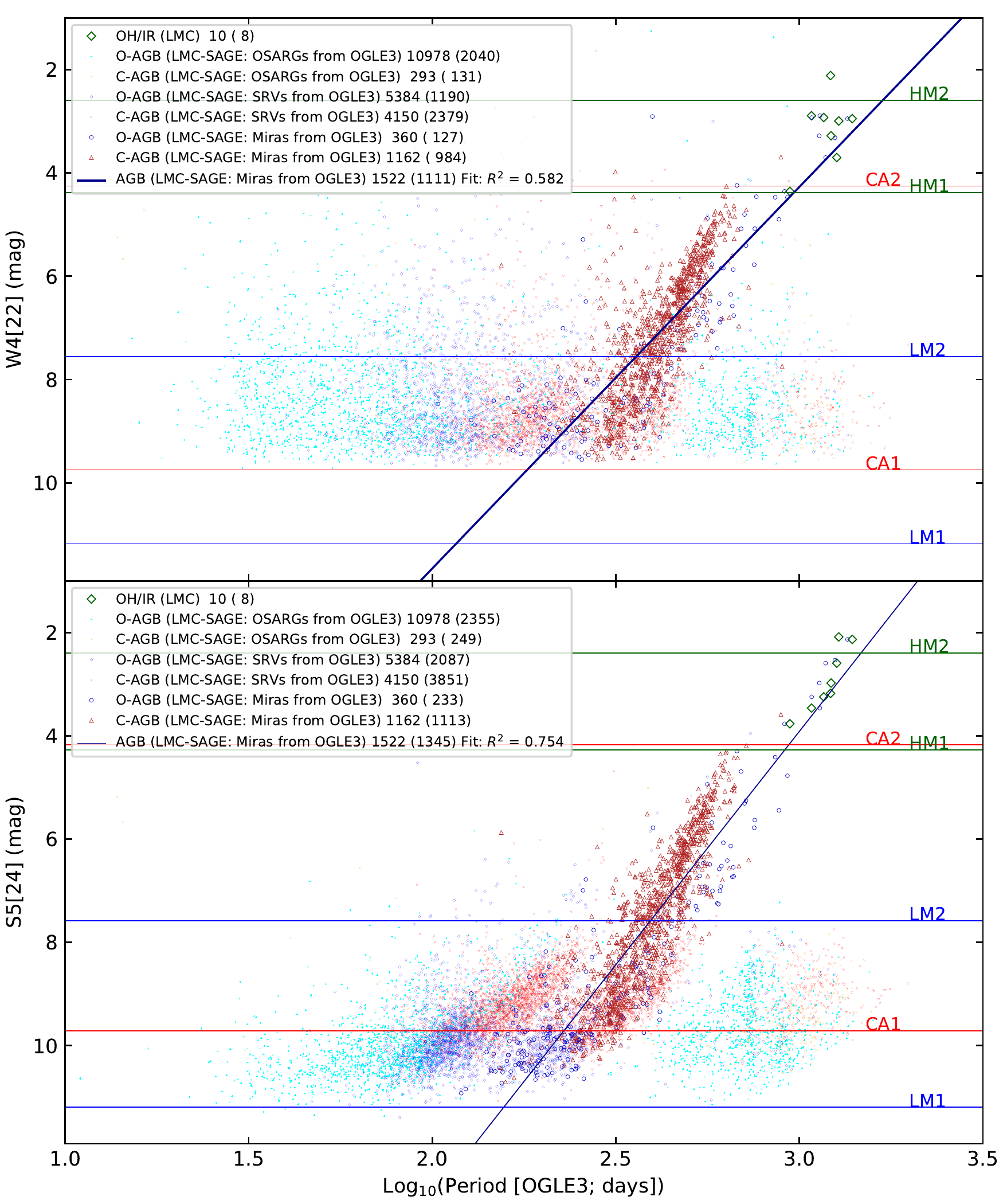}{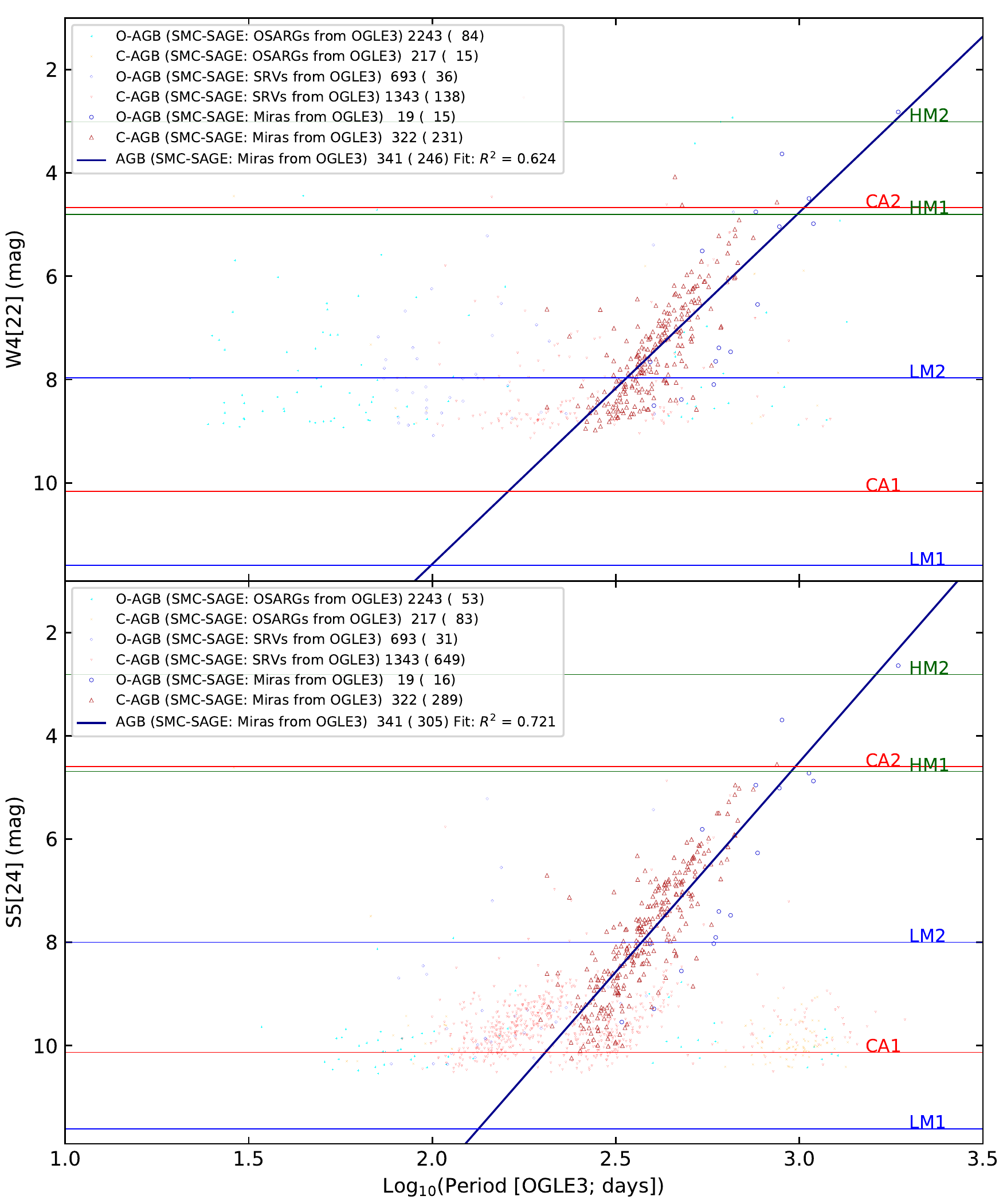}\caption{Period-magnitude relations (at MIR bands) for
all AGB stars in the SAGE samples (LMC in the left and SMC in the right panel).
The periods for OH/IR stars in the LMC are from \citet{goldman2017}.
For each class, the number of objects is shown.
The number in parenthesis denotes the number of the plotted objects with good quality observed flux data.
The coefficient of determination ($R^2$) of the linear relationship for Mira variables is shown.
The horizontal lines indicate model magnitudes for LMOA (blue), C-AGB (red), and HMOA (green) stars (see Table~\ref{tab:tab5}).}
\label{f18}
\end{figure*}

\subsection{WISE-Spitzer 2CDs\label{sec:wise-spitzer}}

Figures~\ref{f7} and~\ref{f8} show Spitzer-WISE 2CDs. Figure~\ref{f7} shows the
2CDs using S3[5.8]-S5[24] versus W2[4.6]-W3[12] for AGB stars in our Galaxy and
the LMC. Figure~\ref{f8} shows the Spitzer-WISE 2CDs using W3[12]-S5[24] versus
S2[4.5]-S4[8.0] for AGB stars in our Galaxy and the LMC. The S5[24] band can be
useful to investigate the Mg$_{0.9}$Fe$_{0.1}$S dust features around 26 $\mu$m
(see Figure~\ref{f11}) for C-AGB stars.

The upper panels of Figures~\ref{f7} and~\ref{f8} show the two WISE-Spitzer
2CDs for our Galaxy, which are less meaningful than other 2CDs because the
usable Spitzer data are available only for a minor portion of AGB stars in our
Galaxy. For C-AGB stars, the sample number is too small to find any meanings.
Though the sample number is small, the observed data points of Galactic O-AGB
stars are located in the wide range O-AGB model colors ($\tau_{10}$ = 0.001 -
40) on the 2CDs.

The lower panels of Figures~\ref{f7} and~\ref{f8} show the two WISE-Spitzer
2CDs for the LMC. The observations of AGB stars in the LMC show similar
properties on both 2CDs though there are more scatters in the W3[12]-S5[24]
color. This could be partly because the W3[12] and S5[24] fluxes are obtained
at different pulsation phases.

For O-AGB stars, the ratio of HMOA stars with thick dust shells ($\tau_{10} >$
7) for the LMC (SAGE-S: 8.1-29.0 \%; SAGE: 0.1-0.2 \%) is smaller than the one
for our Galaxy (13.8-67.6 \%) (see Table~\ref{tab:tab3}). Observations of C-AGB
stars in the LMC-SAGE-S sample can be reproduced by wide ranges of the C-AGB
dust model colors ($\tau_{10}$ = 0.001 - 5) using AMC, SiC, and
Mg$_{0.9}$Fe$_{0.1}$S dust grains. For C-AGB stars, the effect of the
Mg$_{0.9}$Fe$_{0.1}$S and SiC dust, which shows deviations from the pure AMC
model, is more conspicuous on the 2CD using the W3[12]-S5[24] color (see
Figure~\ref{f8}).

Figure~\ref{f9} shows the two WISE-Spitzer 2CDs for AGB stars in the SMC. For
C-AGB stars, the ratio of C-AGB stars with thick dust shells ($\tau_{10} >$ 1)
for the SMC (SAGE-S: 2.5-2.7 \%; SAGE: 0.1-0.2 \%) is smaller than the one for
LMC (SAGE-S: 13.3-13.9 \%; SAGE: 0.6 \%) (see Table~\ref{tab:tab3}).

\subsection{SiC and MgS features for C-AGB stars\label{sec:SiC}}

For C-AGB stars, the SiC dust features at 11.3 $\mu$m and Mg$_{0.9}$Fe$_{0.1}$S
dust features at 28 $\mu$m can be useful to compare the theoretical models with
the observations (see the lower panel of Figure~\ref{f10}). For the SiC dust
feature at 11.3 $\mu$m, the emission feature becomes stronger as $\tau_{10}$
increase up to $\tau_{10}$ = 0.1, then becomes weaker emission feature, and
then it becomes an absorption feature for $\tau_{10} >$ 1. On IR 2CDs, the
observation of C-AGB stars show similar effects.

Figure~\ref{f14} shows the strength of SiC and MgS line emission features in
the Spitzer IRS spectra versus IR colors (W2[4.6]-W3[12] and S2[4.5]-S4[8.0])
for the C-AGB stars in the (LMC and SMC) SAGE-S samples. The plots in
Figure~\ref{f14} show the similar effects: the strength become stronger as the
color gets redder up to some point, then it becomes weaker.

For C-AGB stars, the theoretical models using AMC dust with a mixture of SiC
and Mg$_{0.9}$Fe$_{0.1}$S grains can reproduce the observations in much wider
regions on any IR 2CDs (see Figures~\ref{f4} -~\ref{f9}).

\section{Infrared properties of known pulsating variables\label{sec:pul}}

AGB stars are characterized by long-period and large amplitude pulsations. It
is generally believed that more evolved (or more massive) AGB stars would have
the larger pulsation amplitudes, longer pulsation periods, and higher mass-loss
rates (e.g., \citealt{debeck2010}; \citealt{sk2013b}).

In the sample of 4996 Galactic AGB stars, there are 1736 Miras identified from
AAVSO (see Section~\ref{sec:gagb}). In the sample of 33,537 LMC-SAGE AGB stars,
there are 22,327 pulsating variables identified from the OGLE3 (1522 Miras,
9534 SRVs, and 11,271 OSARGs). And in the sample of 5742 SMC-SAGE AGB stars,
there are 4837 pulsating variables known from the OGLE3 (341 Miras, 2036 SRVs,
and 2460 OSARGs) (see Section~\ref{sec:magb-c}). In this section, we
investigate infrared properties of the known pulsating variables in our Galaxy
and the Magellanic clouds (SAGE sample).

\subsection{Period-color relations\label{sec:period-color}}

Figure~\ref{f15} shows period-color relations. It shows K[2.2]-W3[12] and
W2[4.6]-W3[12] colors versus pulsation periods for AGB stars in our Galaxy and
the LMC. The left and right panels show the relation for AGB stars in our
Galaxy and the LMC, respectively. Though there are large scatters, we find that
Mira variables, among all types of variables, show stronger relationship
between IR colors and pulsation periods.

During the AGB phase, the more evolved stars with longer pulsation periods
would have thicker dust envelopes and redder IR colors. Compared with our
Galaxy, we find that Mira variables in the LMC show larger coefficients of
determination ($R^2$) for both IR colors, which mean higher strength of the
relationship. This could be because most of the Galactic AGB stars with thick
dust shells are not listed in the AAVSO catalog (see Section~\ref{sec:gagb}).
The AAVSO catalog is mainly based on optical observations, which suffer severe
extinctions due to the Galactic disk.

\subsection{Period-magnitude relations\label{sec:pmr}}

Compared with AGB stars in our Galaxy, it is easier to study the
period-magnitude relation (PMR) for the AGB stars in the Magellanic Clouds
because they share similar distances. Among all types of pulsating variables,
it is known that Mira variables show a stronger relationship between the IR
fluxes and pulsation periods (e.g., \citealt{sus09}).

Figure~\ref{f16} shows the relation for Miras in the LMC-SAGE sample at NIR and
MIR bands. Though the relation shows very large scatters at the 2MASS bands
(J[1.2], H[1.7], and K[2.2]) and shorter wavelengths, the Miras show a strong
linear relationship when the wavelength is longer than about 3 $\mu$m (W1[3.4],
S2[4.5], and S3[5.8]). We find that Mira variables in the LMC show fairly large
coefficients of determination ($R^2$=0.6-0.85) of the linear relationship at
the wavelength bands in the range 3-24 $\mu$m.

Figure~\ref{f17} shows the PMRs at NIR and MIR bands (W2[4.6], S4[8.0], and
W3[12]) for SAGE AGB sample stars in the Magellanic Clouds that are identified
as pulsating variables from OGLE3. The plots show the objects of different
variable types and chemical classifications. Again, we find that the Mira
variables in the LMC and SMC show fairly strong linear relationships.

Figure~\ref{f18} shows the PMRs at longer wavelength bands (W4[22] and S5[24]).
In the two plots, low brightness O-AGB stars are relatively deficient compared
with the three plots at shorter wavelengths (see Figure~\ref{f15}). This would
be because of the lower sensitivity of the detectors (W4[22]: 5.4 mJy, S5[24]:
0.11 mJy; see Section~\ref{sec:photdata}). Because of the even lower
sensitivity, the W4[22] band cannot detect more low brightness objects (dimmer
than about 9.5 mag). Therefore, the upper panel is more deficient in low
brightness O-AGB stars (mostly OSARGs). We expect that there would be as many
low brightness O-AGB stars as those in Figure~\ref{f18} if we had detectors
with higher sensitivities.

\section{Magnitude distributions at MIR bands for AGB stars in the Magellanic Clouds\label{sec:br1}}

Magnitude distributions at MIR bands for different classes of AGB stars can be
useful to study the nature of the galaxy. Compared with Galactic AGB stars, it
is much easier to investigate the magnitude distributions for the AGB stars in
the Magellanic Clouds because they share similar distances and they are
relatively freer from interstellar extinctions. In Section~\ref{sec:pmr}, we
presented PMRs at IR bands for known pulsating variables. In this section, we
present the brightness distributions at MIR bands for all AGB stars in the SAGE
samples of the LMC and SMC and compare them with the theoretical model
magnitudes for typical AGB stars.

\begin{figure}
\centering
\smallplot{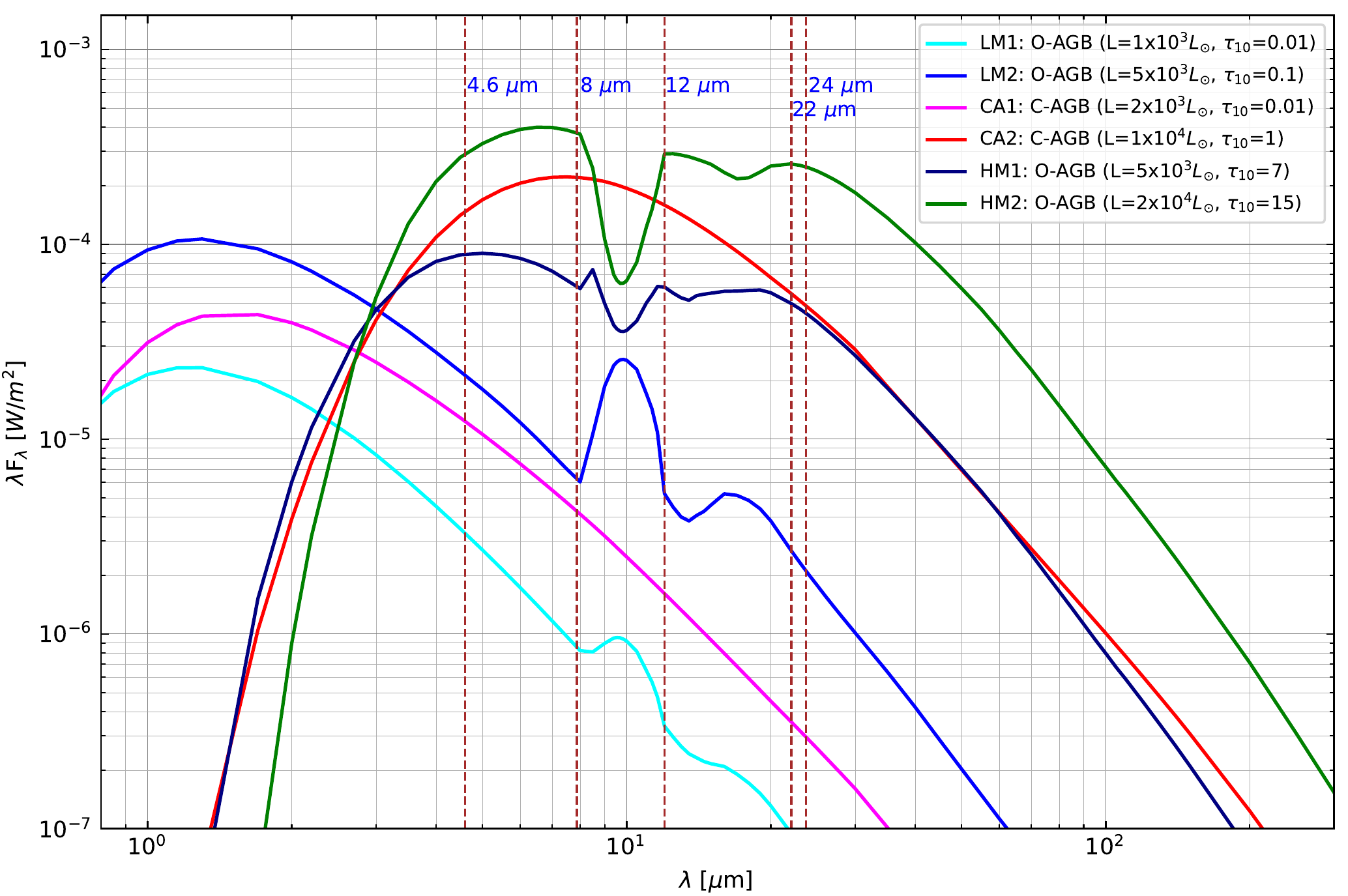}\caption{Model SEDs of the six model for LMOA, C-AGB, and HMOA stars (see Table~\ref{tab:tab5})
using the distance of 1 pc.}
\label{f19}
\end{figure}

\begin{figure*}
\centering
\smallplottwo{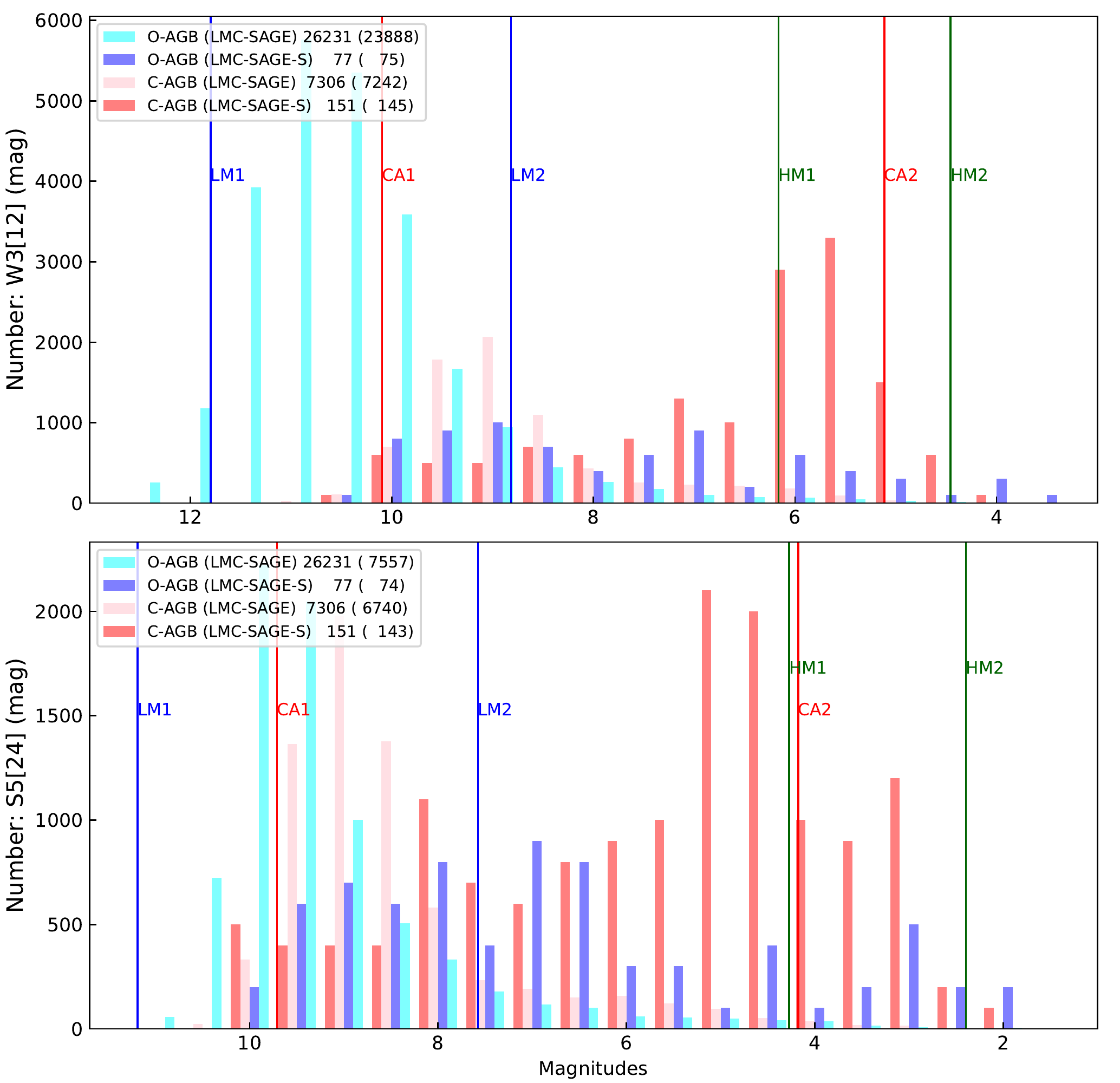}{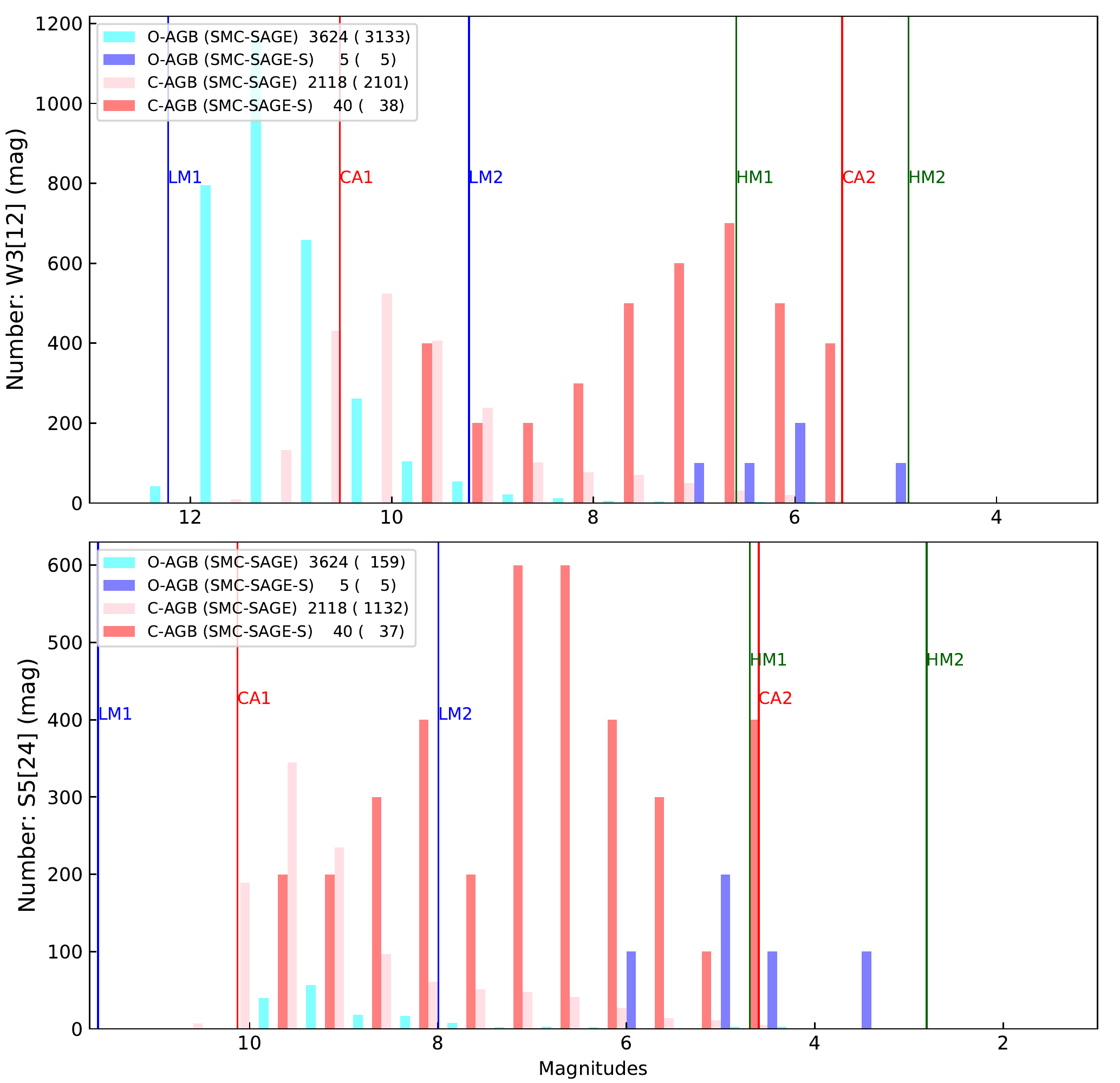}\caption{Magnitude distributions at MIR bands for
all AGB stars in the SAGE samples (LMC in the left and SMC in the right panel).
For each class, the number of objects is shown.
The number in parenthesis denotes the number of the observed objects.
The numbers of the observed objects in the SAGE-S samples have been multiplied by 100 for the plots.
The vertical lines indicate model magnitudes for LMOA (blue), C-AGB (red), and
HMOA (green) stars (see Table~\ref{tab:tab5}).} \label{f20}
\end{figure*}

\subsection{Theoretical model magnitudes\label{sec:modelm}}

We may obtain the theoretical model magnitude at given wavelength band from the
model SED using the ZMF at the wavelength band (see Table~\ref{tab:tab2}).
Because the shape of the model SED is independent on the luminosity of the
central star when all other parameters are fixed, the DUSTY code
(\citealt{ivezic1997}) calculates the model SED only in relative scale (see
Section~\ref{sec:agbmodels}). To calculate the model SED in absolute scale for
given luminosity and distance, we use the RADMC-3D code
\url{http://www.ita.uni-heidelberg.de/~dullemond/software/radmc-3d/}) in
conjunction with the DUSTY code. The RADMC-3D models using the same scheme as
used by \citet{sk2013a} and the same model parameters as the DUSTY model
produce almost identical results.

We obtain the model magnitudes from the model SED in absolute scales for given
luminosity of the central star and distance. For the RADMC-3D models, we use
the same model parameters as the DUSTY model used in this work (see
Section~\ref{sec:agbmodels}) except for the luminosity of the central star.
Table~\ref{tab:tab5} lists the parameters for six models of typical AGB star
with various luminosity of the central star. Figure~\ref{f19} shows the model
SEDs for the six models for LMOA stars, C-AGB stars, and HMOA stars in absolute
scales when we use the distance of 1 pc.

In Figures~\ref{f17} and~\ref{f18}, the horizontal lines indicate the model
magnitudes for the six different models for LMOA, C-AGB, and HMOA stars (see
Table~\ref{tab:tab5} and Figure~\ref{f19}). We assume that the distances of the
LMC and SMC are 49.97 and 60.6 kpc, respectively.

\begin{table}
\caption{Six models for typical AGB stars\label{tab:tab5}}
\centering
\begin{tabular}{llllll}
\hline \hline
Model &Class &Dust$^1$ &$\tau_{10}$ & $T_*$ (K) & $L_* (10^{3} L_{\odot}$) \\
\hline
LM1  & LMOA  &silicate  &0.01 & 3000  & 1 \\
LM2  & LMOA  &silicate  &0.1  & 2500  & 5 \\
CA1  & C-AGB &AMC       &0.1  & 2500  & 2 \\
CA2  & C-AGB &AMC       &1    & 2000  & 10 \\
HM1  & HMOA  &silicate  &7    & 2000  & 5 \\
HM2  & HMOA  &silicate  &15   & 2000  & 20 \\
\hline
\end{tabular}
\begin{flushleft}
\scriptsize
$^1$See Section~\ref{sec:agbmodels} for details. For all models, $T_c$ = 1000 K.
\end{flushleft}
\end{table}

\subsection{Magnitude distributions at MIR bands\label{sec:br2}}

Figure~\ref{f20} shows magnitude distributions at two MIR bands (W3[12] and
S5[24]) for all AGB stars in the SAGE samples (LMC and SMC). We choose the two
MIR bands (W3[12] and S5[24]) because they can be good measures of the
luminosity for the objects with thick dust shells. The vertical lines indicate
the model magnitudes for the six different models for LMOA, C-AGB, and HMOA
stars (see Table~\ref{tab:tab5}).

Table~\ref{tab:tab6} lists the percentages of bright stars at MIR bands. The
weighted averaged percentages of bright O-AGB stars (brighter than model HM2)
in the LMC and SMC are 1.76 \% and 0.07 \%, respectively. And the weighted
averaged percentages of bright C-AGB stars (brighter than model CA2) in the LMC
and SMC are 8.33 \% and 0.97 \%, respectively.

Based on the magnitudes at wavelength bands in the range 4-24 $\mu$m (see
Figures~\ref{f17} -~\ref{f20}), we may roughly divide the AGB stars in the LMC
and SMC into three groups: (1) a large group of low brightness O-AGB stars
(LMOA stars) including most of the OSARGs, (2) a large group of intermediate
brightness C-AGB stars, and (3) a small group of bright O-AGB stars (HMOA
stars) including OH/IR stars. See also Table~\ref{tab:tab7}.

We find that the LMC is deficient in the O-AGB stars that are bright at MIR
bands and the SMC is more deficient in those stars. Though it can not be
confirmed in this work because it requires distance information for the large
sample of Galactic AGB star, we expect that there would be abundant O-AGB stars
that are bright at MIR bands in our Galaxy. 1520 OH/IR stars are identified in
our Galaxy (see Section~\ref{sec:ohir}), from which many of them are known to
be as bright as 3$\times10^{4} L_\odot$ at maximum phases (e.g.,
\citealt{suh2004}; \citealt{sk2013a}). We expect that there would be more
bright O-AGB stars including many extreme OH/IR stars in the third group if we
could make similar plots (see Figures~\ref{f17} -~\ref{f20}) for AGB stars in
our Galaxy.

\begin{table}
\scriptsize
\caption{Percentages$^1$ of bright AGB stars at MIR bands. See Figure~\ref{f20}.\label{tab:tab6}}
\centering
\begin{tabular}{lllll}
\hline \hline
Class               & W3[12]        & S5[24]      & Average & WA$^2$\\
\hline
C-AGB (LMC-SAGE)    & 0.29 (7242)  & 0.65 (6740)   & 0.46  & 8.33 \\
C-AGB (LMC-SAGE-S)  & 6.21 (145)   & 18.18 (143)   & 12.15 & - \\
C-AGB (SMC-SAGE)    & 0.10 (2101)  & 0.18 (1132)   & 0.12  & 0.97 \\
C-AGB (SMC-SAGE-S)  & 0.00 (38)    & 2.70 (37)     & 1.33  & - \\
\hline
O-AGB (LMC-SAGE)    & 0.042 (23888) & 0.066 (7557) & 0.048 & 1.76 \\
O-AGB (LMC-SAGE-S)  & 6.67 (75)     & 4.05 (74)    & 5.37  & - \\
O-AGB (SMC-SAGE)    & 0.064 (3133) & 0.63 (159)    & 0.091 & 0.07 \\
O-AGB (SMC-SAGE-S)  & 0.00 (5)     & 0 (5)         & 0.00  & - \\
\hline
\end{tabular}
\begin{flushleft}
$^1$The percentages of the C-AGB (O-AGB) stars that are brighter than model CA2 (HM2) (see Table~\ref{tab:tab5}).
$^2$The weighted averaged percentages for the LMC and SMC, which assumes that the numbers
of observed objects in the SAGE-S samples are multiplied by 100.
The number in parenthesis denotes the number of the observed objects plotted in Figure~\ref{f20}.
\end{flushleft}
\end{table}

\section{Discussion: AGB stars in our Galaxy and the Magellanic Clouds\label{sec:agba}}

Compared with our Galaxy, we find that the LMC and SMC are deficient in O-AGB
stars with thick dust shells (or large dust optical depths) on any IR 2CDs (see
Section~\ref{sec:comparison}). The weighted averaged percentages of HMOA stars
with thick dust shells ($\tau_{10} >$ 7) for our Galaxy (17.2 \%) is larger
than the ones for the LMC (3.4 \%) and SMC (1.8 \%) SAGE sample stars (see
Table~\ref{tab:tab4}). This could be because the high-mass star formation is
less active in the Magellanic Clouds than in our Galaxy. O-AGB stars with thick
dust shells or OH/IR stars are generally considered to be more massive O-AGB
stars. Up to now, 1520 OH/IR stars are identified in our Galaxy, 10 OH/IR stars
are identified in the LMC, and no OH/IR star in the SMC is identified yet (see
Section~\ref{sec:ohir}).

Compared with our Galaxy, we find that the LMC and SMC are deficient in C-AGB
stars with thick dust shells on the IR 2CDs (see Section~\ref{sec:comparison}).
The weighted averaged percentages of C-AGB stars with thick dust shells
($\tau_{10} >$ 1) for our Galaxy (17.1 \%) is larger than the ones for the LMC
(9.4 \%) and SMC (1.3 \%) SAGE sample stars (see Table~\ref{tab:tab4}). This is
in accord with \citet{ventura2016} who studied infrared properties of C-AGB
stars in the Magellanic Clouds found that the infrared colors of C-AGB stars in
the LMC are redder compared to their counterparts in the SMC.

It is believed the stars with initial masses in the intermediate mass range can
become C-AGB stars (see Section~\ref{sec:intro}). Compared with the LMC, the
SMC is more deficient in C-AGB stars with thick dust shells. Also
\citet{nanni2019} found a group of C-AGB stars with high mass-loss rates in the
LMC that is not present in the SMC. This could be because the initial masses of
C-AGB stars in the LMC are larger than those in the SMC
(\citealt{ventura2016}).

In the study of the magnitude distributions at MIR bands for AGB stars in the
Magellanic Clouds, we find that the LMC is deficient in AGB stars that are
bright at MIR bands and the SMC is more deficient in those stars (see
Section~\ref{sec:br2} and Table~\ref{tab:tab6}). Brighter AGB stars at MIR
bands are generally considered to be more massive AGB stars with thick dust
shells.

Compared with our Galaxy, the LMC looks to be deficient in O-AGB and C-AGB
stars that are bright at MIR bands and have thick dust shells. And the SMC
looks to be more deficient in those stars. This could be because Magellanic
Clouds are more metal poor than our Galaxy and the LMC is more metal rich than
the SMC. It is known that the metallicity, gaseous content, and historical star
formation rate of the LMC lies midway between those of our Galaxy and the SMC.
The LMC's metallicity is about 50 \% that of the Sun while the SMC's is only
about 20 \% (e.g., \citealt{madden2013}; \citealt{hofner2018}). In a galaxy
with a higher metallicty (and a lower gas-to-dust ratio), the star formation in
a higher mass range would be more active and the galaxy would have a higher
ratio of AGB stars with thick dust shells, which are bright at MIR bands.

Table~\ref{tab:tab7} summarizes overall distributions of AGB classes in our
Galaxy and the Magellanic Clouds based on our studies of the IR 2CDs (see
Section~\ref{sec:comparison}) and magnitudes distributions at MIR bands (see
Section~\ref{sec:br1}).

\begin{table}
\scriptsize
\caption{Overall distributions of AGB classes based on the IR 2CDs and IR magnitudes \label{tab:tab7}}
\centering
\begin{tabular}{llllll}
\hline \hline
Class  &Luminosity &Dust & Our Galaxy	& LMC & SMC  \\
(Mass$^1$) &($10^4$ $L_{\odot}$) &($\tau_{10}$) &	&  &   \\
\hline
LMOA  & less luminous &silicate  & AB 	& AB	&	AB	\\
(0.5-1.55) &(0.1-0.5) &(0.001-3)    &     &   &     \\
LMOA:C$^2$  & - &$<$ 0.1  & $^4$27.1 \%	& 79.5 \%	&  82.7 \% 	\\
LMOA:M$^3$  & L $<$ LM2  &-  & AB$^5$ 	& $^6$83.0 \%	& $^6$85.0 \%	\\
\hline
C-AGB      & luminous &AMC       & AB	&	DH 	&	MDH \\
(1.55-4)   &(0.2-1)   &(0.001-5)   &          &               & \\
C-AGB:C$^2$     & -  & $>$ 1  & 17.1 \% & 9.4 \% & 1.3 \% \\
C-AGB:M$^3$      & L $>$ CA2     & -   & AB$^5$      & 8.33 \%	&	0.97 \% \\
\hline
HMOA   & more luminous  &silicate  &AB	& DH & MDH \\
(4-10)      &(0.5-3) &(3-40) &   &  & \\
HMOA:C$^2$    & - & $>$ 7   &	 17.2 \%   & 3.4 \%	&	1.8 \% \\
HMOA:C$^2$    & - & $>$ 15  &  7.0 \%	&	0.44 \%	&	0.048 \% \\
HMOA:M$^3$    & L $>$ HM2   &  -    & AB$^5$ & 1.76 \%	&	0.07 \% \\
\hline
\end{tabular}
\begin{flushleft}
Acronyms; AB: abundant, DH: deficient in high-mass stars, MDH: more deficient in high-mass stars.
$^1$the mass range in $M_{\odot}$ (see Section~\ref{sec:intro}). $^2$Based on
the IR 2CDs (see Section~\ref{sec:comparison} and Table~\ref{tab:tab4}). $^3$Based on the magnitude
distributions at MIR bands (see Section~\ref{sec:br1} and Table~\ref{tab:tab6}). $^4$This could be due
to a selection effect (see Section~\ref{sec:cnumber}). $^5$This can not be
confirmed in this work because it requires distance information for the large sample of the Galactic AGB
stars (see Section~\ref{sec:br2}). $^6$Obtained from Figure~\ref{f20} at the MIR band W3[12] because the S5[24] band would
not detect all of the dim AGB stars with thin dust shells (see Section~\ref{sec:pmr}).
\end{flushleft}
\end{table}

\section{summary\label{sec:sum}}

We have investigated infrared properties of AGB stars in our Galaxy and the
Magellanic Clouds using various infrared observational data and theoretical
models. We have used catalogs for the sample of 4996 AGB stars in our Galaxy
and about 39,000 AGB stars in the Magellanic Clouds from the available
literature.

For each object in the sample, we have cross-identified the 2MASS, WISE, and
Spitzer counterparts. To compare the physical properties of O-AGB and C-AGB
stars in our Galaxy and the Magellanic Clouds, we have presented IR 2CDs by
using the 2MASS, WISE, and Spitzer photometric data.

The IR 2CDs are useful to compare IR properties of AGB stars in our Galaxy and
the Magellanic Clouds with theoretical models. For AGB stars in our Galaxy, the
IR 2CDs using Spitzer photometric data are less useful because the reliable
data are available only for a small number of objects.

We have performed radiative transfer model calculations for AGB stars using
various possible parameters of central stars and spherically symmetric dust
shells. We have compared the various theoretical models with the observations
of AGB stars on the IR 2CDs.

All AGBs stars in our Galaxy and the Magellanic Clouds look similar in dust
properties. The theoretical dust shell models can roughly explain the
observations of AGB stars in our Galaxy and the Magellanic Clouds on various IR
2CDs using dust opacity functions of amorphous silicate and amorphous carbon.
For LMOA stars, the silicate dust with a mixture of amorphous alumina
(Al$_2$O$_3$) and Fe-Mg oxides can explain wider regions on the IR 2CDs. For
C-AGB stars, AMC dust with a mixture of SiC and Mg$_{0.9}$Fe$_{0.1}$S grains
can reproduce the observations in much wider regions on the 2CDs.

The observed K[2.2]-W3[12] colors for AGB stars are bluer than the theoretical
dust shell models, which do not consider gas-phase radiation processes. This
`bluing' effect looks to be stronger for AGB stars in the LMC than for those in
our Galaxy. The `bluing' effect could be due to circumstellar (or interstellar)
molecules and/or inadequate dust opacity, but further investigations on the NIR
spectra of AGB stars are necessary to clarify it.

Compared with our Galaxy, we have found that the Magellanic Clouds are
deficient in O-AGB and C-AGB stars with thick dust shells on the IR 2CDs. AGB
stars with thick dust shells are generally considered to be more massive (or
more evolved). Compared with the LMC, the SMC is more deficient in the AGB
stars with thick dust shells. This could be because the high-mass star
formation is less active in the Magellanic Clouds than in our Galaxy. It is
known that the Magellanic Clouds are more metal poor than our Galaxy and the
LMC is more metal rich than the SMC.

We have investigated period-magnitude relations for known pulsating variables
in the LMC and SMC. The Mira variables show a strong linear relationship at the
wavelength bands in the range 3-24 $\mu$m.

We have investigated the magnitude distributions at MIR bands for AGB stars in
the LMC and SMC. The LMC and SMC look to be deficient in bright O-AGB and C-AGB
stars at MIR bands. Compared with the LMC, the SMC is more deficient in the
bright AGB stars. Again, this could be because they are more metal poor and
less active in high-mass star formation compared with our Galaxy.


\acknowledgments

I thank the anonymous referee for constructive comments and suggestions. This
work was supported by the National Research Foundation of Korea (NRF) grant
funded by the Korea government (MSIT; Ministry of Science and ICT) (No.
NRF-2017R1A2B4002328). This research has made use of the VizieR catalogue
access tool, CDS, Strasbourg, France. This research has made use of the NASA/
IPAC Infrared Science Archive, which is operated by the Jet Propulsion
Laboratory, California Institute of Technology, under contract with the
National Aeronautics and Space Administration.

\bibliographystyle{aasjournal}
\bibliography{reference.bib}

\begin{thebibliography}{}
\expandafter\ifx\csname natexlab\endcsname\relax\def\natexlab#1{#1}\fi
\providecommand{\url}[1]{\href{#1}{#1}}

\bibitem[Begemann et al.(1994)]{begemann1994} Begemann, B., Dorschner, J., Henning, T., Mutschke, H., \& Thamm, E. 1994, ApJ, 423, L71
\bibitem[Begemann et al.(1997)]{begemann1997} Begemann, B., Dorschner, J., Henning, T., et al. 1997, ApJ, 476, 199
\bibitem[Beichman et al.(1988)]{beichman1988} Beichman, C. A., Neugebauer, G., Habing H., Clegg, P. E., \& Chester, T. C. 1988, IRAS Catalogs and Atlases: Explanatory Supplement, NASA RP-1190 (Washington: NASA)
\bibitem[Bl\"{o}cker et al.(2000)]{bloecker2000} Bl\"{o}cker, T., Herwig, F., \& Driebe, T. 2000, MmSAI, 71, 711
\bibitem[Boyer et al.(2011)]{boyer2011} Boyer, M. L., Srinivasan, S., van Loon, J. T., et al. 2011, AJ, 142, 103
\bibitem[Chen et al.(2001)]{chen2001} Chen, P. S., Szczerba, R., Kwok, S., \& Volk, K. 2001, A\&A, 368, 1006
\bibitem[Cohen et al.(2003)]{cohen2003} Cohen, M., Wheaton, W. A., \& Megeath, S. T. 2003, AJ, 126, 1090
\bibitem[Cutri et al.(2003)]{cutri2003} Cutri, R. M., Skrutskie, M. F., Van Dyk, S., et al. 2003, The IRSA 2MASS All-Sky Point Source Catalog, NASA/IPAC Infrared Science Archive
\bibitem[De Beck et al.(2010)]{debeck2010} De Beck, E., Decin, L., de Koter, A., et al. 2010, A\&A, 523, A18
\bibitem[Draine et al.(2007)]{draine2007} Draine, B. T., Dale, D. A., Bendo, G., et al. 2007, ApJ, 663, 866
\bibitem[Gehrz et al.(2007)]{gehrz2007} Gehrz, R. D., Roellig, T. L., Werner, M. W., et al. 2007, RScI, 78, 011302
\bibitem[Goldman et al.(2017)]{goldman2017} Goldman, S. R., van Loon, J. Th., Zijlstra, A. A., et al. 2017, MNRAS, 465, 403
\bibitem[Goldman et al.(2018)]{goldman2018} Goldman, Steven R., van Loon, J. Th., G\'{o}mez, J. F., et al. 2018, MNRAS, 473, 3835
\bibitem[Gonneau et al.(2016)]{gonneau2016} Gonneau, A., Lancon, A., Trager, S. C., et al. 2016, A\&A, 589, A36
\bibitem[Gonz\'{a}lez-L\'{o}pezlira(2018)]{gonzalez2018} Gonz\'{a}lez-L\'{o}pezlira, R. A. 2018, ApJ, 856, 170
\bibitem[Groenewegen et al.(1995)]{groenewegen1995} Groenewegen, M. A. T., van den Hoek, L. B., \& de Jong, T. 1995, A\&A, 293, 381
\bibitem[Groenewegen \& Sloan(2018)]{groenewegen2018} Groenewegen, M. A. T., \& Sloan, G. C. 2018, A\&A, 609, A114
\bibitem[Henning et al.(1995)]{henning1995} Henning, T., Begemann, B., Mutschke, H., \& Dorschner, J. 1995, A\&AS, 112, 143
\bibitem[H{\"o}fner \& Olofsson(2018)]{hofner2018} H{\"o}fner, S., \& Olofsson, H.,  2018, A\&A Rev., 26, 1
\bibitem[Hony et al.(2002)]{hony2002} Hony, S., Waters, L. B. F. M., \& Tielens, A. G. G. M. 2002, A\&A, 390, 533
\bibitem[Iben \& Renzini(1983)]{iben1983} Iben, I., \& Renzini, A. 1983, ARA\&A, 21, 271
\bibitem[Ivezi\'{c} \& Elitzur(1997)]{ivezic1997} Ivezi\'{c}, A., \& Elitzur, M. 1997, MNRAS, 287, 799
\bibitem[Jarrett et al.(2011)]{jarrett2011} Jarrett, T. H., Cohen, M., Masci, F., et al. 2011, ApJ, 735, 112
\bibitem[Jones et al.(2014)]{jones2014} Jones, O. C., Kemper, F., Srinivasan, S., et al. 2014, MNRAS, 440, 631
\bibitem[Jones et al.(2017)]{jones2017} Jones, O. C., Woods, P. M., Kemper, F. et al. 2017, MNRAS, 470, 3250
\bibitem[Kraemer et al.(2017)]{kraemer2017} Kraemer, K. E.; Sloan, G. C.; Wood, P. R.; Jones, O. C.; Egan, M. P, 2017, ApJ,.834,.185
\bibitem[Kwon \& Suh(2012)]{ks2012} Kwon, Y.-J., \& Suh, K.-W. 2012, JKAS, 45, 139
\bibitem[Lan{\c c}on \& Wood(2000)]{lancon2000} Lan{\c c}on, A., \& Wood, P. R. 2000, \aaps, 146, 217
\bibitem[Le Bertre(2005)]{LeBertre2005} Le Bertre, T., Tanaka, M., Yamamura, I., Murakami, H., \& MacConnell, D. J. 2005, PASP, 117, 199
\bibitem[Loup et al.(1993)]{loup1993} Loup, C., Forveille, T., Omont, A., \& Paul, J. F. 1993, A\&AS, 99, 291
\bibitem[Madden et al.(2013)]{madden2013} Madden, S. C., R\'{e}my-Ruyer, A., Galametz, M., et al. 2013, PASP, 125, 600
\bibitem[Meixner et al.(2006)]{meixner2006} Meixner, M., Gordon, K. D., Indebetouw, R., et al. 2006, AJ, 132, 2268
\bibitem[Nanni et al.(2019)]{nanni2019} Nanni, A., Groenewegen, M. A. T., Aringer, B., et al. 2019, MNRAS, 487, 502
\bibitem[P\'{e}gouri\'{e}(1988)]{pegouri1988} P\'{e}gouri\'{e}, B. 1988, A\&A, 194, 335
\bibitem[Riebel et al.(2012)]{Riebel2012} Riebel, D., Srinivasan, S., Sargent, B.,\& Meixner, M. 2012, ApJ, 753, 71
\bibitem[Siess(2006)]{siess2006} Siess, L. 2006, A\&A, 448, 717
\bibitem[Sloan et al.(2016)]{sloan2016} Sloan, G. C., Kraemer, K. E., McDonald, I., et al. 2016, ApJ, 826, 44
\bibitem[Soszy{\'n}ski et al.(2009)]{sus09} Soszy{\'n}ski, I., Udalski, A., Szyma{\'n}ski, M.~K., et al. 2009,  AcA, 59, 239
\bibitem[Soszy{\'n}ski et al.(2011)]{sus11} Soszy{\'n}ski, I., Udalski, A., Szyma{\'n}ski, M.~K., et al. 2011,  AcA, 61, 217
\bibitem[Srinivasan et al.(2016)]{Srinivasan2016} Srinivasan, S., Boyer, M. L., Kemper, F., et al. 2016, MNRAS, 457, 2814
\bibitem[Suh(1999)]{suh1999} Suh, K.-W. 1999, MNRAS, 304, 389
\bibitem[Suh(2000)]{suh2000} Suh, K.-W. 2000, MNRAS, 315, 740
\bibitem[Suh(2002)]{suh2002} Suh, K.-W. 2002, MNRAS, 332, 513
\bibitem[Suh(2004)]{suh2004} Suh, K.-W. 2004, ApJ, 615, 485
\bibitem[Suh(2014)]{suh2014} Suh, K.-W. 2014, JKAS, 47, 219
\bibitem[Suh(2015)]{suh2015} Suh, K.-W. 2015, ApJ, 808, 165
\bibitem[Suh(2016)]{suh2016} Suh, K.-W. 2016, JKAS, 49, 127
\bibitem[Suh(2018)]{suh2018} Suh, K.-W. 2018, JKAS, 51, 155
\bibitem[Suh \& Kwon(2011)]{sk2011} Suh, K.-W., \& Kwon, Y.-J. 2011, MNRAS, 417, 3047
\bibitem[Suh \& Kwon(2013a)]{sk2013a} Suh, K.-W., \& Kwon, Y.-J. 2013a, ApJ, 762, 113
\bibitem[Suh \& Kwon(2013b)]{sk2013b} Suh, K.-W., \& Kwon, Y.-J. 2013b, JKAS, 46, 235
\bibitem[Suh \& Hong(2017)]{sh2017} Suh, K.-W., \& Hong, J. 2017, JKAS, 50, 131
\bibitem[Th. Posch et al.(2001)]{thposch2002} Th. Posch, F., Kerschbaum, H., Mutschke, J., et al. 2002, A\&A, 393, L7
\bibitem[Ventura et al.(2016)]{ventura2016} Ventura, P., Karakas, A. I., Dell'Agli, F., et al. 2016, MNRAS, 457, 1456
\bibitem[Watson et al.(2019)]{watson2019} Watson, C., Henden, A. A., \& Price, A. 2019, yCat, 102027W, 0
\bibitem[Wright et al.(2010)]{wright2010} Wright, E. L., Eisenhardt, P. R. M., Mainzer, A. K., et al. 2010, AJ, 140, 1868
\end{thebibliography}



\end{document}